\documentclass[12pt,a4paper,oneside]{article}
\usepackage{graphicx,amssymb,amsmath,color}
\usepackage[linkcolor={blue},citecolor={red},colorlinks=true]{hyperref}
\usepackage{cite}
\usepackage{nccmath}
\usepackage{relsize}
\usepackage{enumerate}
\usepackage[margin=3 cm]{geometry}
\usepackage{slashed}
\usepackage{multirow}
\usepackage[normalem]{ulem}
\usepackage[dvipsnames, x11names]{xcolor}
\usepackage[compat=1.0.0]{tikz-feynman}
\usepackage[labelfont=bf]{caption}

\usepackage{cleveref}
\usepackage{cancel}

\definecolor{flag}{rgb}{1.0, 0.03, 0.0}

\def\mt{\tilde{m}}

\def\kt{\tilde{k}}
\def\kp{k_\perp}

\def\zr{{\zeta_R}}
\def\zl{{\zeta_L}}

\def\beq{\begin{equation}}
\def\eeq{\end{equation}}
\def\bea{\begin{eqnarray}}
\def\eea{\end{eqnarray}}

\def\bit{\begin{itemize}}
\def\eit{\end{itemize}}

\def\l{\left}
\def\r{\right}
\def\c{\chi}

\def\baa{\begin{array}}
\def\eaa{\end{array}}

\def\d{\partial}

\def\simgt{\mathrel{\lower2.5pt\vbox{\lineskip=0pt\baselineskip=0pt
           \hbox{$>$}\hbox{$\sim$}}}}
\def\simlt{\mathrel{\lower2.5pt\vbox{\lineskip=0pt\baselineskip=0pt
           \hbox{$<$}\hbox{$\sim$}}}}
\newcommand{\vev}[1]{ \langle {#1} \rangle }

\def\bfc{\begin{figure}\begin{center}}
\def\efc{\end{center}\end{figure}}
\def\nn{\nonumber\\}

\definecolor{chromeyellow}{rgb}{1.0, 0.65, 0.0}
\definecolor{darkcoral}{rgb}{0.8, 0.36, 0.27}
\definecolor{cadmiumgreen}{rgb}{0.0, 0.42, 0.24}

\begin{document}

\begin{flushright}
\hspace{3cm} 
SISSA  13/2023/FISI
\end{flushright}
\vspace{.6cm}
\begin{center}

{\Large \bf 
{Quantisation Across Bubble Walls and Friction}}
\vspace{0.5cm}

\vspace{1cm}{Aleksandr Azatov$^{a,b,c,1}$, Giulio Barni$^{a,b,2}$, Rudin Petrossian-Byrne$^{d,3}$, Miguel Vanvlasselaer$^{e,4}$ }
\\[7mm]
 {\it \small
$^a$ SISSA International School for Advanced Studies, Via Bonomea 265, 34136, Trieste, Italy\\
$^b$ INFN - Sezione di Trieste, Via Bonomea 265, 34136, Trieste, Italy\\[0.1cm]
$^c$ IFPU, Institute for Fundamental Physics of the Universe, Via Beirut 2, 34014 Trieste, Italy\\[0.1cm]
$^d$ Abdus Salam International Centre for Theoretical Physics, Strada Costiera 11, 34151, Trieste, Italy\\[0.1cm]
$^e$ Theoretische Natuurkunde and IIHE/ELEM, Vrije Universiteit Brussel,
\& The International Solvay Institutes, Pleinlaan 2, B-1050 Brussels, Belgium\\[0.1cm]}
\end{center}

\bigskip \bigskip \bigskip

\centerline{\bf Abstract}

\begin{quote}
We quantise from first principles 
field theories living on the 
background of a bubble wall in the planar limit with 
particular focus on the case of 
spontaneous breaking of gauge symmetry.  
Using these tools,
we compute the average momentum 
transfer from transition radiation: the 
soft emission of radiation by an energetic 
particle passing across the 
wall, with a particular focus on the longitudinal 
polarisation of vectors. We find these to be comparable to transverse polarisations in symmetry-breaking transitions with mild super-cooling, and dominant in broken to broken transitions with thin wall.
Our results have phenomenological applications for the expansion of bubbles during first order phase transitions. 
Our general framework allows for the robust calculation of any particle processes of interest in such translation breaking backgrounds. 


\end{quote}

\vfill
\noindent\line(1,0){188}
{\scriptsize{ \\ E-mail:
\texttt{$^1$\href{mailto:aleksandr.azatov@NOSPAMsissa.it}{aleksandr.azatov@sissa.it}}, \texttt{$^2$\href{mailto:giulio.barni@NOSPAMsissa.it}{giulio.barni@sissa.it}, \texttt{$^3$\href{mailto:rpetross@NOSPAMictp.it}{rpetross@ictp.it},
\texttt{$^4$\href{mailto:miguel.vanvlasselaer@NOSPAMvub.be}{miguel.vanvlasselaer@vub.be}}}
}}}

\newpage

\newpage
{\hypersetup{linkcolor=black}
\tableofcontents}

\section{Introduction}

First-order phase transitions (FOPTs) were proposed long ago as potentially occurring during the hot big-bang phase of the universe
\cite{old_FOPT_Kirzhnits_1972,old_FOPT_Weinberg1974,Witten:1984rs}. In the past, it was entertained that even within the known Standard Model (SM) of particle physics there might have been as many as two: chiral symmetry breaking/confinement in QCD at temperatures $T\sim 150 \ \rm MeV$ and the spontaneous breaking of electroweak (EW) symmetry at $T \sim 160 \ \rm GeV$. Both are now understood to be smooth crossovers \cite{Bonati:2013tqa,Laine:1998jb}.
In fact, it is interesting to note that from the current laws of physics, there is no conclusively established meta-stable vacuum \textit{for any temperature} at zero chemical potential
\footnote{The closest thing that we are aware of is the instability in the Higgs effective potential for central values of SM parameters when extrapolated to very large field range \cite{Degrassi:2012ry}. However, this is sensitive to possible - though unknown - UV physics, over many orders of magnitude, so that we certainly cannot count it as `conclusive'.}.

By contrast, FOPTs are ubiquitous in beyond the SM (BSM) theories. This is due firstly to a vast richness of important phenomenological consequences, among which baryogenesis\cite{Kuzmin:1985mm, Shaposhnikov:1986jp,Nelson:1991ab,Carena:1996wj,Cline:2017jvp,Long:2017rdo,Bruggisser:2018mrt,Bruggisser:2018mus,Morrissey:2012db,Azatov:2021irb, Baldes:2021vyz, Chun:2023ezg}, the production of heavy dark matter\cite{Falkowski:2012fb, Baldes:2020kam, Azatov:2021ifm,Baldes:2021aph, Asadi:2021pwo, Baldes:2022oev, Baldes:2023fsp}, primordial black holes\cite{10.1143/PTP.68.1979,Kawana:2021tde,Jung:2021mku,Gouttenoire:2023naa,Lewicki:2023ioy} and gravitational waves (GW)\cite{Witten:1984rs,Hogan_GW_1986,Kosowsky:1992vn,Kosowsky:1992rz,Kamionkowski:1993fg} to name a few. 
In particular, the EW phase transition is easily made first order in many BSM models\cite{Grojean:2004xa,Delaunay:2007wb,Ellis:2018mja,Ellis:2019oqb,Bruggisser:2018mus,  Espinosa:2007qk, Beniwal:2017eik,Barger:2007im,Espinosa:2011ax,Kozaczuk:2019pet, Kurup:2017dzf,Azatov:2022tii } and the consequent out-of-equilibrium dynamics (in conjunction with $B$ violation in the SM) still make for an attractive theory of baryogenesis. 
Secondly, our currently most compelling picture of physics at the highest energy scales seems to suggest \textit{landscapes} of countless meta-stable vacua. From this perspective, FOPTs may even be expected during the post-inflationary era \cite{GarciaGarcia:2016xgv}.
Finally, and perhaps most importantly from a phenomenological perspective, the advent of gravitational wave detectors has re-energised interest in these violent phenomena with the prospect of upcoming experiments possibly detecting a stochastic gravitational wave background relic \cite{Witten:1984rs, Espinosa:2010hh,Jinno:2017fby}. Thus even
FOPTs occurring in potential hidden sectors decoupled from the SM and its thermal history become of interest \cite{Fairbairn:2019xog,Breitbach:2018ddu}.

A FOPT proceeds through the nucleation and subsequent expansion of bubbles of new phase,   {pushed by the free-energy density difference} $\Delta V$ between phases. 
If friction from the surrounding matter can be ignored, the bubble wall interpolating between the two phases will continue to expand with constant proper acceleration until they collide with each other, with most of the vacuum energy released thus going into kinetic energy of the walls. 
This scenario is known as \textit{runaway}. If instead friction causes a pressure $\mathcal{P}$ which manages to equilibrate the driving force $\mathcal{P}\simeq\Delta V$, a constant subluminal terminal velocity is reached and energy is efficiently transferred to the medium. 
All phenomenological consequences listed above, for example, the strength and spectral shape of the stochastic GW signal,  depend crucially on the bubble velocity and which of the two regimes is realised. To this end, it becomes
important to understand precisely the dynamics of an expanding domain wall in medium\footnote{In this work by domain wall we simply mean any bubble wall interpolating between different phases of a theory in the planar limit.}.
The analysis of bubble-medium interactions is a complicated problem which is largely still under investigation \cite{Moore:1995ua, Moore:1995si,Laurent:2022jrs, Laurent:2020gpg, DeCurtis:2022hlx, DeCurtis:2023hil,Mancha:2020fzw, Balaji:2020yrx, Ai:2021kak, Ai:2023see,Dine:1992wr,Bodeker:2009qy, Bodeker:2017cim, Hoeche:2020rsg, Azatov:2020nbe, Gouttenoire:2021kjv}. Friction is expected in 
general to be an involved function of bubble velocity $v_w$ and the surrounding degrees of freedom (d.o.f.). A distinction can be made however between low $v_w$, when a fluid description is most appropriate, 
and the ultra-relativistic regime, $\gamma_w \equiv 1/\sqrt{1-v_w^2} \gg L_w\Gamma_{\rm int.}$, where $L_w$ is the wall thickness in its rest frame and $\Gamma_{\rm int.}$ is the interaction rate between particles in medium \cite{BarrosoMancha:2020fay}. 
We will focus on the latter regime in this work, where the wall can be said to be interacting with individual particles.

\begin{figure}
    \centering
    \includegraphics[width=\textwidth]{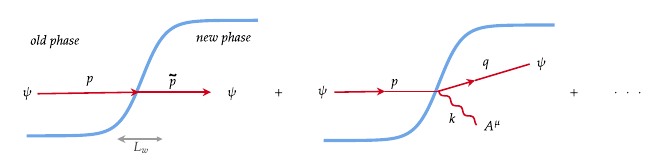}
    \caption{Diagrams corresponding to LO and NLO processes contributing to friction on a moving bubble wall. The emission of vectors with changing mass is generally the dominant $1\rightarrow 2$ process for friction and the subject of this work. Tilde distinguishes objects in the new phase.}
    \label{fig:cartoon}
\end{figure}

A particle hitting the wall from the old phase can undergo many processes, which can be organised in terms of a perturbative expansion in couplings of the theory defined in the background of the wall profile, as sketched in \cref{fig:cartoon}. 
The spontaneous breaking of translation symmetry means that momentum perpendicular to the wall is no longer conserved. 
The average momentum lost $\langle \Delta p \rangle$ times the flux of incoming particles is then the pressure opposing the bubble's expansion. It is most convenient to work in the rest frame of the wall.
At leading order (LO) incoming particles either cross the wall or reflect. It is easy to show that when reflections can be neglected \footnote{Although these can be important and even dominant for intermediate relativistic $\gamma_w$ \cite{GarciaGarcia:2022yqb}. }, 
\begin{align}
    \mathcal{P}^{\rm LO}_{\gamma_w\rightarrow \infty} \simeq (\gamma_w n v_w) \langle \Delta p\rangle \simeq  (\gamma_w n v_w) \Delta m^2/2  {\langle p^0\rangle}  \ ,
    \label{eq:LOpressure}
\end{align}
where $  {\langle p^0\rangle \propto \gamma_wT}$ is the (boosted) incoming particle's energy in the wall frame, $\Delta m \equiv \sqrt{\mt^2-m^2}$ the change in mass between phases, and $n$ the number density in the plasma frame. This LO pressure is independent of $\gamma_w$, scaling like $\propto T^2 \tilde{v}^2$ in the case of a thermal bath, where $\tilde{v}$ is the vev in the 
broken phase 
\footnote{This lead to the so-called Bodeker-Moore (BM) criterion $\mathcal{P}^{\rm LO} < \Delta V$, for the wall to 
become relativistic. Under the assumption of pressure monotonically increasing with $\gamma_w$, the BM criterion was used also as a rough
\emph{runaway} condition.}  
\cite{Dine:1992wr,Bodeker:2009qy}.

Later, the same authors analyzed the next-to-LO (NLO) $1 \rightarrow 2$ processes in the same ultra-relativistic regime and found that, despite paying the price of the coupling, the emission of soft vector bosons that gain mass during the transition leads to a friction pressure scaling like $\mathcal{P}^{\rm NLO}  \propto \gamma_w$ \cite{Bodeker:2017cim}, eventually dominating over the LO effect. 
This soft emission is known as \emph{transition radiation}. 
While the original \cite{Bodeker:2017cim} 
focused on particles emitted forward into the wall (to the right in \cref{fig:cartoon}), the authors of \cite{Vanvlasselaer:2020niz,Gouttenoire:2021kjv} considered also reflected emission (to the left in \cref{fig:cartoon}) and argued it
was larger by a factor of four. 

However, all studies after \cite{Bodeker:2017cim}  only considered the emission of transverse vector
polarisations, ignoring the effects of longitudinal ones. 
The analysis of these modes is complicated by the rearrangement of particle degrees of freedom across a gauge symmetry breaking transition \cite{Farrar:1994vp}, which has naturally been the case of greatest interest. Moreover, it is well known that amplitudes involving NGBs can give spurious divergences without proper care. Recently it was shown that 
LO  effects from longitudinal 
modes can have a large impact on pressure \cite{GarciaGarcia:2022yqb}. It thus becomes of interest to properly account for their contribution at NLO.
In addition, a weakness of the treatments used so far is the frequent reliance on WKB approximations, which are known to break down for the soft momenta dominating the emission phase space.


In this paper, we approach the calculation of transition radiation by quantising field theories in the translation-breaking background of a domain wall from first principles. A complete orthonormal basis is constructed out of `left' and `right' mover energy eigenstates 
\footnote{Throughout this paper, the reader should associate `right-moving' with positive $z-$momentum and `left moving' with negative $z-$momentum particles.}, each wavemode having `reflected' and `transmitted' parts. We then carefully relate these to \textit{in} and \textit{out} 
\footnote{To be understood in the $S$ matrix language.} asymptotic eigenstates of 4-momentum.
In the case of vectors, we show that the degrees of freedom across the wall are most conveniently described in terms of `wall polarisations' $\tau$ and $\lambda$ rather than the conventional transverse and longitudinals, as already pointed out in \cite{Farrar:1994vp}. The advantage is that $\tau_{1,2}$ and $\lambda$ are not mixed with each other in the presence of the domain wall.
The two sets coincide only for zero transverse momentum $\vec{k}_\perp = 0$ (normal incidence on the wall), where rotations around the direction of propagation are a symmetry 
\footnote{Starting from $\vec{k}_\perp = 0$, the general $\tau$ and $\lambda$ polarisation vectors can be obtained by general transverse Lorentz boost - a good symmetry of the theory. Thus orthogonality is obvious. In general, they are also distinguished by 
whether in unitary gauge the $z-$ component of the vector $A^\mu$ is zero or not. See \cref{sec:Global_degrees_of_freedom}.}.
Moreover, in the case of gauge symmetry breaking, $\lambda$ smoothly interpolates between a Higgs d.o.f. on the symmetric side and a third massive vector d.o.f. on the broken side.
We show how to perform calculations using this basis consistently and avoid 
divergences which seem to appear in a naive analysis.

Although we explain how to (numerically) compute $\vev{\Delta p}$ for a general wall profile, we dedicate most of the work to approximating it in a way that is independent of the particular shape, with the rough scale $L_w$ playing the only role, while commenting on the sensitivity thereon. 
For the IR of the emitted spectrum $k^z \lesssim L_w^{-1}$ the wall appears effectively as a step function. This limit is particularly interesting theoretically (as well as phenomenologically important, as mentioned already) since everything can be computed analytically and relatively simply. 
 For wavelengths shorter than the wall width $k^z > L_w^{-1}$, the WKB approximation becomes applicable. The integral over the phase space thus splits into two contributions and the averaged momentum exchange very schematically takes the  form
\begin{align}
\label{eq:first}
   \langle \Delta p \rangle \sim \int^{k^z<L_w^{-1}} {d^3k \:\Delta p} \; |\mathcal{M}^{\rm step} |^2 \;+\;  \int_{k^z>L_w^{-1}} {d^3k \:\Delta p} \;|\mathcal{M}^{\rm wkb}|^2 \ .
\end{align}
where the $\mathcal{M}$ are matrix elements for emission calculated using the respective approximations. As the incoming flux scales like $\gamma_w$, we then have
\begin{align}
    \mathcal{P} \propto \gamma_w \langle \Delta p \rangle  \ .
\end{align}

As a warm up, we study transition radiation in a theory with two scalars and observe some surprises. Though we find that the pressure from the emission of one scalar by the other always saturates at large velocities $\mathcal{P}^{\rm NLO, \rm scalars}_{\gamma_w \to \infty} \propto \gamma_w^0$, we find also that there can be an intermediate regime of linear growth $\mathcal{P}^{\rm NLO, \rm scalars}_{\rm intermediate  } \propto \gamma_w$. 
For scalars we find that 
the WKB contribution (second term in \cref{eq:first}) dominates the momentum transfer in the asymptotic $\gamma_w$ limit.

\begin{figure}
    \centering
    \includegraphics[width=.48\textwidth]{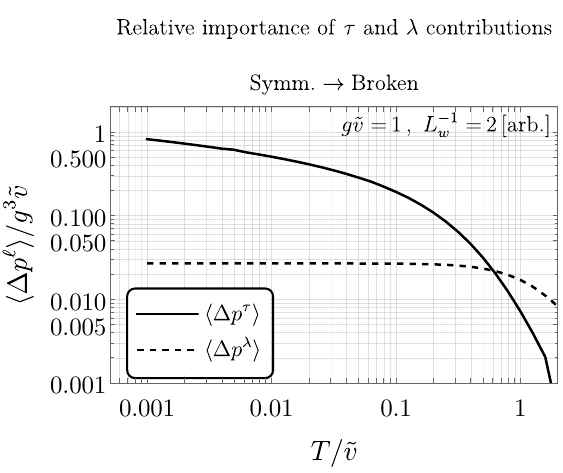}\includegraphics[width=.48\textwidth]{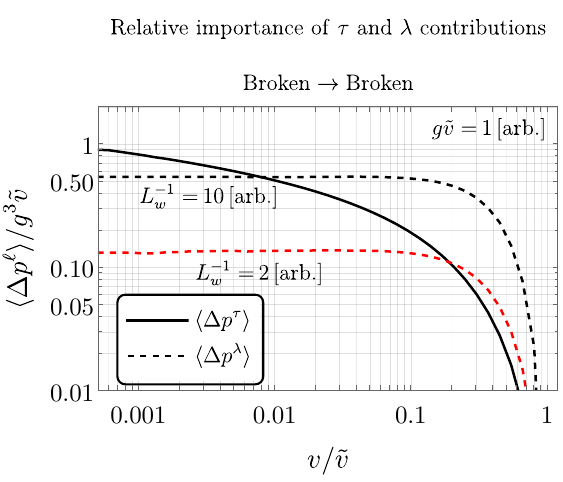}
    \caption{Comparison between the averaged exchanged momentum from 
    transition radiation due to the emission of $\tau$ and $\lambda$ vector polarisations, in the limit of 
    large incoming particle energy $p_0 \to \infty$. 
     \textbf{Left}: Symmetric $\to$ Broken transition at finite temperature $T$, shown as a function of temperature over Higgs vev $\tilde{v}$. $T$ enters via thermal masses (for details see the discussion in \cref{sec:Phase_space_vector}). Results do not change significantly in the limit $L_w \rightarrow 0$. \textbf{Right}: Broken $\to$ Broken transition. While the $\tau$ contribution does not change, $\lambda$ emission can easily become dominant. We highlight the sensitivity on the wall thickness. [arb.] means arbitrary units.
    }
    \label{fig:punchline1}
\end{figure}

In the case of spontaneous breaking of gauge
symmetry we find that 
the total friction from vector emissions scales as $\propto \gamma_w \log\frac{gT}{\tilde v}$ for $T/\tilde{v} \ll 1$, where $\tilde v$ is the Higgs' vev, in line with literature. We provide an updated fitted formula in \cref{eq:limit-pinf,eq:ThermalPressureFittedFinal}.
The logarithmic enhancement appears only for the $\tau$ polarisations, and is dominated by the step function contribution (the first term in the eq. \eqref{eq:first}), 
however we also find that effects of the $\lambda$ polarisations can lead to significant corrections for mild supercooling ($\frac{\tilde v}{ T} \sim {\rm few}$). We compare the relative importance in \cref{fig:punchline1} (Left). The curves are only very weakly dependent on $L_w$.
{  This and all of the rest of the figures in the paper are in natural units with some arbitrary scale [arb.].}

As a side application, we also compute the transition radiation when the bubble wall connects two vacua with broken gauge symmetry but different vevs $v$ and $\tilde{v}$.
 In this case,
 the contribution to friction from the longitudinal vector 
 emission scales as $\mathcal{P}^{\rm NLO, \rm vectors}_{ \gamma_w \to \infty} \propto \gamma_w L_w^{-1}$ (see \cref{fig:punchline1}, Right) and can dominate over the transverse for thin wall.
 
\begin{figure}
    \centering
    \includegraphics[width=.48\textwidth]{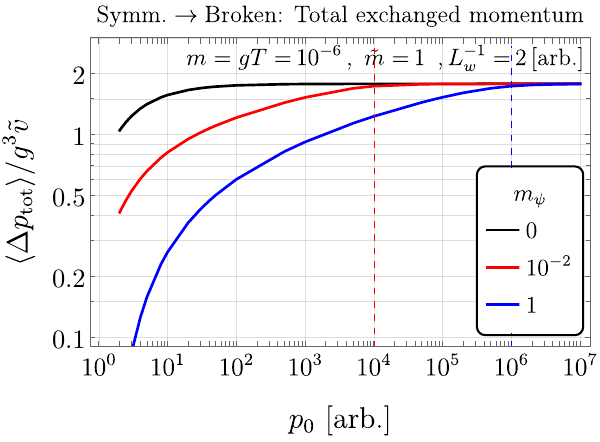}\includegraphics[width=.48\textwidth]{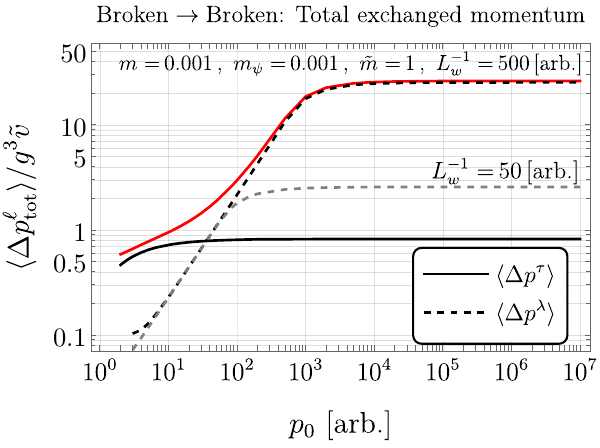}
    \caption{Total averaged momentum exchange as a function of incoming particle energy $p^0$. \textbf{Left}: Symmetric $\to$ Broken for different values of the mass of the \emph{emitter} particle $\psi$. The asymptotic regime is reached around $p^0 \sim m_\psi \tilde m/m$ (dashed lines), with intermediate $\log(p^0/m_\psi)$ growth. \textbf{Right}: Broken $\to$ Broken with thin walls. 
    Here the saturation value is reached around $p_0 \sim L_w^{-1} {\rm Max}[1, m_\psi /m]$. Again we highlight the sensitivity to wall width. For small enough $L_w$ we find an inter-relativistic regime with 
    averaged exchange momentum growing linearly, which translates to pressure scaling like $\gamma_w^2$. 
    }
    \label{fig:punchline 2}
\end{figure}

Aside from the asymptotic $p^0\rightarrow \infty$ limit, we are also able to explore regimes with intermediate - though large - $\gamma_w$. 
For symmetric $\to$ broken transition we find that the saturating value is reached at energies dependent on the mass of the emitter particle, as shown in fig. \ref{fig:punchline 2} (Left). 
In the case of the broken $\to$ broken transition we find that there is an intermediate regime where the pressure scales as
 $\mathcal{P}^{\rm NLO, \rm vectors}_{ \rm intermediate } \propto \gamma_w^2$ (right panel of \cref{fig:punchline 2}).

The paper is organised as follows: in the section \ref{sec:scalar} we work through a toy model with only scalars, introducing various elements of the calculation. In \cref{sec:Spontaneously broken Gauge theories}
we quantise an Abelian Higgs model in the presence of a symmetry breaking 
domain wall and present the results for transition radiation of vectors in \cref{sec:trans}. We summarise in \cref{sec:summary}.

\paragraph{Summary of notation:}
In the rest of this paper, we will adopt the following conventions:
\begin{enumerate}
    \item We treat the bubble wall in the planar limit, where it is one dimensional and centred around $z=0$. 
    \item  We use a hybrid notation for four-vector Lorentz indices: \\ $\mu = (n,z) \equiv (0,1,2,z)$. Coordinates are $x^\mu = (x^n,z)\equiv (t,\vec{x}_\perp,z)$.
    \item Similarly for momenta $k^\mu = (k^n, k^z) \equiv (k_0, \vec{k}_\perp,k^z)\equiv  (k_0, \vec{k})$. \\ Also $E^2=k^nk_n = k_0^2 - \kp^2$, where $\kp=|\vec{k}_\perp|$.
    \item We define the change in mass across the wall $\Delta m \equiv \sqrt{\mt^2-m^2}$.
    \item $L_w$ is the thickness of the wall.  
    \item $\gamma_w$ ($v_w$) is the boost factor (the velocity) of the wall.
    \item The momenta $p,q,k$ will always be used as in \cref{fig:cartoon} and we define:
    \begin{fleqn}[0pt]
    \begin{align}
        \begin{split}
            \Delta p \equiv p^z-q^z-k^z \ , \quad \Delta p_r \equiv p^z-q^z+k^z \ , \\
            \Delta \tilde{p} \equiv p^z-q^z-\kt^z \ , \quad \Delta \tilde{p}_r \equiv p^z-q^z+\kt^z \ .
        \end{split} 
        \label{eq:Deltap_notations}
    \end{align}
    \end{fleqn}
\end{enumerate}


\section{Simple example: scalars}
\label{sec:scalar}

In this section, as a warm up for the more complex case of gauge theories, we go through the quantisation of a scalar field theory in the presence of a domain wall and derive results for transition radiation for the case of one scalar emitting another. This toy example is sufficient to highlight many features of calculations in a spatially dependent background.

Consider two different scalar fields $\phi,\psi$, the first of which feels the wall and has different mass depending on the phase, while the second for simplicity does not. 
The Lagrangian we consider is the following
\begin{align}
    \label{eq:2scalarlag}
\mathcal{L}={1 \over 2}(\d \phi)^2+{1 \over 2}(\d \psi)^2-{1\over 2} m^2_\phi(z) \phi^2- {1 \over 2} m_\psi ^2 \psi^2- y(z) \frac{1}{2}\psi^2 \phi,
\end{align}
where $m(z)$ interpolates between $m^2(z)|_{z\to -\infty}= m^2=const $ and  $m^2(z)|_{z\to +\infty}= \tilde m^2=const$. Similarly $y(z)$ goes from $y$ to $\tilde y$. The profiles change on the scale of the wall width $L_w$ around $z=0$.
The interactions in eq. \eqref{eq:2scalarlag} are not the most general, but are designed to mimic the vector case when $y=const$.
The process that we will be studying is $\psi\to \psi \phi$, which would be forbidden by kinematics if it was not for the breaking of  $z-$momentum. 
\paragraph{Section summary:} In section \ref{sec:basis} and \ref{sec:quantisation}, we quantise the free theory, focusing on the $\phi$ field
\footnote{ The quantisation of $\psi$, as it does not feel the wall, is instead completely standard.}
, by defining a complete basis of solutions that solve its equations of motion. In section \ref{sec:new_basis}, we define a new basis that corresponds to \textit{out-going} eigenstates of momentum. Later, in section \ref{sec:amplitude}, we calculate the amplitude for the $\psi\to \psi \phi$ transition in the step wall approximation, valid when $k^z L_w \lesssim 1$.
In section \ref{sec:Phase_space_int}, we present
the proper domain for the phase space 
integration over the final state. In \cref{sec:emission_energ}, we complete the emission spectrum discussing the calculation of the amplitude in the (opposite) WKB regime $k^z L_w \gtrsim 1$. In \cref{sec:procedure}, we summarise and present master formulae for the calculation of the averaged momentum transfer $\vev{\Delta p}$.
We conclude by discussing results for $\vev{\Delta p}$ and pressure $\mathcal{P}^{\rm NLO}$ in  \cref{sec:momentumTransferScalar,sec:PressureScalar} respectively. 

\subsection{Complete basis}
\label{sec:basis}
The quantisation of modes in the presence of a background profile arises in many corners of physics. A very similar task appears for example in the quantisation of field theory in black hole spacetimes. We found the treatment in \cite{Shkerin:2021zbf} particularly useful. In the simple example of eq. \eqref{eq:2scalarlag} above, the discussion is relevant for $\phi$, which satisfies
\bea
(\partial^2 + m_\phi^2(z))\phi=0 \, ,
\eea
with a $z-$dependent mass term.
To perform second quantisation we need to first find a convenient basis of solutions of this equation. Far away from the wall the solutions are plane waves. A convenient choice of complete orthonormal basis is given in terms of `right' and `left' moving solutions, which are defined by their boundary conditions as follows \footnote{Recall that the index $n$ designates $0, 1, 2$ and not the $z$ direction.}
\begin{align}
    \label{eq:R_movers}
&&  \phi_{R,k}  = e^{-i k_n x^n    } \chi_{R,k}(z)  \equiv
    e^{-i  k_n x^n  }
    \begin{cases}
            e^{i k^z z} +  r_{R,k}  e^{-i k^z z}  \ , & z\to -\infty\\
            t_{R,k} e^{i \tilde{k}^z z}  \ , &   {z\to +\infty} 
    \end{cases}
    \quad \text{(Right)}
\end{align}
with $k^0>m$
    and
\begin{align}
    \label{eq:L_movers}
 \phi_{L,k}  = e^{-i k_n x^n    } \chi_{L,k}(z)  \equiv
    e^{-i  k_n x^n  }\sqrt{ k^z \over \tilde  k^z}
   \begin{cases}
           t_{L,k} e^{-i k^z z}  \ , & z\to-\infty \\
           r_{L,k}  e^{i \tilde{k}^z z}  +  e^{-i \tilde{k}^z z} \ , & z\to +\infty 
    \end{cases} 
   \quad \text{(Left)}
\end{align}
with $k^0>\mt$ and we take $k^z,\kt^z$ to be strictly positive \footnote{As is well known, the basis formed by just $\phi_{R,k}$ but allowing $k^z$ to take both signs (and $\kt^z = {\rm sign}(k)(\sqrt{k^2 - \Delta m^2 }$) is also complete but not orthogonal and therefore less convenient. For example, the algebra of creation and annihilation operators would be more complicated.}. 
The factor  $ \sqrt{ k^z/ \tilde{k}^z} $ 
is included in \cref{eq:L_movers} to ensure appropriate normalisation (see below, \cref{eq:LRorthonormal}).
In the limit of no domain wall the $r_L,r_R$ ($t_L,t_R$) coefficients are zero (one) and 
$\phi_L,\phi_R$ correspond simply to the plane waves with $\mp k^z$ momenta. 
The momentum along $z$ is not conserved across the wall; however, asymptotically far it becomes constant and  fixed by the relations
\bea 
\label{eq:kz kt}
k^z\equiv \sqrt{k_0^2-\kp^2-m^2},~~\tilde k^z\equiv \sqrt{k_0^2-\kp^2-\tilde m^2} \ . 
\eea
In general, we need to solve the 
equations of motion to find the expression of the coefficients 
$r_{L,R},t_{L,R}$. Consequently, they will depend on the explicit form of the mass variation $m^2_\phi(z)$. 
However, here it will be sufficient to consider the \emph{step wall} ansatz for  the mass $m_\phi^2(z)$
\bea
m_\phi^2(z)=m^2+ \Delta m^2 \Theta(z) \, , \qquad \Delta m^2 \equiv \tilde m^2 - m^2 \,.
\eea
using the Heaviside Theta function. 
The form of the coefficients 
for the scalar case under consideration can be obtained by matching $\phi$ and its first derivative at the origin 
$z=0$, where the step wall lies. They take the form
\bea 
 r_{R,k} = \frac{k^z - \tilde{k}^z}{k^z + \tilde{k}^z} \ ,  \quad t_{R,k} = \frac{2 k^z  }{k^z +  \tilde{k}^z} \ .
 \label{eq:scalar_coef}
\eea 
These expressions are specific 
to the step-wall assumption. However, the general treatment that we present here will
hold for general $r_k,t_k$ coefficients and could be easily adapted to a smooth wall case. Modes with $m<k_0<\tilde{m}$ decay exponentially on the right of the wall and are automatically included as right-movers. For these, $\kt$ is purely imaginary with magnitude
\begin{align}
\label{eq:ktImaginaryMag}
    |\kt^z|^2 = \Delta m^2 - k_z^2  \qquad \text{for} \qquad 0 < k^z < \Delta m \ .
\end{align}

In a similar fashion, for the left moving solution we find 
\bea 
r_{L,k} = - {r}_{R,k} =\frac{\tilde{k}^z - k^z}{k^z + \tilde{k}^z} \ ,  \quad t_{L,k} = \frac{\tilde{k}^z}{k^z}t_{R,k} = \frac{2 \tilde{k}^z  }{k^z +  \tilde{k}^z} \ .
\eea 
and we explicitly note the condition $k_0 > \tilde{m}$, to avoid the inclusion of solutions growing exponentially at infinity. 
The left and right moving modes are orthonormal in the sense that
\begin{equation}
\label{eq:LRorthonormal}
    \int_{-\infty}^\infty dz \; \chi_{I,k} \; \chi_{J,q}^* = 2\pi \delta_{IJ}\delta(k^z-q^z) \ , \qquad I,J \in \{R,L\} \ .
\end{equation}
Computing integrals such as \cref{eq:LRorthonormal}in the step function case requires the identity
\begin{equation}
\label{eq:HalfDelta}
	\int_{-\infty}^0 e^{i \beta z} dz = {\rm PV} \left( \frac{1}{i\beta} \right) + \pi \delta(\beta) \ ,
\end{equation}
and its complex conjugate (which gives the integral from $0$ to $\infty$). In eq. \eqref{eq:LRorthonormal} the principle value (PV) pieces vanish as soon as we specify the relation between $\tilde k^z$ and $k^z$, i.e. $\tilde k_z^2=k_z^2+m^2-\tilde m^2$.
Notice we can discard terms proportional to $\delta(k^z+q^z)$ due to the strictly positive definition of $k^z,q^z$ in our definition. 
If explicitly computing things like the Hamiltonian and operator algebra (see next subsection) it is also useful to know the other inner products:
\begin{align}
   & \int_{-\infty}^\infty dz \; \chi_{R,k} \; \chi_{R,q} =
-      \int_{-\infty}^\infty dz \; \chi_{L,k} \; \chi_{L,q}
     = 2\pi { k^z - \tilde{k}^z  \over  k^z + \tilde{k}^z } \delta(k^z-q^z) ,\nn 
&     \int_{-\infty}^\infty dz \; \chi_{R,k} \; \chi_{L,q} = 4\pi { \sqrt{k^z\tilde{k}^z } \over  k^z + \tilde{k}^z }\delta( k^z- q^z) .
 \label{eq: chiLchiL}
\end{align}

Finally, we would like to comment that, in general, bound states may also appear in the spectrum, in addition to the scattering states studied above, if the function $m^2_\phi(z)$ is non-monotonic and has minima in the vicinity of the domain wall. These are of the form $\phi_{\rm b}\propto e^{-i k^n x_n }\chi_{\rm b}(z)$, with $\chi_{\rm b}$ exponentially decaying for $|z|\rightarrow \infty$ and should be included in the upcoming expansion \cref{eq:fieldExpansion}.

\subsection{Quantisation}
\label{sec:quantisation}
Now that we have a complete orthonormal basis of eigenstates in the presence of the wall, we can proceed to quantise the theory. The field $\phi$ can be expanded in the form  
\footnote{We use normalisation conventions in line with \cite{Peskin:1995ev}.}
\begin{align}
\label{eq:fieldExpansion}
    \phi (x, t) = \sum_{I=R,L}\int \frac{d^3 k}{(2\pi)^3 \sqrt{2 k_0}}  \left( a_{I,k} \phi_{I,k} + a^\dagger_{I,k} \phi^*_{I,k}  \right)  ,\nn  
    {\rm where } \quad \phi_{L,k}\equiv 0 \quad \text{for} \quad E < \tilde{m} \ ,
\end{align}
where $d^3k \equiv dk^z d^2k_\perp$, we recall $E \equiv \sqrt{k_0^2 - k_\perp^2}$  and $k^z$ runs between $[0,\infty)$. We choose to label states by their quantum numbers outside the wall\footnote{This is more convenient than labeling with respect to $\kt^z$ since this becomes imaginary for the branch $0<k^z<\Delta m$.}.
Note we have trivially extended the definition of the left moving modes to the region $ E < \tilde{m}$  for convenience.
 
Using \cref{eq:LRorthonormal,eq: chiLchiL}, one can show that
\begin{align}
\label{eq:aIp from phi}
    a_{I,p} = \int {dz \over \sqrt{2p_0}} e^{i p_0 t} \chi^*_{I,p}(i \pi + p_0 \phi) \ ,
\end{align}
where $\pi \equiv\d_t\phi$.
Promoting Poisson brackets of $\phi$ and its conjugate momentum $\pi$ to canonical commutation relations gives the familiar commutation algebra
\begin{align}
\label{eq:SHOalgebraScalar}
   \begin{split}
        [a_{I,k},a_{J,q}^\dagger] &= (2\pi)^3 \delta(\vec{k}-\vec{q})\delta_{IJ} \ , \\
[a_{I,k},a_{J,q}]&=[a_{I,k}^\dagger,a_{J,q}^\dagger]=0 \ , \qquad I,J \in \{R,L\} \ .
   \end{split}
\end{align}
We can define two types of states
\begin{align}
    \left| k_R \right\rangle &\equiv \sqrt{2 k_0} \; a_{R,k}^\dagger  \left| 0 \right\rangle \ , \\
    \left| k_L \right\rangle &\equiv \sqrt{2 k_0} \;  a_{L,k}^\dagger \left| 0 \right\rangle  \ ,
\end{align}
which should be thought of as independent external states in any process. The space of physical states is thus the \emph{Fock} space defined by arbitrary powers of $a_{R,k}^\dagger$ and $a_{L,k}^\dagger$ acting on the vacuum.

\subsection{Out-going eigenstates of momenta}
\label{sec:new_basis}
In the previous subsections we chose to quantise the orthonormal basis $\{ \phi_{R,k} \:, \phi_{L,k} \}$ and defined associated one-particle states $\left| k_R \right\rangle$ and $ \left| k_L \right\rangle$. As we explain in more detail in \cref{app:Wavepackets} by the use of wave-packets, these should be thought of as describing incoming particles with definite $z-$momenta $k^z$ and $ -\tilde k^z$ respectively at $t\to -\infty$, but at $t\to +\infty$ they correspond to a superposition between a transmitted and reflected particle.
As a consequence, the functions $\phi_{R,L}$ are eigenstates of momenta only at $t\to -\infty$. They are well-suited for processes with asymptotic \textit{in-state} $\phi$ particles.

\begin{figure}
    \centering
    \includegraphics[width=.48\textwidth]{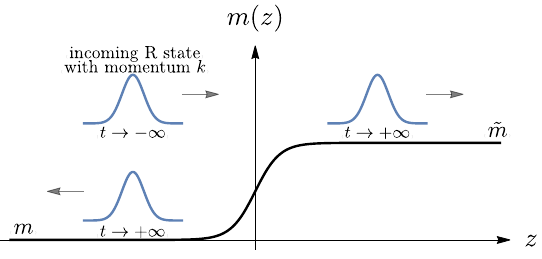}\includegraphics[width=.48\textwidth]{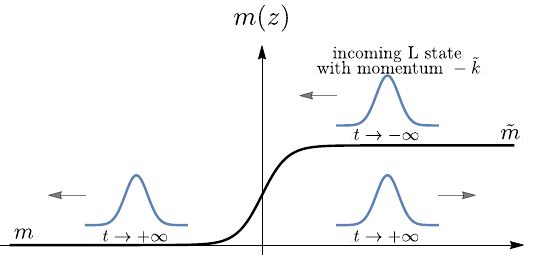}\\
    \includegraphics[width=.48\textwidth]{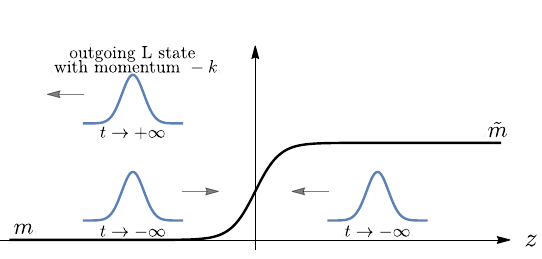}\includegraphics[width=.48\textwidth]{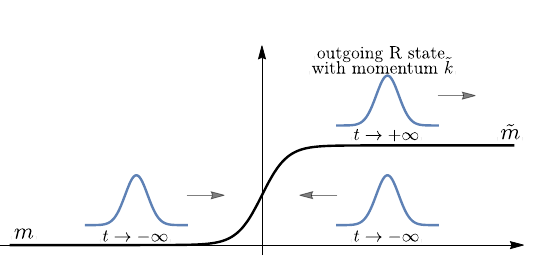}
    \caption{ Summary of asymptotic external states with definite $4-$momentum, to be used in calculating any process in the background of a  domain wall (or any localised $z-$dependent potential). The upper panels represent in-state particles incoming from $z = -\infty$ and $z=\infty$. In the plane wave limit they correspond to states $ \left| k_R \right\rangle$ and $ \left| k_L \right\rangle$ with wavefunctions \cref{eq:R_movers,eq:L_movers} respectively. The lower panels represent out-state particles travelling out towards $z = -\infty$ and $z=\infty$. They correspond to states \cref{eq:koutL,eq:koutR} with wavefunctions \cref{eq:LoutWavefn,eq:RoutWavefn} respectively.
    }
    \label{fig:AsymptoticStates}
\end{figure}

On the other hand, in this work we will be interested
in the momentum transfer to the wall, so it is more convenient to have a $\phi$ particle emitted as an asymptotic \textit{out-state}  with well-defined momentum at $t\to \infty$. 
A complete orthonormal basis of such late-time eigenstates of momentum is given by
\begin{align}
\label{eq:LoutWavefn}
\phi^{\rm out}_{L,k} &\equiv e^{-i k_n x^n}\zeta_{L, k}(z) =e^{-i k_n x^n}\chi_{R, k}^*(z)
=e^{-i k_n x^n}\l(r_{R, k}^* \; \chi_{R,k}+t_{R, k}^* \sqrt{\frac{\tilde k^z}{k^z}} \; \chi_{L,k} \r) \ , \\
\label{eq:RoutWavefn}
\phi^{\rm out}_{R,k}&\equiv e^{-i k_n x^n}\zeta_{R, k}(z)=e^{-i k_n x^n}\chi_{L,k}^*(z)=e^{-i k_n x^n}\l(r_{L, k}^* \; \chi_{L,k}+t_{L, k}^* \sqrt{\frac{ k^z}{\tilde k^z}} \;\chi_{R,k}\r) \ , 
\end{align}
where in the last equalities we related them to the basis of \cref{sec:basis}.
We emphasise again that in our notation $k^z,\tilde k^z>0$ always.
$\phi^{\rm out}_{L,k}$ and $\phi^{\rm out}_{R,k}$ should be thought of as describing an outgoing final state particle with $-k^z$  and   $+\tilde k^z$ momentum respectively. 
Recall that the function $\chi_{L,k}$ vanishes for $k^z<\Delta m$ and the corresponding $\Theta$ functions are implicit.
At $t\to -\infty$ they are both superpositions of incoming particles from $z = \pm \infty $  and do not have well defined momentum.
In practice we need to calculate the amplitudes with  $\zeta_{L,R}=\chi^*_{R,L}$ wave functions.
At the level of states, we have
\begin{align}
\label{eq:koutL}
| k^{\rm out}_L \rangle &= r_{R,k}^*|k_R\rangle + t_{R,k}^* \sqrt{\tilde{k}^z / k^z}\; |k_L\rangle \; \Theta(k^z-\Delta m)  \ , 
\\
\label{eq:koutR}
| k^{\rm out}_R \rangle  &=  t_{L,k}^* \sqrt{k^z/\tilde{k}^z}\; |k_R\rangle + r_{L,k}^* |k_L\rangle  \ , 
\end{align}
where we explicitly remind ourselves that when $0<k^z<\Delta m $ the left mover state does not exist.
The different asymptotic states are illustrated in the \cref{fig:AsymptoticStates}. 

We emphasise that both bases can be used to quantise the theory. In our present paper however we will consider only outgoing $\phi$ particles so that the basis $\{ \phi^{\rm out}_{R,k},\phi^{\rm out}_{L,k}\}$ is actually more convenient. From now on we drop the label `out` and we will refer to $R$ ($L$) emission meaning using the mode functions $\{\zr, \zl\}$, if not stated otherwise.

\subsection{Amplitudes}
\label{sec:amplitude}
We now finally turn to compute the amplitude for the process $\psi\to \psi \phi$ in the background of a domain wall. We have not discussed the quantisation of $\psi$ since it does not feel the wall directly and there are no complications with respect to the standard theory.
In the previous sections we argued that there are two processes we have to consider separately: the emission of a left and right moving $\phi$ particle, with respective wavefunctions $\zeta_{L} (\zeta_{R})$. Having quantised the free theory, the treatment of perturbative interactions proceeds as standard, by defining an $S$-matrix in terms of the interaction Hamiltonian $\mathcal{S}={\rm T \; exp}\left( - i \int d^4x \mathcal{H}_{\rm Int} \right) $ where T here denotes time ordering. We have the amplitudes of interest
\begin{align}
\label{eq:MatrixElement3_def_scalars}
    \langle  k_I^{\rm out} \; q  |\: \mathcal{S} \:| p \rangle &\equiv (2\pi)^3 \delta^{(3)}(p^n-k^n-q^n) i\mathcal{M}_{I} \overset{\rm tree}{=} - i\int   d^4x  \langle   k_I^{\rm out} \; q  | \  \mathcal{H}_{\rm Int} \ | p \rangle \\
    \text{with} \quad I & =  L,R \nonumber \ ,
\end{align}
where $p$ and $q$ stand for the initial and final one particle states for the $\psi$ field and their respective $4-$momenta,
and the last equality is up to leading order in perturbation theory (tree level). Notice we have defined the matrix element $\mathcal{M}$ as closely as possible to standard theory. Of course, we cannot extract a $z-$momentum conserving delta function but rather $\mathcal{M}$ still contains the integral over $z$.

For the theory of scalars of \cref{eq:2scalarlag}, we have $\mathcal{H}_{\rm Int} = -i y  \psi^2(x) \phi(x) $ and we can now proceed to explicitly computing amplitudes.
In the case of a $\zl$ mode emission (where the emitted scalar has $-k^z$ momentum), the amplitude takes the form
\begin{align}
\label{eq:Lemission}
   \mathcal{M}_{L} &=  \int_{-\infty}^\infty  dz \; y(z) e^{i(p^z-q^z)z}\l[\zeta_{L,k}=\chi^*_{R,k}(z) \r]^*   \nonumber \\
   &=\left[ \frac{-iy}{p^z-q^z+k^z} + r_{R,k}\frac{-iy}{p^z-q^z-k^z} - t_{R,k}\frac{-i\tilde{y}}{p^z-q^z+\tilde{k}^z} \right]  \ .
\end{align} 
Instead, in the case of $\zr$ mode emission (where the emitted scalar has $+\tilde k^z$ momentum): 
\begin{align}
\label{eq:Remission}
\mathcal{M}_R &=   \int_{-\infty}^\infty  dz \; y(z) e^{i(p^z-q^z)z}\l[\zeta_{R,k}=\chi^*_{L,k}(z)\r]^* \nonumber \\
   &= \sqrt{\frac{ k^z}{\tilde k^z}}\left[t_{L,k} \frac{-iy}{p^z-q^z-k^z} - \frac{-i\tilde {y}}{p^z-q^z-\tilde{k}^z} - r_{L,k}\frac{-i\tilde {y}}{p^z-q^z+\tilde{k}^z} \right] \ ,
\end{align}
where the square root factor comes from the normalisation condition in \cref{eq:L_movers}. To compute the total friction from $\psi \rightarrow \psi \phi$ we must sum the contributions from both processes. 
Then we can compute the amplitude squared for the emission of a right/left mover, under the assumption that $y=const.$, we obtain
\begin{align}
    \label{eq:amplitudeMscalarsR}
|\mathcal{M}_R|^2 &=  y^2{4 k^z \tilde{k}^z  (k^z-\tilde k^z)^2 \over (\tilde{k}_z^2-(p^z-q^z)^2)^2(k^z-p^z+q^z)^2} \ , \\
\label{eq:amplitudeMscalarsL}
|\mathcal{M}_L|^2 &=   y^2{4 k_z^2  \over (k_z^2-(p^z-q^z)^2)^2} 
\begin{cases}
{(k^z-\kt^z)^2 \over (\kt^z + p^z -q^z)^2} \ , & k^z > \Delta m \ , \\
{\Delta m^2 \over \Delta m^2 - k_z^2 + (p^z-q^z)^2 } \ , & k^z < \Delta m \ ,
\end{cases} 
\end{align}
where for $L$ emission we distinguished between the two branches corresponding to $\kt^z$ purely real and imaginary, and used \cref{eq:ktImaginaryMag} to simplify in the latter case.


\subsection{Phase Space integration}
 \label{sec:Phase_space_int}
 We can now compute the average exchanged momentum due to transition radiation, $\langle \Delta p\rangle$ due to a single incoming particle. We integrate over the whole allowed phase space of the final two particles, 
 weighting the amplitude squared for $\psi$ to emit $\phi$ by the momentum lost. We add separately contributions from left and right movers with their respective slightly different phase space.
 $\langle \Delta p\rangle(p) $ is in general a function of the four-momentum $p$ of the incoming $\psi$.
From $\langle \Delta p\rangle $ we 
  can compute the total friction pressure by 
 integrating over the incoming flux. 

For simplicity we take $\vec{p}_\perp=0$ and parameterise the kinematics of the $1 \to 2$ process as follows
\bea 
\label{eq:kinem-param}
&p^\mu=(p_0, 0, 0, \sqrt{p_0^2-m_\psi^2}), \qquad &q^\mu=(q_0, -k_\perp, 0, q^z),\notag \\
&k^\mu=(k_0, k_\perp, 0, k^z), \qquad &\tilde k^\mu=(k_0, k_\perp, 0 , \tilde k^z),
\eea 
where $\kt^z$ is defined in eq. \eqref{eq:kz kt} and $q^z=\sqrt{q_0^2-k_\perp^2-m_\psi^2}$, with $q_0=p_0-k_0$ from energy conservation. 
Note we have used the cylindrical symmetry
of the set up to make the second spatial component of $k^\mu$ zero.
 $\langle \Delta p\rangle $ is given by the sum of left and right contributions
\begin{align}
\label{eq:averaged_p 1}
    \langle \Delta p \rangle&=\langle \Delta p_R \rangle+\langle \Delta p_L \rangle \nonumber\\
    &\equiv \int d\mathbb{P}_{\psi \rightarrow \psi \phi_R}  \underbrace{(p^z-q^z-\tilde k^z)}_{\Delta p^z_R}+
    \int d\mathbb{P}_{\psi \rightarrow \psi \phi_L}\underbrace{(p^z-q^z+k^z)}_{\Delta p^z_L}  \ .
\end{align}
where $d\mathbb{P}_{\psi \rightarrow \psi \phi_I}$ is the differential probability. The first and second terms on the RHS of \cref{eq:averaged_p 1}, will be different in their lower limits on the $k^z$ integrals. The second term, left-mover emission, contains also the modes $k^0<\mt$ which are exponentially decaying inside the wall. In \cref{sec:phase_space_derivation} we show that
\begin{align}
    \label{eq:PSI-RIGHT}
 \int d\mathbb{P}_{\psi \rightarrow \psi \phi_I} \Delta p^z_I= \int_{k^{z,I}_{\rm min}}^{k^z_{\rm max}} \dfrac{dk_z}{2\pi} {1 \over 2k_0}\int_0^{k_{\perp, \rm max}^2}\ \dfrac{dk_\perp^2}{4 \pi}\cdot \dfrac{1}{2p^z}\left[\dfrac{1}{2|q^z|} |\mathcal{M}_I|^2 \Delta p^z_I\right]_{q^z=\pm q_k^z},
\end{align}
where $I=R,L$ and  
$\left[...\right]_{q^z=\pm q_k^z}$ is intended to be the sum over $q^z=\pm q_k^z$. The contribution with $q^z=-q_k^z$  
corresponds to the reflection of the incoming particle $\psi$, a branch missing in the previous literature.
Of course, in the ultra-relativistic regime, it is expected that this should be highly suppressed \footnote{Notice that the sign change in $\Delta p^z_R$ means that a process with $q^z<0$ contributes more to momentum 
exchange, but it is the 
amplitude which is generally suppressed.}. 
The limits of integration of \cref{eq:PSI-RIGHT} are found demanding the reality of the $q^z$ momentum, obtaining for the $(R)$ modes
\bea
\label{eq:kperp max}
({\rm Right}): && k^{z,R}_{\rm min}\equiv \Delta m\leq   k^z \leq k^{z}_{\rm max}\equiv \sqrt{(p_0-m_\psi)^2- m^2},\nn
&&0\leq  k_\perp^2 \leq k_{\perp, \rm max}^2\equiv{1 \over 4p_0^2}( p_0^2 +k_z^2+m^2-m_\psi^2)^2-k_z^2-m^2.
\eea
For the $(L)$ modes, the only difference is  
\bea 
({\rm Left}): \quad k^{z,L}_{\rm min}\equiv 0&\leq  k^z& \leq k^{z}_{\rm max} \ .
\eea 
Following this discussion, in general, there will be four 
contributions: 
$$
\langle \Delta p^{q^z < 0}_{L}\rangle, \langle \Delta p^{q^z > 0}_{L}\rangle, \langle \Delta p^{q^z < 0}_{R}\rangle, \langle \Delta p^{q^z > 0}_{R}\rangle.$$ 
However, we explicitly checked 
that in all cases of interest, the contributions with $q^z<0$ are largely subdominant and we will ignore them completely in the rest of this paper. 

\subsection{Emission in the WKB  regime}
\label{sec:emission_energ}
So far we have been treating the bubble wall as a step function. This is a good approximation if the $z$ momentum of the emitted particle is 
less than the inverse scale over which the background (in the case of \cref{eq:2scalarlag}, the mass) changes significantly, i.e. $k^z \lesssim L_w^{-1}$. How can we 
proceed if the particles' momentum becomes comparable or 
larger than the width of the wall?
First of all, if we know the shape of the potential exactly, we can solve for the 
left and right mover solutions as we did above for the step-wall, and 
proceed with these functions in precisely the same way as before. In principle, this can always be done numerically. 
However, we will argue that even in the 
case when we do not know the exact shape 
of the potential, we can still obtain reliable results.

Let us consider a 
particle hitting the wall with 
momentum  $k^z\gg L_w^{-1}$. Reflection will be suppressed and the WKB approximation, which has been used extensively in the literature \cite{Bodeker:2017cim, Azatov:2020nbe, Gouttenoire:2021kjv}, becomes applicable. The approximate form of the z-dependent wavefunctions are now $\chi \approx \sqrt{k^z(z_0)/k^z(z)}e^{\pm i\int_{z_0}^zk^z(z')dz'}$, with $z_0$ some reference position. 
Having the approximate solutions to the $\chi_{R,L} (\zeta_{R,L})$ basis functions, we proceed in the same way as in section \ref{sec:amplitude}.
 Thus, in practice this means separating the phase space into two regions:
\begin{align}
\begin{split}
        \label{eq:space-sep}
&\hbox{region  (1)}~~~~~~~~k^{z}_{\phi}\ll L_w^{-1},~~~ \hbox{step wall, \quad  }\\
&\hbox{region  (2)}~~~~~~~~k^{z}_{\phi}\gg L_w^{-1}~~~~ \hbox{WKB. \quad  }
\end{split}
\end{align}
In the WKB regime the amplitude for a general $1 \to$ many right-movers\footnote{We will see later on how left mover emission is negligible for us in the WKB regime.} process, allowing for all masses to vary, can be schematically written as follows
\bea
{\cal M}_{\rm wkb}&=&\int_{-\infty}^{+\infty} d z \,V(z) \exp \l[i\int_0^z\Delta p(z') dz'\r]\prod_i\sqrt {\frac{p^i_z(0)}{p^i_z(z)}}\ , \nn
\Delta p(z')&\equiv& p^z(z')-\sum_i p^z_i( z')=p^z(z')-\sum_i\sqrt{(p^i_0)^2-(p^i_\perp)^2-m_i^2(z')} \ ,
\eea
where $i$ sums over final state particle momenta.
Naively, computing this integral requires knowledge of the functions $m_i(z)$. 
However, these are changing only in the vicinity of the wall, while outside they quickly reach the asymptotic constant values. This means we can split the amplitude into two pieces
\bea 
\mathcal{M}^{\rm wkb} = &&\underbrace{\int_{-\infty}^{0} dz V(-\infty) e^{i \Delta p_z(-\infty) z} +e^{i\int_0^{L_w} d z' \Delta p(z')}\int^{\infty}_{0} dz V(+\infty) e^{i \Delta p_z(+\infty)z}}_{\mathcal{M}_{\rm outside}}  \nn
+&&\underbrace{\int^{L_w}_{0} dz V(z) e^{i \int_0^z d z'\Delta p( z') } }_{\mathcal{M}_{\rm inside}} \ ,
\eea 
where the assumption is that things are varying only between $z\in [0,L_w]$.
In the WKB regime, all the momenta of the particles are 
much larger than the inverse width of the wall $p^z L_w\gg 1$  so 
the overall modification of momenta $p(+\infty)-
p(-\infty)\ll p(\infty)$ is much less than its absolute 
value (if the wall is not too thick $m(z) L \lesssim 1$), this is why we have approximated $\prod\sqrt{p/p(z)}\to 1$. 
Similarly $\Delta p(+\infty)-\Delta 
p(-\infty)\ll \Delta p(\infty)$, then from basic properties of Fourier transformations the amplitude 
\bea
\label{eq:cut-fourier}
{\cal M}\to 0 ,\hbox{  if  } \Delta p_z^{\rm max} L_w \gg 1  \ ,
\eea
with $\Delta p_z^{\rm max}={\rm Max}[\Delta p_z(\pm \infty)]$. The physics behind this relation is very simple: the wall of the width $L_w$ can 
lead to the momentum loss $\Delta p_z$ at 
most $L_w^{-1}$. This is expected since the processes with $\Delta p_z \gg L_w^{-1}$ happen at distances $\sim \Delta p_z^{-1}$, much shorter than the typical wall width. However, at such small distances we recover translational symmetry along the $z$ direction and transition radiation must be forbidden (we checked these statements for various wall ansatzes in appendix \ref{app:suppression}).
From these arguments, we can see that independently of the wall ansatz the particle emission will be dominated by the region $\Delta p_z L_w \lesssim 1$. Then we can approximate the amplitude as follows
\begin{align}
    \mathcal{M}^{\rm wkb} = &\int_{-\infty}^{0} dz V(-\infty) e^{i \Delta p_z(-\infty) z} +\int^{\infty}_{0} dz V(+\infty) e^{i \Delta p_z(+\infty)z} \nonumber \\ 
     + &\int^{L_w}_{0} dz V(z) e^{i \int d \tilde z\Delta p(\tilde z) } \ .
\end{align}
 Performing the $z$ integrals  for the first two terms  is trivial and, using \cref{eq:HalfDelta}, we get
\bea
\mathcal{M}^{\rm wkb}\approx 
\frac{V(-\infty)}{i\Delta p_z(-\infty)}-\frac{V(+\infty)}{i\Delta p_z(+\infty)}+ 
\int^{L_w}_{0} dz V(z) e^{i \int d \tilde z\Delta p(\tilde z) } \ 
\eea
The last term scales very roughly as $V(z\in [0,L_w]) L_w$  then assuming $V(z\in [0,L_w])\sim V(\infty )\sim V(-\infty)$ we can see it will be suppressed  by the condition $\Delta p L_w \lesssim 1$.
Thus we arrive at the Bodeker-Moore formula \cite{Bodeker:2017cim} for reduced matrix element
\bea
\mathcal{M}^{\rm wkb \; red.}= 
\frac{V(-\infty)}{i\Delta p_z(-\infty)}-\frac{V(+\infty)}{i\Delta p_z(+\infty)} \ .
\eea 
Now one can take this formula and perform the phase space integration. However, we would like to emphasise a simple but important point. Since we have ignored the contribution inside the wall, there is no guarantee that the matrix element will be suppressed in the region with $\Delta p_z L_w \gg 1$.  In all of our calculations we always:
\bit
\item impose $\Delta p_z L_w < 1$ $-$ Fourier decomposition properties
\item verify that ${\cal M}_{\rm inside} \ll {\cal M}_{\rm outside}  $ $-$ applicability of BM approximation (we will see that satisfying this inequality turns out to be non-trivial for longitudinal vector bosons).
\eit

Finally, from this discussion it is clear that we should not worry about left emission in the WKB regime since for left movers with $k^z > L_w^{-1}$ the total loss of momenta $\Delta p_z=p^z-q^z+k^z > L_w^{-1}$, meaning these processes must be strongly suppressed and we can safely ignore them.

\paragraph{Scalars example}
Let us apply this very generic discussion to the case of scalar radiation. Then the  matrix element will be given by
\bea
\label{eq:BM}
{\cal M}^{\rm wkb \; red.}= \frac{-iy}{p^z-q^z-k^z} - \frac{-i y}{p^z-q^z-\tilde{k}^z} \, \ ,
\eea
for the contribution outside of the wall. The contribution inside the wall (which we ignore) scales roughly as 
\bea
\frac{{\cal M}^{\rm inside}_{\rm wkb}}{{\cal M}^{\rm wkb \; red.}}\sim \Delta p_z L_w\ ,
\eea
which is always less than one. We conclude that the neglected corrections coming from inside of the wall contributions are indeed negligible for scalars.

\subsection{Procedure for the momentum transfer calculation: summary}
\label{sec:procedure}
In this section we summarise 
 the previous results and give a concise 
prescription for the momentum 
transfer calculation independent of the wall shape details. There are three contributions:
\bea
 \vev{\Delta p_L^{\rm step}} \ , ~~\vev{\Delta p_{R}^{\rm step}} \ ,~~
\vev{\Delta p^{\rm wkb}} \ ,
\eea
where the  first two  
correspond to emission to the left and right in the step wall regime and the last one to 
the emission to the right in the WKB regime. These are given explicitly by the following phase space integrals:
\begin{align}
    \vev{\Delta p_L^{\rm step}}=& 
\int_{0}^{k^z_{\rm max}} \dfrac{dk_z}{2\pi} {1 \over 2k_0}\int_0^{k_{\perp, \rm max}^2}\ \dfrac{dk_\perp^2}{4 \pi}\cdot \dfrac{1}{2p^z}\left[\dfrac{1}{2|q^z|} |\mathcal{M}_L|^2  (p^z-q^z+k^z)\right]
\Theta (L_w^{-1}-k^z) \ ,\nn
\vev{\Delta p_R^{\rm step}}=
&\int_{\Delta m}^{k^z_{\rm max}} \dfrac{dk_z}{2\pi} {1 \over 2k_0}\int_0^{k_{\perp, \rm max}^2}\ \dfrac{dk_\perp^2}{4 \pi}\cdot \dfrac{1}{2p^z
}\left[\dfrac{1}{2|q^z|} |\mathcal{M}_R|^2 (p^z-q^z-\tilde k^z)\right]\Theta(L_w^{-1}- k^z) \ ,\nn
\label{eq:all contributions}
\vev{\Delta p^{\rm wkb}}=
&\int_{\Delta m}^{k^z_{\rm max}} \dfrac{dk_z}{2\pi} {1 \over 2k_0}\int_0^{k_{\perp, \rm max}^2}\ \dfrac{dk_\perp^2}{4 \pi}\cdot \dfrac{1}{2p^z}\left[\dfrac{1}{2|q^z|} |\mathcal{M}^{\rm wkb \; red.}|^2 (p^z-q^z-\tilde k^z)\right] \nn
&\times\Theta\l( k^z-L_w^{-1}\r)\Theta\left(L_w^{-1}-(p^z-q^z-\tilde k^z)\right) \ ,
\end{align}
where the limits $k^z_{\rm max},k_{\perp, \rm max}^2$ were defined in eq. \eqref{eq:kperp max} but we repeat them here for the reader's convenience,
\bea
&&k_{z}^{\rm max}\equiv \sqrt{(p_0-m_\psi)^2- m^2}\ , \nn
&&k_{\perp, \rm max}^2\equiv{1 \over 4p_0^2}( p_0^2 +k_z^2+m^2-m_\psi^2)^2-k_z^2-m^2 \ .
\eea
and also recall that $\Delta m^2 \equiv \mt^2-m^2$.
$\mathcal{M}_L,\mathcal{M}_{R}$ are the amplitudes for the process calculated using the step wall ansatz (see section \ref{sec:amplitude}) and  $\mathcal{M}^{\rm wkb \; red.}$ is the amplitude in the WKB approximation, without the contribution inside the wall,
calculated following the discussion in section \ref{sec:emission_energ}.
Note the presence of various Theta functions imposing cuts on phase space. For the $\vev{\Delta p_{{L,R}}^{\rm step}}$ cases, these ensure that $k^z  < L_w^{-1}$, i.e. a step wall approximation is valid. 
Similarly  $\Theta (k^z -L_w^{-1})$ for the WKB regime.
For the latter, the second constraint $\Theta\left(L_w^{-1}-(p^z-q^z-\tilde k^z)\right)$ imposes that the
momentum transfer not surpass the inverse wall width (see discussion near \cref{eq:cut-fourier}). 
In practice, all constraints can be implemented by cutting the integration limits, as is shown in  Table \ref{tab:integration limits}.
The $\Theta (\pm(L_w^{-1}-k^z))$ are easily implemented by cutting the $k^z$ integration appropriately. For the WKB regime, the extra constraint $p^z-q^z-\kt^z < L^{-1}_w$ amounts to cutting also the $k_\perp^2$ integration as follows
\begin{align}
\label{eq:kcut_forWKB}
   k_\perp^2 \leq  
    \begin{cases}
        k_{\perp, \rm wkb}^{2} \ , & k^z < k^z_* \\
        k_{\perp, \rm max}^{2} \ , & k^z > k^z_*
    \end{cases}
    \qquad\quad \text{(WKB red.)}
\end{align}
where 
\begin{align}
    \label{eq: further WKB}
  k_{\perp, \rm wkb}^{2} &\equiv  {1 \over 4p_0^2}\left(p_0^2 + k_z^2+ m^2-m_\psi^2-(p^z-\tilde k^z-L_w^{-1})^2 \right)^2-  k_z^2- m^2 \ , \\
k^z_*&\equiv\sqrt{(p^z-L_w^{-1})^2+\Delta m^2} \ . 
\end{align}
To derive this, note that if $p^z-\kt^z-L_w^{-1}<0$, the positivity of $q^z$ means we are done with no extra condition. If instead $p^z-\kt^z-L_w^{-1}> 0$, squaring the constraint and solving for $\kp^2$ gives  \cref{eq: further WKB}.

\color{black}
\begin{table}[]
    \centering\footnotesize
\begin{tabular}{ |c||c|c|c|c| }
 \hline
 \multicolumn{5}{|c|}{\textbf{Phase space integration limits}} \\
 \hline
  & $L-$step&$R-$step&\multicolumn{2}{|c|}{WKB red.}\\
 \hline
$k_z$& $\biggl[ 0, {\rm Min}[L_w^{-1},k^z_{\rm max}] \biggr]$    &$\biggl[ \Delta m, {\rm Min}[L_w^{-1},k^z_{\rm max}] \biggr]$& $\biggl[ L_w^{-1}, {\rm Max}[L_w^{-1},k^z_*] \biggr]$&$\biggl[\ {\rm Max}[L_w^{-1},k^z_*], k^z_{\rm max}\ \biggr]$\\[.3cm] \hline
 $k_\perp^2$&   $[0, k_{\perp, {\rm max}}^2]$  & $[0, k_{\perp, {\rm max}}^2]$   &$[0, k_{\perp, \rm wkb}^2]$&$[0, k_{\perp, {\rm max}}^2]$\\
 \hline
\end{tabular}
 \caption{We report here all phase space integration limits for each emission contribution after explicitly taking into account constraints imposed by Theta functions in \cref{eq:all contributions}. The WKB regime is divided into two regions having imposed the constraint $\Delta p_z L_w<1$, as explained in section \ref{sec:procedure}. All definitions Max \& Min functions are included to also capture low $p^0$ or large $L_w^{-1}$. 
 }
    \label{tab:integration limits}
\end{table}

\subsection{Momentum transfer from scalar emission}
\label{sec:momentumTransferScalar}
\begin{figure}
    \centering
    \includegraphics[width=.48\textwidth]{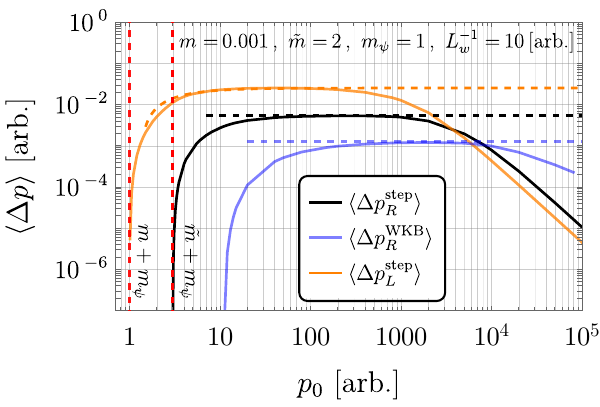}
     \includegraphics[width=.48\textwidth]{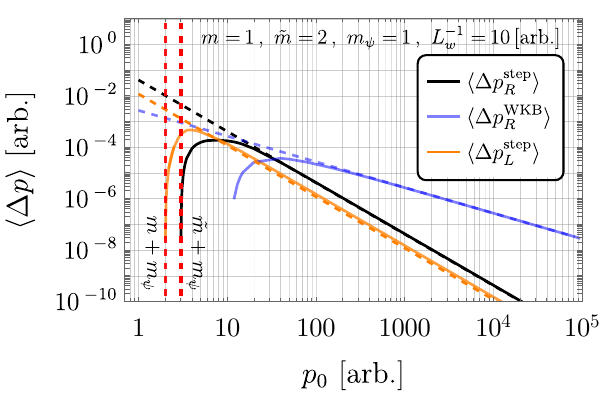}\\
      \includegraphics[width=.48\textwidth]{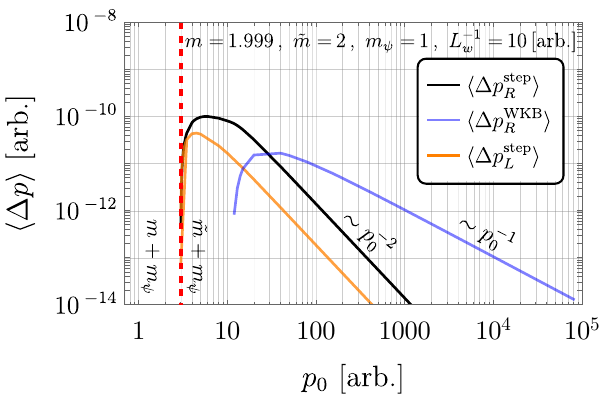}
      \includegraphics[width=.48\textwidth]{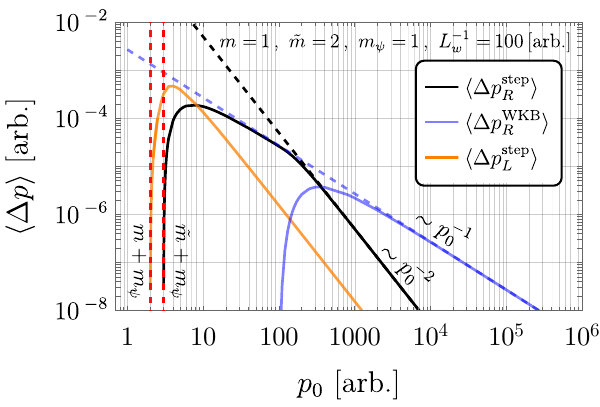}
    \caption{Numerical results for the averaged momentum transfer to the wall for each of the three contributions in eq. \eqref{eq:all contributions} as a function of incoming particle energy. Black and orange curves stand for right and left emission of soft quanta (using step wall), while blue is for the energetic quanta (using WKB). The vertical red dashed lines are the lower bounds on the minimal energy the incoming particle can have to emit right and left modes. 
    We show four cases of interest, that respectively are: $\mt \gg m$, $\mt \sim O(1) m$, $\mt - m \ll 1$ and $L_w^{-1} \gg \mt$. The horizontal and oblique dashed lines are analytical estimates in \cref{eq:WKBScalars} ($\vev{\Delta p^{\rm wkb}_{\zeta_R}}$), in \cref{eq:ScalarL step} ($\vev{\Delta p^{\rm step}_{\zeta_L}}$) and in \cref{eq:ScalarR step} ($\vev{\Delta p^{\rm step}_{\zeta_R}}$). 
    At the highest scales, the WKB contribution always dominates and falls as $p_0^{-1}$ (while both step wall contributions fall as $p_0^{-2}$). In the limit $L_w \rightarrow 0$, $R$ emission also falls off as $p_0^{-1}$. Most interestingly, for large hierarchy $m_\psi / m$, $\vev{\Delta p}$ is constant until $p_0 \lesssim \mt m_\psi / m$. 
    }
    \label{fig:PSI}
\end{figure}

Using the expressions summarised in \cref{sec:procedure},
we calculate the momentum transfer from scalar emission $\psi \rightarrow \psi \phi$ in our toy model \cref{eq:2scalarlag}. As explained above, the total averaged momentum transfer is the sum of three separate contributions
$\vev{\Delta 
p^{\rm step}_{L,R}}$, $\vev{\Delta 
p^{\rm wkb}_{R}}$.
Numerical integration is relatively straightforward and representative results are shown in \cref{fig:PSI}. Analytical expressions can be derived, with some details given in \cref{app:Scalars}, and are presented when deemed useful.  
There are several parameters in the problem so which of the three contributions dominates is a function of different 
hierarchies. 
Generically we find that at large energies  $p^0 \rightarrow \infty$ the WKB 
contribution always dominates and falls off as $1/p^0$. 
Computing the phase space integrals in this asymptotic limit, we obtain a very good approximation
\begin{align}
\label{eq:WKBScalars}
&\left.\langle \Delta p^{\rm total} \rangle\right|_{p_0 \gg L^{-1}_w {\rm Max}[1,\:m_\psi/m]} \approx \langle \Delta p_{R}^{\rm wkb}\rangle \approx    {y^2\mt  \over 32 \pi^2 m_\psi^2}  \ \times\\
&
 {p_0^{-1} \over 2 \mt} 
    \left[2 \mt^2 \ln\left(\frac{m}{\mt}\right) 
      + \frac{2 (m^2 + \mt^2) m_{\psi}^2-m^2 \mt^2  }{S(m)}
         \ln\left[D(m)\right] 
      + S(\mt)\ln\left[D(\mt)\right]\right] \nonumber \ ,
\nonumber 
\end{align}
where
\begin{align}
\quad D(m) \equiv \frac{m^2 - 2 m_{\psi}^2 + S(m)}{2 m_{\psi}^2} \ , \quad S(m) \equiv i m \sqrt{4 m_{\psi}^2-m^2}  \nonumber \ .
\end{align}
On the other hand, at low and intermediate relativistic energies, the step function contributions typically dominate. This behaviour is amplified in two independent regimes. For very thin wall $L_w\rightarrow 0$ the WKB contribution naturally only turns on at higher energies $p_0 \gg L^{-1}_w$, with $\langle \Delta p_{R}^{\rm step}\rangle$ temporarily dominating in its place (see bottom-right panel of \cref{fig:PSI}). Nonetheless, the trend is still reasonably approximated by interpolating backward the asymptotic result of \cref{eq:WKBScalars}.

More interesting is the case when the initial mass of $\phi$ is very light $m\ll \mt,m_\psi$. Each contribution to momentum transfer becomes constant for an inter-relativistic plateau as can be observed in the leftmost panel of \cref{fig:PSI}. 
Moreover, it is actually $\langle \Delta p_{L}^{\rm step}\rangle$ that dominates, with a value
\begin{align}
\label{eq:delta p L scalar} 
   \vev{ \Delta p^{\rm total} } \approx \langle \Delta p_{L}^{\rm step}\rangle &\approx 
     {y^2 \mt \over 8\pi^2 m_\psi^2} \qquad \text{for} \quad p_0 \lesssim \mt m_\psi / m \ .
\end{align}
Further details can be found in \cref{app:Scalars}.

\subsection{Pressure from scalar emission}
\label{sec:PressureScalar}
The pressure induced on the bubble wall from scalar emission is finally obtained by integrating over the incoming flux
\bea
{\cal P}=\int \frac{d^3 p}{(2\pi)^3}f_\psi (T,\gamma_w)\times\frac{p^z}{p_0} \vev{\Delta p^{\rm total}} \ ,
\eea
where $f_\psi$ is the phase space distribution of incoming $\psi$ particles in the frame of the wall.
Let us first focus  on the interesting regime  when $m\ll m_\psi$, identified in \cref{eq:delta p L scalar}, where $\vev{\Delta p}^{\rm total}$ is constant for a plateau lasting till $p_0 \lesssim \tilde m m_\psi/m$. 
In the ultra-relativistic limit $p^z/p_0\to 1$, and the integration gives simply
\bea
{\cal P}^{\rm scalar}= n_\psi(T,\gamma_w)\vev{\Delta p^{\rm total}}= \gamma_w n_\psi(T) \vev{\Delta p^{\rm total}} \ , \qquad p_0 \lesssim \frac{\tilde m m_\psi}{m} \ .
\eea
To go from the second equality to the third we used that the number density of $\psi$ in the wall frame is boosted with respect to the FRW frame number density. Thus, even scalar emission can cause friction pressure that grows with $\gamma_w$.
Combining the plateau and asymptotic contributions for scalar emission, we have approximately, for an incident thermal population,
\bea
\label{eq:scalarpressure}
{\cal P}^{\rm scalar}\approx 
\begin{cases}
  \frac{ T^3}{8\pi^4} {y^2 \mt \over  m_\psi^2}\times \gamma_w  \ , &  \qquad \gamma_w \lesssim \frac{\tilde m m_\psi}{m T}\\
  \frac{  T^2}{64\pi^3}{y^2 \tilde m^2  \over m_\psi m} \ , & \qquad \gamma_w \gtrsim \frac{\tilde m m_\psi}{m T}
\end{cases} \ ,
\eea
where the second expression was obtained by expanding \cref{eq:WKBScalars} for small $m$ and simply multiplying by   {$n_\psi(T)\approx T^3/\pi^2$}. The asymptotic constant pressure can be compared with the LO contribution from a thermal population of $\phi$ particles crossing the wall as per \cref{eq:LOpressure}, which gives $\sim T^2 \mt^2$. We observe that the NLO contribution is more important when 
${y^2   \over 64\pi m_\psi m} \gtrsim 1$. 

 We remind the reader that our results for scalar emission pressure were obtained for a particularly simple choice of interactions in the Lagrangian \cref{eq:2scalarlag} (with $y=\rm const$) and we highlighted here a particular regime of parameters. Our main goal was to mimic as much as possible the vector radiation to be discussed in the next section.  We will return to whether the pressure in \cref{eq:scalarpressure} can be phenomenologically relevant in future work.

\section{Spontaneously broken gauge theories}
\label{sec:Spontaneously broken Gauge theories}

We now proceed to the phase transitions related to the spontaneous breaking of gauge symmetry and the emission of vector bosons.
The procedure will in essence be exactly the same as what we presented for the case of the 
scalar emission. We will quantise the theory of a gauge field in the background of a domain wall interpolating between a symmetric and broken phase. As should be expected, the extra difficulty 
will involve dealing with gauge-fixing, spin and the change of degrees of freedom due to the spatially-dependent rearrangement of the vacuum.

For simplicity, we will consider the Abelian Higgs model of a charged complex scalar $H$, whose potential $V(\sqrt{2}|H|)$ is responsible for the spontaneous breaking of the $U(1)$ gauge symmetry. A second scalar field $\psi$ charged under the same $U(1)$ will play the role of matter; its potential is trivial. The Lagrangian is 
\begin{align}
    \begin{split}
\label{eq:AbelianHiggsmodel}
    \mathcal{L} &= -\frac{1}{4} F_{\mu \nu} F^{\mu \nu}+| D_\mu H|^2 -V\left(\sqrt 2 |H|\right) + | D_\mu \psi|^2 -m_\psi^2 |\psi|^2  ,  
\end{split}
\end{align}
where we are using the convention $D_\mu \psi= (\d_\mu +ig A_\mu)\psi$ and $A_\mu$ is the vector gauge field. We will in general not need to commit to a specific potential but will simply assume that it has two minima at $\sqrt 2 |H| = v,\tilde v$, where $v \rightarrow 0$ corresponds to the symmetric phase. We will quantise the theory in the background of a domain wall $\sqrt 2\langle |H| \rangle = v(z)$. 
We will also be interested, both as a computational tool and as a phenomenological case in its own right, in imagining a distorted or more general class of potential with non-zero $v < \tilde v$. We will call this scenario a \emph{broken to broken} phase transition, in opposition to the more familiar case of \emph{symmetric to broken} phase transitions.

To work with the theory described by \cref{eq:AbelianHiggsmodel}, one has to make two independent choices: what field coordinates to use for $H$, such as Cartesian or polar, and what gauge to impose. The value of each choice is determined by the particular application. Much of the following pages will be dedicated to arguing for the most convenient choices for our application.

If we are interested in studying the geometry of the vacuum manifold, polar coordinates $H={1 \over \sqrt{2}}(h+v) e^{i {\theta}}$ are most convenient. The potential depends only on the modulus. In the symmetric phase however, this coordinate choice is singular.
On the other hand,  Cartesian coordinates are well-defined everywhere
\bea 
H={1 \over \sqrt{2}} (h_1+i h_2) \equiv {1 \over \sqrt{2}} \left(h + v(z) + i h_2 \right) \ , 
\eea 
where we have expanded around the background solution $h_1=v(z)$.
The Lagrangian for the Higgs and gauge fields becomes
\begin{align}
\begin{split}
\label{eq:Lag_expanded_quandratic_AH}
     \mathcal{L}_{A,H} &= -{1 \over 4}F_{\mu\nu}F^{\mu\nu} 
     + \frac{1}{2} (\partial h)^2 + \frac{1}{2} (\partial h_2)^2 - g A^\mu \left[ h_2 \partial_\mu v(z)- v(z) \partial_\mu h_2 \right] \\ 
     & \quad  + {1 \over 2} g^2v^2(z) A^2 - \frac{1}{2}\d_1^2V(v(z))h^2 - \frac{1}{2}\d_2^2V(v(z)) h_2^2+ \dots \ , 
\end{split} 
\end{align}
up to quadratic terms, where $\d_i\equiv \d/\d h_i$, so that the last two terms are the $z$-dependent mass terms of $h$ and $h_2$.

 At a non-zero minimum of the potential, $h_2$ becomes massless and is the would-be Nambu-Goldstone boson (NGB). As always for gauge theories, when $v\neq0$ a mixing term appears between this goldstone boson and the gauge boson. However, in the context of a varying background, there is also an extra mixing proportional to $\d_z v$. When $v$ is a constant, the mixing can be eliminated completely while also gauge fixing by adding the so called $R_\xi$ gauge term
\bea 
\delta \mathcal{L}_{g.f.}=-{1 \over 2 \xi}(\d_\mu A^\mu -\xi gv h_2)^2 \ ,
\eea 
and integrating by parts.
For $v=v(z)$, adding this same (now $z-$dependent) gauge-fixing term does not get rid of mixing entirely but localises it to the region of the wall
\begin{align}
\begin{split}
\label{eq:Lag_quandratic_Rxi}
    \mathcal{L}_{A,H} + \delta \mathcal{L}_{g.f.} =& -{1 \over 4}F_{\mu\nu}F^{\mu\nu} -{1 \over 2 \xi}(\d_\mu A^\mu)^2
     + \frac{1}{2} (\partial h)^2 + \frac{1}{2} (\partial h_2)^2 \\ 
     & -2g h_2 A^z \d_z v(z) + {1 \over 2} g^2v(z)^2 A^2 - \frac{1}{2}\d_1^2V(v(z))h^2 \\
     &- \frac{1}{2}\left[\d_2^2V(v(z))  + \xi g^2 v^2(z) \right] h_2^2+ \dots \ .
\end{split} 
\end{align}

\subsection{Particle content in the asymptotic regions}
We briefly remind ourselves of the spectrum of the theory in the asymptotic regions $v=0 $ and $\tilde v$ at $z\rightarrow -\infty$ and $z \rightarrow \infty$ respectively, before discussing the full interpolating space.

\paragraph{Symmetric phase:}
The theory \cref{eq:Lag_quandratic_Rxi} around the symmetric point minimum $v=0$ describes two scalars $h_{1,2}$, with equal mass by symmetry
\begin{align}
    m_{h,s}^2 \equiv \d_1^2V(0) = \d_2^2V(0)
\end{align}
The gauge-fixing affects only the Maxwell equations of motion for the massless vector $A^\mu$. Whatever the value of $\xi$, we can identify two physical transverse (in the sense that $A_\mu k^\mu$=0) degrees of freedom with polarisation vectors given by
\begin{align}
    \label{eq:tradTransverse}
\epsilon^\mu_{T_1}=(0,0,1,0) \ ,  \qquad \epsilon^\mu_{T_2}=\frac{1}{\sqrt{k_\perp^2+k_z^2}}(0,k^z,0,-k_\perp) \ ,
\end{align}
for $k^\mu=(k_0, k_\perp,0,k^z)$.
The only (well-known) subtlety involves imposing on the Hilbert space a constraint to project out unphysical states, the so-called Gupta-Bleuler condition \cite{tong2006lectures}.


\paragraph{Broken phase(s):}
At a symmetry breaking minimum $\d_2^2V(\tilde v)=0$ and $h_2$ describes the would-be NGB.
A particularly convenient choice is `unitary gauge', corresponding to $\xi \rightarrow \infty$ in which the NGB decouples completely, making manifest the spectrum. We are left with a single massive scalar (the Higgs) $h$ with mass squared equal to $\partial_1^2V(\tilde v)$ and 
a massive vector boson $A_\mu$ with mass 
\bea
\mt \equiv g \tilde v\ ,
\eea
and satisfying the Proca equation
\begin{align}
    \partial_\mu F^{\mu\nu}+ \mt^2 A^\nu = 0 \ .
\end{align}
This reduces to a Klein-Gordon equation for each component of $A^\mu$ supplemented by the Lorentz condition: 
\begin{align}
 \implies \; & \partial^2 A^\mu + \mt^2 A^\mu = 0 \ ,  \qquad \partial_\mu A^\mu = 0 \ .
 \label{eq:StandardLorenz}
\end{align}
Solving this is straightforward and one adds to the transverse polarisations of \cref{eq:tradTransverse} a third longitudinal one parallel to $3-$momentum
\begin{align}
 \label{eq:tradLongitudinal}
    \begin{split}
\epsilon_{L}^\mu=\l(\frac{k_0^2-m^2}{k_0},k_\perp,0, k^z\r)\frac{k_0}{m\sqrt{k_0^2-m^2}}.
\end{split}
\end{align}

\subsection{Global degrees of freedom}
\label{sec:Global_degrees_of_freedom}
Here will analyse the fields defined over the entire region and identify the appropriate global modes to quantise, where by global we simply mean they are good across the wall.

In principle, one could choose a convenient value of $\xi$ in \cref{eq:Lag_quandratic_Rxi} and push ahead with quantisation. However, we would have to deal with mixing when solving for the mode functions, as well as taking care to impose a non-trivial Gupta-Bleuler like condition on physical states. Luckily, we will argue that even when asymptotically approaching the symmetric point as $z \rightarrow - \infty$, it is possible to work with unitary gauge $\xi \rightarrow \infty$ with impunity. This approach was already made at the classical level in ref.\cite{Farrar:1994vp}, and we will re-derive and tweak some of their results using a slightly different language, before quantising.

In unitary gauge, the $h_2$ degree of freedom decouples and the theory becomes
\begin{align}
\begin{split}
\label{eq:Lag_quandratic_Unitary_gauge}
    \mathcal{L}^{U.G.}_{A,H}  = &-{1 \over 4}F_{\mu\nu}F^{\mu\nu} 
      + {1 \over 2} g^2 v^2(z) A^2 +  \frac{1}{2} (\partial h)^2 - \frac{1}{2}\d_1^2V(v)h^2 + \dots \ , 
\end{split} 
\end{align}
and the equations of motion for \cref{eq:Lag_expanded_quandratic_AH} reduce to just two uncoupled equations
\bea
&&\Box h= - V''(v) h \, ,
\\
\label{eq:system_eq_unit}
&&\partial_\nu F^{\mu \nu}=   g^2 v^2(z) A^\mu \,.
\eea
The first one is the equation of motion for the physical Higgs boson $h$ and we will not have any more to say about it here. The second will be the focus of our attention. While $v(z)>0$, the theory is always in a broken phase and $A^\mu$ describes a massive vector. Unitary gauge is then manifestly a valid choice. 
Note that the usual transversality condition for massive vector bosons in this case becomes
\bea
\label{eq:Generalised_Lorentz}
\d_\mu \d_\nu F^{\mu\nu} =  \d_\mu (g^2v^2 A^\mu)=0 \, , \\
\implies \d_n A^n + \d_z A^z = - {\d_z v^2  \over v^2}A^z\ ,
\label{eq:Generalised_Lorentz_condition}
\eea
which, in the presence of the domain wall, generalises the standard Lorentz condition for massive electrodynamics in \cref{eq:StandardLorenz}.
The constraint above ensures this vector field has three polarisation degrees of freedom. Subbing this back into \cref{eq:system_eq_unit} we get
\begin{align}
\label{eq:Constrained_EOM}
    \partial^2 A^\mu +\partial^\mu \left[\left( \d_z v^2 \over v^2 \right) A^z\right] + g^2 v^2 A^\mu = 0 \ .
\end{align}
The general Fourier mode  can be written as
\bea
A^\mu_{k_0,\kp} =e^{-i k_n x^n}\sum_{l} a_l \;\chi_{l,k^n}^\mu  (z) \, ,
\eea
where $l$ runs over three indices, $a_l$ are some constant Fourier coefficients and the functions $\chi_{l,k_0,\kp}^\mu  (z)$ have to be found by solving \cref{eq:Constrained_EOM}. One might be tempted to define left and right moving $\chi^\mu$ modes by fixing the incoming piece at infinity in terms of conventional transverse and longitudinal modes \cref{eq:tradTransverse,eq:tradLongitudinal}.
However, under such a choice, transmitted and reflected pieces would contain also other polarisations. Fundamentally, this is because
for non-zero $\kp$, rotations around $\vec{k}$ are not a symmetry and conventional spin is not conserved. We now construct more convenient `wall polarisations', which instead do not mix.

\paragraph{$\tau$ polarisations: } It is useful to define what we call $\tau-$polarisations by the condition $A^z=0$, since for these we recover the Lorentz condition $\d_\mu A^\mu = 0$, which in Fourier space reduces to 
\begin{align}
   k_{n} \chi_{\tau}^{n} \equiv  k_0 \chi ^0_\tau -  k_\perp \chi ^\perp_\tau = 0\ ,
\end{align}
and has solutions in terms of two \textit{constant} vectors 
\begin{align}
    \chi_\tau^\mu &= \epsilon_{\tau_{1,2}}^\mu \; \chi_{\tau_{1,2}}(z) \ ,
\end{align}
where
\begin{align}
\epsilon_{\tau_1}^\mu =& (0, 0,1,0), \qquad \epsilon_{\tau_2}^\mu = (k_\perp, k_0, 0, 0)/\sqrt{k_0^2 - k_\perp^2}\ ,
\end{align}
and the equations of motion \cref{eq:Constrained_EOM} become the Klein-Gordon-like
\begin{align}
    \label{eq:Eom_chiTau}
  {\left[-E^2-\partial_z^2 + g^2 v^2(z) \right] \chi_{\tau_{1,2}}(z) = 0}  \ ,
\end{align}
where we remind the reader of the definition $E = \sqrt{k_0^2-k_\perp^2}$. 
Note that $\epsilon_{\tau_1}^\mu$ is one of the standard transverse polarisations, but $\epsilon_{\tau_2}$ is not, since it has a non-zero time component, and is not orthogonal to three momentum. The wave equation to solve across the wall for $\tau$ d.o.f. is thus identical to the scalar case studied in \cref{sec:scalar}.

\paragraph{$\lambda$ polarisation:} It remains to solve for the remaining degree of freedom with $A^z \neq 0$. The wave equation to solve for is obtained by setting $\mu=z$ in \cref{eq:Constrained_EOM}. We note already that it is 
significantly more complicated than what we found for $\tau$. Whatever the solution for $A^z$, the 
other components of the vector are fixed. Requiring orthogonality with $\chi_\tau^\mu$ implies the form 
$\chi_{\lambda}^\mu(z) = \left(- i k^{n} \alpha(z) , \chi^z_\lambda \right)$ where we recall $n=0,\perp$.
The generalised Lorentz condition in \cref{eq:Generalised_Lorentz_condition} immediately leads to the relation
 \bea
 \label{eq:alpha_Def}
 \alpha(z) = {\partial_z \left( v^2 \chi^z_\lambda \right)\over  E^2 \; v^2} \ .
\eea
Plugging this back into the  \cref{eq:Constrained_EOM}, we obtain the equation in terms of the $\chi^z$ only
\begin{align}
\label{eq:eq_JustAz}
    -E^2 \chi^z_{\lambda} - \d_z \left(\frac{1}{v^2} \d_z (v^2 \chi^z_{\lambda})\right) + g^2 v^2(z) \chi^z_{\lambda} = 0 \ .
\end{align}
We can get rid of the linear in derivative term if we introduce a new function $\lambda(z)$
\bea
\c_\lambda^z={E \over  {g v(z)}}\lambda(z) \ ,
\eea 
and eq. \eqref{eq:eq_JustAz} becomes Schrodinger-like 
\begin{align}
    \label{eq:EOM_lambda}
\left(-E^2-\d_z^2 +  U_\lambda(z) \right)\lambda=0 \ ,
\end{align}
with effective  potential 
\begin{align}
\label{eq:Potential_lambda}
\begin{split}
   U_\lambda(z) = & \; g^2 v^2(z)  - v\;\d_z \left( \d_z v \over v^2 \right) \\ 
   = & \; g^2 v^2(z)+2
   \left( \d_z v \over v\right)^2-{\d_z^2 v  \over v} \ .  
\end{split}
\end{align}
The solutions in terms of $\lambda$, unlike $\chi_\lambda^z$, satisfy the usual orthogonality relations
\bea
\int d z \lambda_k(z)\lambda_{q}^*(z)=2 \pi\delta(k^z-q^z) \ .
\eea
It is easy to prove $-$ and we do so explicitly in \cref{sec:LambdaProof} $-$ that for an interpolating solution $v(z)$ of a completely arbitrary Higgs potential $V$, we have $(v'/v)^2,v''/v \rightarrow m_{h,s}^2$ as $z\rightarrow - \infty$ so that $U_\lambda(z)$  is always finite even if $v =  0$. More precisely
\begin{align}
\label{eq:Ulambda Limits}
    \lim\limits_{z\to -\infty} U_\lambda(z)  = 
\begin{cases}
     g^2 v^2 \ ,    &   v \neq 0\\
    m_{h,s}^2  \ , &   v = 0
\end{cases} \
\equiv m_\lambda^2 \ .
\end{align}
We see that $\lambda$ is the perfect cross-wall field. It interpolates between one of the massive (Higgs) degrees of freedom on the symmetric $v=0$ side $z\rightarrow -\infty$ and a third component of the massive vector in the broken region $z\rightarrow \infty$. If instead $v \neq 0$, $\lambda$ simply interpolates between the different mass vectors.

Let us look more closely at the $A^\mu$ vector formed by the $\lambda$ field. 
Using the equations of motion forces the  on-shell relation:
\bea
\label{eq:epslambda-der}
\c^z_\lambda&=&\frac{E^2 \d_z\alpha}{E^2-g^2 v^2(z)}\Rightarrow \nn
\chi_\lambda^\mu &=& \left( - i k^n \alpha(z) , \chi_\lambda^z \right) =   (-i k^n,  \d^z) \;\alpha(z)  +  {g^2 v^2(z) \over E^2} \chi_\lambda^z (0  ,  0 , 0  , 1)\nn
&=&(-i k^n,  \d^z) \;\alpha(z)+  {g v(z) \over E} \lambda(z) (0  ,  0 , 0  , 1)
\eea
So that the vector can be written as a total derivative plus a term sub-leading in energy. This form will turn out to be very useful in calculating the amplitudes for physical processes. Far from the wall, when $v\to const$, we can introduce the polarisation vector $\epsilon_\lambda^\mu$ such that
\bea
\label{eq:epslambda-pol}
\chi_\lambda^\mu&\propto& \epsilon_\lambda^\mu,~~~\epsilon^{\lambda}_\mu\epsilon^{\lambda^\mu}=-1\nn
\epsilon_\lambda^\mu &=& \left( k^n {k^z \over E^2} \ , 1 \right) \times \frac{E}{g v} \nn
&=& {k^z \over E g v}\;k^\mu + {gv  \over E}(0\ ,0 \ ,0 \ , 1)\ .
\eea
We emphasise again that these $\lambda$ and $\tau$ differ from the conventional transverse and longitudinal polarisations. Far from the wall, all polarisations satisfy the same equation of motion and one can use any linear combination of either basis to decompose the vector field. 
We can relate the wall polarisations to the conventional longitudinal and transverse ones in \cref{eq:tradTransverse,eq:tradLongitudinal}
via the rotation matrix
\bea
\l(\baa{c}\epsilon_{T_1}\\
\epsilon_{T_2}\\
\epsilon_{L}\eaa\r)=
\l(\baa{ccc}1 &0&0 \\
0& \frac{k_0 k^z}{E\sqrt{k_0^2-m^2}}&-\frac{k_\perp m}{E\sqrt{k_0^2-m^2}}\\
0& \frac{k_\perp m}{E\sqrt{k_0^2-m^2}}&\frac{k_0 k^z}{E\sqrt{k_0^2-m^2}}\\
\eaa\r)
\l(\baa{c}\epsilon_{\tau_1}\\
\epsilon_{\tau_2}\\
\epsilon_{\lambda}\eaa\r) \ .
\eea
In the case of a very large $k^z, E\gg k_\perp, m$ the mixing angle between $\tau,\lambda$ transverse and longitudinal scales as $m/E$. 
We can see that the two bases of polarisations are exactly the same for the 
case $k_\perp=0$. This is expected since  for this configuration of momenta $T_1,T_2$ 
polarisations have zero components in the $z$ direction. Using the unbroken part of the Lorentz symmetry (boosts in 
$x-y$ direction) we can obtain the polarisations for generic momenta, which 
indeed agrees with the $\tau,\lambda$ basis derived before.
Our goal in this paper is to calculate the total pressure acting on the domain wall and for this, we sum the contributions from all polarisations. We perform all of the calculations in the $\tau, \lambda$ basis, without even reporting the results for $T,L$ polarisations.

\begin{figure}
    \centering
    \includegraphics[width=.48\textwidth]{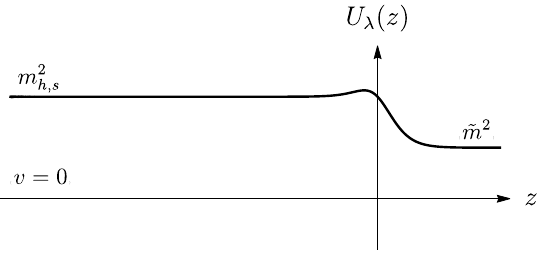} \includegraphics[width=.48\textwidth]{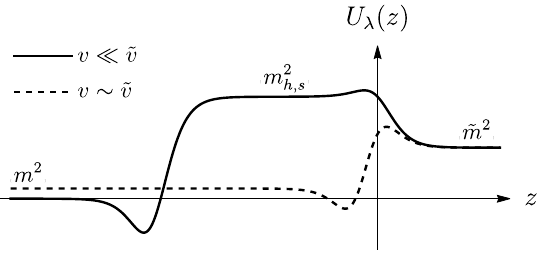}
    \caption{The potential for the $\lambda$ degree of freedom, which is defined everywhere. Recall $m = g v$ and $\mt= g \tilde v$, plus thermal corrections. As $v\rightarrow 0$ a growing plateau develops ending at $z \sim m_{h,s}^{-1}\ln\left(v/\tilde v\right)$ with value $m_{h,s}^2$ - the symmetric side mass of the Higgs in the case $v=0$ exactly. These curves were drawn using the explicit profile \cref{eq:Tanh_Broken_to_Broken}. 
    \label{fig:LambdaPotential}}
\end{figure}

\subsubsection{$v\to 0$  limit}
\label{sec:v=0}
All of the previous discussion 
applied most manifestly for the case when  the vev of the symmetry 
breaking field is $ v \neq 0$. What happens in 
the case when the domain wall on one side approaches a vacuum
where the gauge symmetry is unbroken?
We saw in the previous section that the potential of the $\lambda$ mode has a property that as $v\to 0$ then $U_\lambda(z)\to m_{h,s}^2$, which together with our expectations from the Higgs phenomena hints that on the unbroken side $\lambda$ should correspond to the would-be NGB
\bea
\lambda_{z\to -\infty}\to h_2 \,.
\eea
To understand this matching better, let us look at the $\chi^\mu_\lambda$ vector in the limit $v\to 0$
\bea
\label{eq:chimu}
\chi^\mu_\lambda&=&\left( - i k^n \alpha(z) , \chi_\lambda^z \right)=\l( \frac{-i k^n \d_ z (v \lambda)}{g E v^2},\frac{E}{g v}\lambda\r)\nn
&=&\frac{\lambda}{g v }\l( -\frac{i k^n}{E}\l[\frac{v'}{v}+\frac{\lambda'}{\lambda}\r],E\r)_{v\to 0}
=\frac{e^{-i k x}}{g v}\l(\frac{k ^n}{E}\l[-i m_{h,s}+ k^z\r], E\r),
\eea
where we have used that $\lambda$ becomes a plane wave far from the wall and $v'/v\to m_{h,s}$.
Note that the factor $(-i m_{h,s}+k^z)/E$ is a pure phase if the $\lambda$ dof is on shell. Let us see whether we can build exactly the same vector but from the field $h_2$. Indeed if we consider the vector
\bea
\d^\mu\l(\frac{h_2}{g v}\r)=-\frac{e^{-i k x}}{g v}\l(k^n, k^z+i \frac{v'}{v}\r)=\frac{-i E e^{-i k x}}{g v(k^z-i m_{h,s})}\l(\frac{k ^n}{E}\l[-i m_{h,s}+ k^z\r], E\r).
\eea
So we can see, comparing with \cref{eq:chimu}, that the two vectors $\chi_\lambda^\mu$ and $\d_\mu (h_2/gv)$ are exactly the same apart from the constant phase factor, so indeed $\lambda$ field in the $z\to -\infty$ limit corresponds to the Goldstone boson.

What about starting from a finite value and taking $v \rightarrow 0$? For concreteness let us consider the potential 
\bea
\label{eq:potential-cusp}
V=\lambda_{h} (|h|-v)^2 (|h|-\tilde{v})^2.
\eea
First of all the potential has a cusp at $|h|=0$ so the limit $v\to 0$ becomes discontinuous. This can be seen also in the form of $U_\lambda(z)$. As $v$ becomes smaller a longer finite plateau develops in the potential with value $\approx m_{h,s}^2$, as is shown in \cref{fig:LambdaPotential}. No matter how small $v$ \textit{eventually} the potential turns down and asymptotes to $g^2 v^2$ as is to be expected. Thus for any finite though tiny $v$ the asymptotic states at $z\rightarrow-\infty $ are those of a massive vector.

\subsection{The step wall case}
\label{sec:stepwall}
In order to proceed further we need to solve the equations of motion. In general, it is a complicated problem depending on the shape of the effective potential $V_{\rm eff}(h)$ at the time of the phase transition. One needs to find the solitonic solution $v(z)$ connecting false and true vacuum and later the wavemodes describing perturbations of each field on this background. In the particular case of the domain wall
\begin{align}
\label{eq:Tanh_Symmetric_to_Broken}
v(z) =  \frac{1}{2} \tilde v \left(1+\tanh \left( z \over L_w\right)\right)\ ,
\end{align}
solutions were found in \cite{Farrar:1994vp} in terms of hypergeometric functions. 
In this paper, we will consider an even simpler case, namely a step function ansatz for the wall. This approximation will of course be valid only if the momentum of the particle during the passage is (much) less than the inverse width of the wall $k,\kt \ll L_w^{-1}$. Typically, this width is controlled by the mass of the Higgs $L_w^{-1}\approx m_{h}$.

The solution of the equations of motion can be written down on each side immediately and the only challenge becomes deriving and implementing matching conditions.  In this section, we report the matching conditions for $\tau$ and $\lambda$ polarisations and write down the corresponding wave functions. We will do so first for the broken to broken case $v>0$. As an explicit example, we can imagine distorting \cref{eq:Tanh_Symmetric_to_Broken}
 to 
 \begin{align}
 \label{eq:Tanh_Broken_to_Broken}
    v(z) \rightarrow  v + \frac{1}{2} \left(\tilde v - v\right) \left(1+\tanh \left( z \over L_w\right)\right) \ .
\end{align}
We then comment on the $v \rightarrow 0$ limit, which is straightforward for $\tau$ degrees of freedom but more delicate for $\lambda$.

\subsubsection{$(\tau)$ polarisations}
\label{sec:tau-pol-step}
For the $\tau$ polarisations, we showed in \cref{sec:Global_degrees_of_freedom} that the equations of motion are exactly the same as for the scalar field $(\Box A^\mu_{(\tau)}=-g^2 v^2 A^\mu_{(\tau)})$ and so for the step wall the matching conditions become:
\bea
\chi_{\tau_{i}}|_{<0}&=\chi_{\tau_{i}}|_{>0} \ , \\ 
\d_ z \chi_{\tau_{i}}|_{<0}&=\d_ z \chi_{\tau_{i}}|_{>0} \ ,
\eea
with $i=1,2$. The reflection and transmission coefficients are thus the same as for scalars and the wave functions become
\begin{align}
\label{eq: chi R}
  \chi_{\tau_{i},R,k}^\mu(z) & =
  \epsilon_{\tau_i}^\mu
   \begin{cases}
            e^{i k^z z} +  r^\tau_{k}  e^{-i k^z z}  \ , & z<0 \\
            t^\tau_{k} \; e^{i \tilde{k}^z z}  \ , & z>0 
    \end{cases} \ ,\\ 
    \label{eq: chi L}
     \chi_{\tau_{i},L,k}^\mu(z) & =  \epsilon_{\tau_i}^\mu
    \sqrt{k^z \over \tilde k^z}\begin{cases}
     \frac{\tilde{k}^z}{k^z }t^\tau_{k} \; e^{-i k^z z}
             \ , & z<0 \\
             e^{-i \tilde{k}^z z} -  r^\tau_{k}   e^{i \tilde{k}^z z}  \ , & z>0 
    \end{cases}\ ,
\end{align} 
where
\begin{align}
    r^\tau_{k} &= \frac{k^z - \tilde{k}^z}{k^z + \tilde{k}^z}\ ,~~~ t^\tau_{k} = \frac{2 k^z  }{k^z +  \tilde{k}^z}  \ , \\
        {k_0} &  {= \sqrt{k_z^2+m_\tau^2+k_\perp^2} \ , \quad \tilde{k}^z = \sqrt{k_0^2-\tilde{m}_\tau^2-k_\perp^2} \ , \quad m_\tau\equiv g v \ , \quad \mt_\tau \equiv g \tilde v} \ . \nonumber
\end{align}
Taking the $v \rightarrow 0$ limit for these degrees of freedom is simple and we approach the $v=0$ case smoothly. 

\subsubsection{$(\lambda)$ polarisation}
\label{sec:lambda-pol-step}

The $\lambda$ modes require more work. Again we first focus on the broken to broken case of $v>0$. Matching conditions are easy to derive by 
integrating the wave equation for $\lambda$ once and twice respectively (most easily done at the level of \cref{eq:eq_JustAz}). These are 
\begin{align}
\label{eq:matching-lambda}
  \left. { \partial_z\lambda \over v(z)}\right|_{<0} &= \left. \partial_z \lambda \over v(z)\right|_{>0} \ , \nn
    \left.v(z)\lambda \right|_{<0} &= \left.  v(z) \lambda\right|_{>0} \ ,
\end{align}
which allows us to write down the expressions for (in-state) `left' and `right' movers:
\begin{align}
\label{eq: lambda R}
  \lambda_{R,k}(z) & =
    \begin{cases}
            e^{i k^z z} +  r^\lambda_{k}  e^{-i k^z z}  \ , & z<0 \\
            t^\lambda_{k} \; e^{i \tilde{k}^z z}  \ , & z>0 
    \end{cases}\\ 
    \label{eq: lambda L}
    \lambda_{L,k}(z) & =  
    \sqrt{k^z \over \tilde k^z}\begin{cases}
     \frac{\tilde{k^z}}{k^z}t^\lambda_{k} e^{-i k^z z}
             \ , & z<0 \\
             e^{-i \tilde{k}^z z} -  r^\lambda_{k}   e^{i \tilde{k}^z z}  \ , & z>0 
    \end{cases}
\end{align} 
where
\begin{align}
\label{eq:rt-lambda}
    r^\lambda_{k} &= \frac{\tilde v^2 k^z - v^2 \tilde{k}^z}{\tilde v^2k^z + v^2\tilde{k}^z} \ , \quad  t^\lambda_{k} = \frac{2 k^z v \tilde v }{\tilde v^2k^z + v^2\tilde{k}^z} \ , \\
        {k_0} &  {= \sqrt{k_z^2+m_\lambda^2+k_\perp^2} \ , \quad \tilde{k}^z = \sqrt{k_0^2-\tilde{m}_\lambda^2-k_\perp^2} \ , \quad m_\lambda\equiv g v \ , \quad \mt_\lambda \equiv g \tilde v} \ . \nonumber
\end{align}
Notice that in the relativistic limit 
\begin{align}
\label{eq:rK_asymp}
    r^\lambda_k \rightarrow {\tilde v^2-v^2\over \tilde v^2+v^2} \ , \qquad \text{for} \quad k^z\gg m_\lambda,\mt_\lambda \ .
\end{align}
so that $\lambda$ maintains a finite reflection probability as long as the 
step function is a valid approximation, as was pointed out in \cite{GarciaGarcia:2022yqb}. 

Interestingly we can see that $r^\lambda_k\to 
1$ in the limit $v\to 0$, i.e. the wall becomes completely non-transparent 
for the $\lambda$ polarisations in 
this limit. 
This in-penetrability of the wall deserves some discussion. 
Consider the explicit form of $U_{\lambda}(z)$ for the case of the $tanh$ profile in  \cref{eq:Tanh_Broken_to_Broken}.
As mentioned  in  section \ref{sec:v=0} and sketched in \cref{fig:LambdaPotential}, in the limit of $v \rightarrow 0$, $U_\lambda(z)$ develops a growing plateau with a height $\approx L_w^{-2}\sim m_{h,s}^2$ and width $L_{\rm plateau }\sim L_w \log \tilde v/v$. In the step function approximation, the first scale is effectively treated as infinite. 
For $v=0$ exactly, since $m_{\lambda}(-\infty)=m_{h,s}$ , obviously there are no oscillating $\lambda_{R,k}$ modes at all in the step regime, while $\lambda_{L,k}$ are completely reflected. 
For $v$ tiny but non-zero the potential \textit{eventually} does instead relax to $m_\lambda^2=g^2v^2$ and both oscillating $\lambda_{L,R}$ solutions exist, though strongly constrained to live on opposite sides of the wall\footnote{One might question the validity of the step wall approximation for $k^z \gtrsim L_w^{-1}$ when the potential function $U_\lambda(z)$ has a very long plateau $v\ll \tilde v$. 
However, the solutions \cref{eq: lambda R,eq: lambda L} capture exactly the qualitative behaviour described.}: $|r_k| = 1 - \mathcal{O}(v^2/\tilde v^2)$.
Back to the $v=0$ exact case, one can also consider theories in which the two scales $L_w$ and $m_{h,s}$ are decoupled (e.g. set $m_{h,s}=0$ as an extreme case). Then again oscillating $\lambda_{L,R}$ modes will exist on both sides of the wall even as $L_w\rightarrow0$. Still, by numerical inspection, we find total reflection ($|r_k|\rightarrow 1$) in the step wall limit, though we leave a proper proof to future work.

In conclusion, calculations in the step function regime for the symmetric to broken transition case ($v=0$) can be computed starting from the $v\neq0$ case, using wavemodes \cref{eq: lambda R,eq: lambda L}, and then finally setting $v\rightarrow0$ in the amplitude at the end. Notice instead that the asymptotic masses in the phase space kinematics will need to be discontinuously changed from $gv$ to $m_{h,s}$.

\color{black}
\paragraph{A comment on bound states:}
We have so far considered `scattering state' solutions to the equations of motion, i.e. those which are plane waves far from the wall. What about bound states? In principle, such states are possible for $\lambda$ polarisation. Ref\cite{Farrar:1994vp} found the existence of one for the case $v=0$ when the potential satisfies some specific constraints.
The form of the potential for $v>0$, for example, as sketched in \cref{fig:LambdaPotential}, suggests that a bound state might generically appear. The mass of these bound states is controlled by the scale $L_w^{-1}$, as is obvious by its absence in the step wall limit. One could in principle calculate in the WKB limit the amplitude for an incoming particle to excite this bound state. We leave this interesting exercise to future work.

\subsection{Quantisation}

Following on from  the previous sections, we can expand the field into a complete basis of eigenmodes of the free theory in the background vev $v(z)$,
\begin{align}
 \begin{split}
        A^\mu &= \sum_{I,\ell}\int \frac{d^3k}{(2\pi)^3\sqrt{2k_0}}  \left( a^{\rm in}_{\ell,I,k}\; e^{-i (k_0 t - \vec{k}_\perp \vec{x})}\;\chi_{\ell,I,k}^\mu (z) + h.c. \right) \\
    &= \sum_{I,\ell}\int \frac{d^3k}{(2\pi)^3\sqrt{2k_0}} \left( a^{\rm out}_{\ell,I,k}\; e^{-i (k_0 t - \vec{k}_\perp \vec{x})} \;\zeta_{\ell,I,k}^\mu (z) + h.c. \right)\ ,
    \label{eq:AfieldQuantisation}
 \end{split}
\end{align}
where $I=R,L$ denote right and left movers,  $\ell=\tau_1,\tau_2,\lambda$ sums over different wall polarisations. The wave modes $\chi^\mu(z)$ are in general constructed via 
\begin{align}
\label{eq:chilambda-quant}
    \chi_{\tau_i,I,k}^\mu &= \epsilon_{\tau_i}^\mu \; \chi_{\tau_i,I,k}(z)  \ , \\
    \chi_{\lambda,I,k}^\mu &=\l( \frac{-i k^n \d_ z (v \lambda_{I,k})}{g E v^2},\frac{E}{g v}\lambda_{I,k}\r) \overset{\rm on \ shell}{=}  \bar{\d}^\mu \left({\partial_z \left( v \lambda_{I,k}^z \right)\over  E g \; v^2}\right) +  {g v(z) \over E} \lambda_{I,k} \;\delta^\mu_z \ , 
\end{align}
where $\bar{\partial^\mu} \equiv (-i k^n,  \d^z)$, 
with the explicit form of scalar fields $\chi_{\tau_i,I,k}(z)$ and $\lambda_{I,k}(z)$ obtained by solving the respective Schrodinger-like wave \cref{eq:Eom_chiTau,eq:EOM_lambda} with appropriate $R,L$-mover boundary conditions. 
In the step wall approximation for the vev $v(z)$, these are given analytically in \cref{eq: chi R,eq: chi L,eq: lambda R,eq: lambda L}. In complete analogy to the case of fundamental scalars\footnote{See \cref{sec:new_basis} and \cref{app:Wavepackets}.},
the modes $\chi_{\ell,I,k}^\mu$ should be thought of as describing \textit{incoming} (early time) eigenstates of momentum (particles) in the plane wave limit with physical $z-$momentum  $k^z >0$ and $-k^z$ for  $R$ and $L$ respectively. Modes describing \textit{outgoing} (late time) eigenstates of momenta $\zeta_{\ell,I,k}^\mu$ are instead obtained via
\begin{align}
    \zeta_{\tau_i,\{L,R\}}^\mu &= \epsilon_{\tau_i}^\mu \; \chi^*_{\tau_i,\{R,L\}}(z) , \qquad \\
        \zeta_{\lambda,\{L,R\}}^\mu &=\l( \frac{-i k^n \d_ z (v \lambda^*_{\{R,L\}})}{g E v^2},\frac{E}{g v}\lambda^*_{\{R,L\}}\r) \overset{\rm on \ shell}{=}  \bar{\d}^\mu \left({\partial_z ( v \lambda_{\{R,L\}}^* )\over  E g \; v^2}\right) +  {g v(z) \over E} \lambda_{\{R,L\}}^* \;\delta^\mu_z \ , 
        \label{eq:Lambda Wave fn recap}
\end{align}
where we have dropped $k$ labels to not clutter the notation. Notice the switch in $L,R$ - labels. Both sets of eigenmodes form a complete orthonormal basis and can be used to expand the field operator in
\cref{eq:AfieldQuantisation}. The associated Fourier operators carry $\rm in$ and $\rm out$ labels to emphasise that they create/annihilate in and out states in the $S-$matrix language
\begin{align}
    \left| k^{\rm in}_{\ell, I} \right\rangle &\equiv \sqrt{2 k_0} \; (a_{\ell, I,k}^{\rm in})^\dagger  \left| 0 \right\rangle \ , \\
    \left| k^{\rm out}_{\ell, I} \right\rangle &\equiv \sqrt{2 k_0} \; (a_{\ell, I,k}^{\rm out})^\dagger  \left| 0 \right\rangle  \ , \qquad I \in R,L \quad \& \quad \ell \in \tau_1, \tau_2, \lambda \ .
\end{align}
Both satisfy the usual algebra \cref{eq:SHOalgebraScalar} upon quantisation.

\paragraph{Ward identity and current conservation:}
We now comment on current conservation in the case of spontaneously broken Lorentz symmetry. 
If the gauge symmetry is preserved, vector bosons can couple only to conserved currents.
This is not the case when it is spontaneously broken, but we may still choose to consider coupling to a conserved current \footnote{For example this is the case for the coupling of $W$ boson to light quarks, in the high energy regime when the quark masses can be approximately neglected.}. 
In the Lorentz invariant theory, the statement of the current conservation can be expressed in terms of amplitudes. Given an arbitrary process with an external vector leg with momentum $k^\mu$, we have the following identity
\begin{align}
\label{eq:Ward_no_wall}
    \mathcal{M}^{(4,J)} \equiv \epsilon_k^\mu \mathcal{M}^{(4,J)}_\mu = (\epsilon_k^\mu +{k^\mu})\mathcal{M}^{(4,J)}_\mu \ \ , \qquad \text{(no wall)}.
\end{align}
where the $(4,J)$ label indicates full $4$-momentum conservation
and the process is mediated by the conserved current $J^\mu$, and $\epsilon_k^\mu$ is the external particle's polarisation vector.  
This Ward identity implies that substituting the latter for the particle's momentum $k^\mu$ makes the amplitude vanish.

In the presence of a domain wall in the $z$ direction, the generalised matrix element $\mathcal{M}^{(3)}$ as defined in \cref{eq:MatrixElement3_def_scalars} includes an integral over $z$ and the polarisation tensor is also a function thereof. The expression of conservation closest to \cref{eq:Ward_no_wall}  is 
\begin{align}
     \mathcal{M}^{(3,J)} &\equiv \int dz \; \chi_{\ell,I,k}^\mu(z)   \mathcal{M}_\mu^{(3,J)}(z) = \int dz \; \left(\chi_{\ell,I,k}^\mu(z) + \bar{\partial}^\mu f(z) \right) \mathcal{M}_\mu^{(3,J)}(z) \ , \\
    &\text{where} \quad \bar{\partial^\mu} \equiv (-i k^n,  \d^z) \ ,  \qquad \qquad \qquad \qquad \qquad \text{(with wall)}  \nonumber 
    \label{eq:Ward_generalised}
\end{align}
and $f(z)$ is an arbitrary function. 

To make this discussion more concrete we consider the coupling of the gauge field to the conserved current made out of $\psi$ fields we introduced in   \cref{eq:AbelianHiggsmodel}:
\bea
{\cal L}= i g A_\mu J^\mu,~~ ~~~J^\mu =i \l(\psi^\dagger \d^\mu \psi -\psi \d^\mu \psi^\dagger\r), ~~~\d^\mu J_\mu=0 \ .
\eea
Then the amplitude (defined in \cref{eq:MatrixElement3_def_scalars}) corresponding to the emission of the $(l,I)$ polarisation from the current will be equal to
\bea
\label{eq:matrix3ward}
   {\langle  k_I^{\rm out} \; q  |\: \mathcal{S} \:| p \rangle \equiv (2\pi)^3 \delta^{(3)}(p^n-k^n-q^n) \int d z \left(\zeta_{l,I,k}^\mu(z)\right)^* (p+q)_\mu e^{i (p^z-q^z) z}} \ ,
\eea
where as usual $p,q$ are the initial and final momentum of the $\psi$ particle.
Note, modulo a numerical factor, the same expression will be valid for the emission of the vector boson from an arbitrary conserved current (not necessarily one made from scalars). The current conservation imposes that any interaction which can be written in the form
\bea
J^\mu \d_\mu f ~~~\Rightarrow ~~~ \langle  final   |\: \mathcal{S}_f \:| initial\rangle =\int d^4 x 
~~\d_\mu f(x) J^\mu(x)=0 \ ,
\eea
has a vanishing matrix element. 
We see now the use of writing the polarisation vector for the $\lambda$ d.o.f. as we did in
\cref{eq:Lambda Wave fn recap}. The dangerous-looking first term is actually a total derivative and can be subtracted when computing amplitudes (see \cref{app:current}). We comment further on this in the next section as well as discuss the case of non-conserved current in \cref{sec:wkb-noncons}.


\subsection{Subtleties with WKB regime}
\label{sec:WKB-vector}

Before we start computing the amplitudes of interest, we highlight some important subtleties related to the calculation in the WKB regime. As we have discussed in section \ref{sec:emission_energ}, our formulas are valid only if the contribution inside the wall can be ignored. Let us check whether this is a reasonable approximation for the vector emission.
Consider the $\lambda$ and $\tau$ cases separately:
\begin{itemize}
\item {\bf{$\tau$ polarisation}}

For the  $\tau $ polarisations we can estimate the contribution to the amplitude  inside and outside of the wall and we find:
\bea
\frac{\mathcal{M}^\tau_{\rm inside}}{\mathcal{M}^\tau_{\rm outside} }\simeq \Delta p_z L_w \, \ . 
\eea
Similarly to the scalar case discussed in the section \ref{sec:emission_energ}, the contribution inside the wall can be safely ignored.
\item {\bf{$\lambda$ polarisation}}

Now let us look at the $\lambda$ polarisation, and  the interactions between the current and $\lambda$ field. Using the expansion for $\chi^\mu_\lambda$ field (see for example \cref{eq:chimu})  we get:
\bea
\label{eq:lambdainter}
g J^\mu A^{\mu(\lambda)}&=&g ( J_n \d_n \alpha -  J_z \c^\lambda_z)\nn
&=& J_n \l[ \frac{-i k_n}{E v(z)^2}\d_z \l(  v(z) \lambda(z)\r)\r]- J_z \frac{E}{v(z)}\lambda(z) \ ,
 \eea
outside of the wall, when $\lambda=e^{i k x}$ this becomes 
\bea
&&J_n \l[ \frac{  k_n k^z}{E v(z)}\r]- J_z \frac{E}{v(z) }\lambda=g J^\mu\cdot \epsilon_{\mu}^{(\lambda)} \ ,
\eea
see \cref{eq:epslambda-pol}. 
Let us consider the domain wall connecting the vacua with broken and restored gauge symmetry. 
In this case, the $\lambda$ polarisation will interact with the $\psi$ particle only on the broken side. Then the amplitude originating from the integration outside of the wall will be equal to
 \bea
 \label{eq:BMlambda}
&&\mathcal{M}_{\lambda}^{\rm wkb\ red.}  \propto -\l(\frac{(p+q)^\mu  \epsilon_\mu^{(\lambda)}(z\to -\infty) }{p^z-q^z-\tilde k^z}\r)  \simeq -\l( \frac{\tilde k^z (p^z+q^z)}{\tilde v E}\r),\nn
&&\tilde v\equiv v(z\to +\infty) \ ,
 \eea
where we have kept only the leading term in energy in the polarisation vector and simplified using the conservation of the current, $p^2=q^2$. We can see that this matrix element is growing with energy and is singular in the limit $\tilde v \to 0$, which are very worrisome properties since the limit $\tilde v\to 0$ corresponds to the no domain wall and therefore no transition radiation, i.e. ${\cal M}\to 0$! 
Let us look now at the contribution coming from the integration inside the wall, using the interaction form of \cref{eq:lambdainter}
\begin{align}
    \label{eq:BMwrong}
\mathcal{M}_{\rm inside}\simeq &\int_0^{L_w} dz e^{-i(p-q)^z z}
(p+q)_n \l[ \frac{-i k_n}{E v(z)^2} \l( \d_z  v(z) + v(z)\d_z \r)\r]e^{-i \int_0^z d\tilde z k^z(\tilde z) } \nonumber \\
&-\int_0^{L_w} dz e^{-i(p-q)^z z}
(p+q)_z \frac{E}{v(z)} e^{-i \int_0^z d\tilde z k^z(\tilde z) } \ .
\end{align}
In the first integral there is a term $\propto \d_z v(z)$, which upon integration will necessarily lead to
\bea
{\cal M}_{inside}\sim \frac{(p+q)_0}{E}\frac{1}{\tilde v} \ ,
\eea
which is of the same size as the contribution outside of the wall. We see that the amplitude $\mathcal{M}^{\rm wkb~red.}_\lambda$ will definitely lead to incorrect results, so how can we proceed?
One possibility would be to take some ansatz for the domain 
wall and then perform full WKB calculation keeping the terms inside the wall, which will lead to correct 
results without bad properties of eqs.
\eqref{eq:BMwrong}$-$\eqref{eq:BMlambda}. However, we can still make progress even without the 
knowledge of the shape of the wall using the following trick. By construction, we have been focusing on the case where the current built out of $\psi$ fields is conserved
\bea 
J_{\mu} =i  (\psi^\dagger \partial_\mu \psi- \psi\partial_\mu \psi^\dagger ) \ ,~~~ \d_\mu J^\mu=0 \ .
\eea 
On the other hand, the  $\lambda$ mode can be written as a complete derivative plus a term subleading in energy (eq. \eqref{eq:epslambda-der})
\bea
A_\mu^{(\lambda)}=\d_\mu \alpha+\frac{g v(z)}{E} \lambda(z) (0,0,0,1) \ .
\eea
The part $\d_\mu \alpha$ does not couple to a conserved current, meaning that 
\bea
g J^\mu A_\mu^{(\lambda)}\to - \frac{g^2 v(z)}{E}J_z  \lambda \ .
\eea
With this simplification, we immediately see that all of the problems with $\lambda$ polarisations are cured
\bea
\label{eq:WKB-simplified}
&&\mathcal{M}_{ \rm outside}\sim \frac{(p+q)^z g^2 \tilde v}{E \Delta p_z} \ ,~~~ \mathcal{M}_{\rm inside}\sim
 L_w  g^2 \tilde v \frac{(p+q)^z}{E} \ ,\nn
&&\frac{\mathcal{M}_{\rm inside} }{\mathcal{M}_{\rm outside}}\sim L_w \Delta p_z \ .
\eea
The contribution inside the wall is suppressed and the matrix element is not growing with energy but vanishes in the limit $\tilde v\to 0$.
\end{itemize}
So far we have been focusing only on the case when the current made out of $\psi$ fields is conserved on both sides of the wall. This is not true generically, and in particular for SM fermions, where the Yukawa interactions will lead to current non-conservation. So how should one proceed in that case? As explained in more detail in \cref{sec:wkb-noncons}, it turns out that, with very minor modifications, a very similar 
trick can be used.

\section{Transition radiation and pressure from vectors}
\label{sec:trans}

We are now ready to calculate transition radiation and the resultant pressure from vector boson emission.  
We are working in the Abelian Higgs theory of \cref{eq:AbelianHiggsmodel} and calculate the average momentum transfer during the radiation of a gauge boson from an incoming $\psi$ particle of energy $p^0$. 
We evaluate the amplitudes of interest in the next section, comment on the final state phase space and masses employed in \cref{sec:Phase_space_vector}, and finally present our results in \cref{sec:Vector Pressure Results}.

\subsection{Amplitudes}
All the relevant amplitudes for the particle process $\psi \rightarrow \psi A$ obtained from \cref{eq:matrix3ward} are reported here. These are $\tau$ and $\lambda$ polarisation emission for left and right movers in the step wall and WKB regimes,
\begin{align}
    \mathcal{M}^{\rm step}_{\tau,L}  &= - i g\epsilon_{\tau_2 }^\mu (p+q)_\mu \left( \frac{1}{\Delta p_r } + \frac{r^\tau_{ k}}{\Delta p} +  \frac{t^\tau_{ k}}{-\Delta \tilde{p}_r} \right) \ ,
\\ 
   \mathcal{M}^{\rm step}_{\tau,R}  &= - i g \epsilon_{\tau_2 }^\mu (p+q)_\mu \sqrt{ k^z \over \tilde k^z} \left[ \frac{\tilde k^z}{k^z} \frac{t^\tau_k}{\Delta p} -   \frac{1}{\Delta \tilde{p}} + \frac{r^\tau_k}{\Delta \tilde p_r}\right] \ ,
\\
    \mathcal{M}^{\rm wkb\ red.}_{\tau}  &  {= - i g \epsilon_{\tau_2 }^\mu (p+q)_\mu \left( \frac{1}{\Delta p } - \frac{1}{\Delta \tilde{p}} \right) }\ ,
\\
     \mathcal{M}^{\rm step}_{\lambda,L} &= - i  
     \frac{g}{E} (p^z+q^z) \left[ g v \left( \frac{1}{\Delta p_r} + \frac{r^\lambda_k}{\Delta p}\right) -
     g \tilde{v} \frac{t^\lambda_k}{\Delta \tilde{p}_r} \right] \ ,
\\
   \mathcal{M}^{\rm step}_{\lambda,R}  &= - i  \frac{g}{E}(p^z+q^z) \sqrt{ k^z \over \tilde k^z} \left[ g v \frac{\tilde k^z}{k^z} \frac{t^\lambda_k}{\Delta p} - g \tilde{v} \left( \frac{1}{\Delta \tilde{p}} - \frac{r^\lambda_k}{\Delta \tilde p_r}\right)\right] \ ,
\\
    \mathcal{M}^{\rm wkb\ red.}_{\lambda} &= - i  \frac{g }{E} (p^z+q^z) \left[{ g v\over \Delta p }-{ g\tilde v\over \Delta\tilde{p} } \right] \ ,
\end{align}
where the scattering coefficients $r^{\tau,\lambda}_k, \: t^{\tau,\lambda}_k$ relevant for the amplitudes in the step wall regime are defined in \cref{sec:tau-pol-step,sec:lambda-pol-step} and the $\Delta p$ factors in denominator are defined in \cref{eq:Deltap_notations}. 
We presented amplitudes for $v\neq 0$. 
However, the symmetry-breaking transition case can be obtained smoothly at this level by sending $v\rightarrow 0$. Note that 
\begin{align}
\label{eq:ML goes to zero}
    \mathcal{M}^{\rm step}_{\lambda,L} \rightarrow 0  \qquad \text{as} \quad v \rightarrow 0 \ ,
\end{align}
in this limit. The discontinuity in asymptotic d.o.f. (and therefore masses) is hidden here inside the kinematic factors and are addressed in the following section.

For $\lambda$ emission we used current conservation to simplify the computation of these amplitudes by subtracting the total derivative piece in the wavemode \cref{eq:Lambda Wave fn recap}, as explained in \cref{sec:quantisation}. For $\mathcal{M}^{\rm step}$ this simplification does not change the final expression since it is exact (in the limit of a step wall). However, we emphasise again that it does for $\mathcal{M}^{\rm wkb\ red.}_{\lambda}$, which is an approximation as described in \cref{sec:emission_energ}, and the subtraction is \textit{necessary} to be consistent with the approximations and avoid unphysical divergences, as explained in \cref{sec:WKB-vector}.

\subsection{Phase space integration for  vector emission}
\label{sec:Phase_space_vector}
In going from the amplitudes above to the averaged exchanged momentum $\vev{\Delta p^\ell}$, where $\ell=\tau, \lambda$, we integrate over final state phase space following the prescriptions and kinematics summarised in \cref{sec:procedure}, using $\mathcal{M}^{\rm step}$ and $\mathcal{M}^{\rm wkb \; red.}$ in their respective regimes of validity. However, there are some important subtleties to discuss compared to the simple theory of scalars of \cref{sec:scalar}, particularly for 
a symmetric to broken transition. In this case, the mass of the vector (and therefore $\tau$ d.o.f.) in the old phase ($z\rightarrow - \infty$) is zero by gauge invariance since $v=0$. As shown explicitly in \cref{sec: Analytics Tau m=0}, in principle we can get finite results working with $m=0$ and integrating over the full phase space as long as the mass of the emitter is kept finite $m_\psi \neq 0$.
However, thermal corrections ought to be important. We should expect our calculation to break down for momenta that are too soft (to be defined precisely), where thermal field theory becomes important. To regularise the log divergence in transverse emission when $m_\psi$ is set to zero, \cite{Gouttenoire:2021kjv} cut the $k_\perp$ integration at the soft thermal scale $\sim gT$. This is roughly equivalent to using `electric' thermal masses\footnote{This is the thermal mass correction in the self-energy of $A^0$. It is the relevant scale for example in the Debye screening of the Coulomb field \cite{Bellac:2011kqa}.} in all asymptotic state kinematics
\begin{align}
\begin{split}
        \label{eq:thermal_mass}
(\tau): \begin{cases}
    \mt = m_\tau(z=+\infty) \approx g\sqrt{ \tilde v^2 + T^2} \ , \\
    m = m_\tau (z=-\infty) \approx gT \ , \\
\end{cases}
\\ 
(\lambda): \begin{cases}
    \mt = m_\lambda(z=+\infty) \approx g\sqrt{\tilde v^2 + T^2} \ ,\\
    m = m_\lambda (z=-\infty)=  m_{h,s}(T) \ .\\
\end{cases}
\end{split}
\quad \text{(symmetric to broken)}
\end{align}
In this work we also impose the IR cut-off in this way, since our primary focus is the proper calculation of $\lambda$ emission, which turns out to not need IR regulation. We note however that \cref{eq:thermal_mass} requires further scrutiny. As is well known, the self energy for $\vec{A}$ receives `magnetic mass' thermal corrections only at two loops from charged matter, of parametric order $\sim g^2 T$. Moreover, we are working in the frame of the wall, so the background plasma is boosted and standard results should be distorted. 
We leave the rigorous inclusion of finite temperature to a follow up study.

In \cref{eq:thermal_mass}, $m_{h,s}$ is the mass of the Higgs d.o.f. in the symmetric phase (see \cref{eq:Ulambda Limits}) which will also be temperature dependent \footnote{We do not discuss the explicit form of it since our results are largely insensitive to it.}.
So now, for example, the integration limits for the $k^z$ integral for $\vev{\Delta p^{\lambda, \rm step}_R}$ in \cref{eq:all contributions} become explicitly $k^z \in \l[\sqrt{g^2\tilde{v}^2 + g^2 T^2-m_{h,s}^2(T)}\ , \ k^z_{\rm max}\r]$
\footnote{When $m_{h,s}(T)<\mt(T)$ we are in line with our set up which assumed $m < \mt $, but if it is the other way around the lower limit of $k^z$ becomes imaginary. 
This is signaling the fact that modes with $0 < \kt <\sqrt{m^2 - \mt^2}$ are now exponentially decaying in the old phase (right) side of the wall. 
Then it is more appropriate to define the step wall regime according to $k< \kt \lesssim L_w^{-1}$ and parameterise the PS integrals in terms of $(\kt^z, k_\perp^2)$. The integration limits for $\kt^z$ for $R-$mover emission would be $\kt^z \in [0, \kt^z_{\rm max}]$, where $\kt^z_{\rm max}=\sqrt{(p_0-m_\psi)^2-\mt^2}$.}.

It is worth stressing that for $\tau$ polarisation what appears in the amplitude and in the kinematics boundaries of the PS are always the masses $m$ and $\tilde m$ as defined in \cref{eq:thermal_mass}. However, for the $\lambda$ 
polarisation the coupling appearing in the amplitude is really the bare $gv(\to 0)$ and does not receive thermal corrections, while in 
the kinematics and PS integration what appears is $m$ and $\tilde m$ as defined as in \cref{eq:thermal_mass}.
So  \cref{eq:ML goes to zero} still holds even at finite $T$. For broken to broken transitions the vector masses are, for both $\lambda$ and $\tau$ fields
\begin{align}
    m\approx g\sqrt{ v^2 + T^2}  \ , \quad \mt \approx g\sqrt{ \tilde{v}^2 + T^2} \ ,\qquad \text{(broken to broken)} \ . 
\end{align} 

In summary, $\vev{\Delta p}$ is computed as in \cref{sec:procedure} with phase space integration limits in Table \ref{tab:integration limits} and asymptotic masses defined here above. In general, there is a total of six contributions:
\bea
\label{eq:contr-list}
\vev{\Delta p_{R}^{\tau,\, {\rm step}}} \ ,&\qquad \vev{\Delta p_{R}^{\tau, {\rm wkb}}} \ , & \qquad \vev{\Delta p_{L}^{\tau,\,  {\rm step}}} \ ,\nn 
\vev{\Delta p_{R}^{\lambda,\,  {\rm step}}} \ ,&\qquad \vev{\Delta p_{R}^{\lambda, {\rm wkb}}} \ ,& \qquad \vev{\Delta p_{L}^{\lambda,\,  {\rm step}}} \ .
\eea 

\subsection{Pressure on the bubble wall}
\label{sec:Vector Pressure Results}

\begin{figure}
    \centering    \includegraphics[width=.48\textwidth]{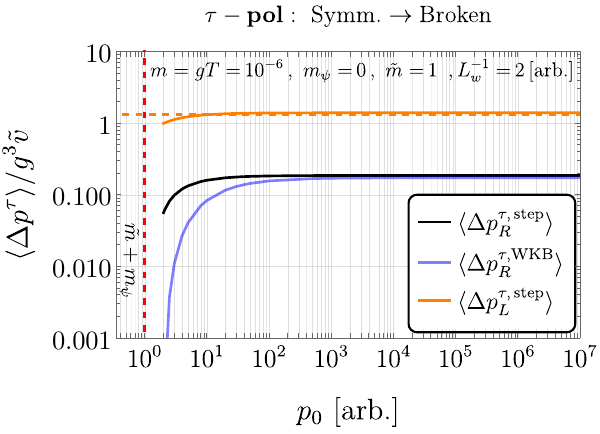}
     \includegraphics[width=.48\textwidth]{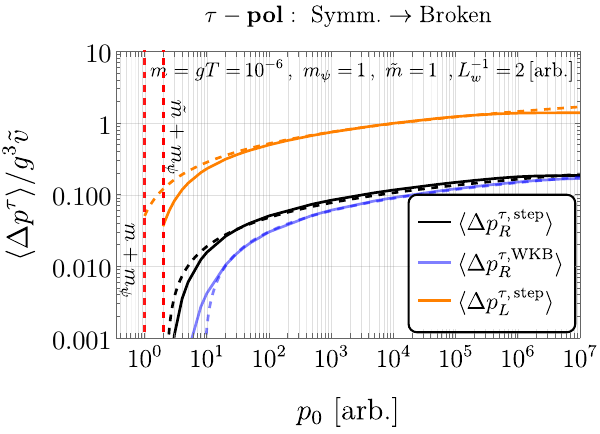}\\
     \includegraphics[width=.48\textwidth]{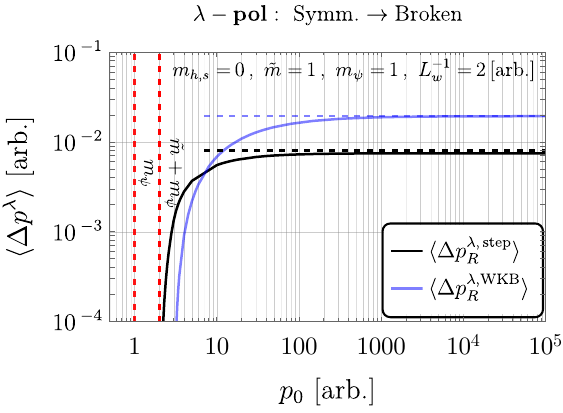}
   \caption{\underline{Symmetric $\to$ Broken}. We present qualitatively the same plot as in \cref{fig:punchline 2} (left panel), this time disentangling all the contributions, and highlighting the effect of a hierarchy $m \ll m_\psi$ for $\tau$ emission. \textbf{Top Left}: Averaged momentum exchange for $\tau$ polarisation contributions. The curves quickly saturate to their constant asymptotic values and the dominant contribution is $L-$mover emission in the step wall regime. Thick lines are numerics and the dashed line is the analytical expression in \cref{eq:DeltaPtauLStep}. \textbf{Top Right}: Same as previous, but for $m / m_\psi \ll 1$. We see the transient inter-relativistic regime scaling as $\log(p_0)$, well described by the $m=0$ analytical formulae (dashed lines) found in \cref{sec: Analytics Tau m=0}. The regime ends around $p^0 \sim \mt m_\psi /m$. \textbf{Bottom}: $\lambda$ polarisation contributions. The result is quite insensitive to the symmetric side $\lambda$ mass $m_{h,s}(T)$ as long as it is $\lesssim \mt$. We plot for $m_\psi = 1\ {\rm [arb.]}$, but varying it does not change much either; unlike for $\tau$, there is no intermediate regime. Dashed lines are analytical formulae in \cref{eq:DeltaPlambdaStep,eq:DeltaPlambdaWKB}.}
   \label{fig: symm to brok}
\end{figure}

\begin{figure}
    \centering
    \includegraphics[width=.48\textwidth]{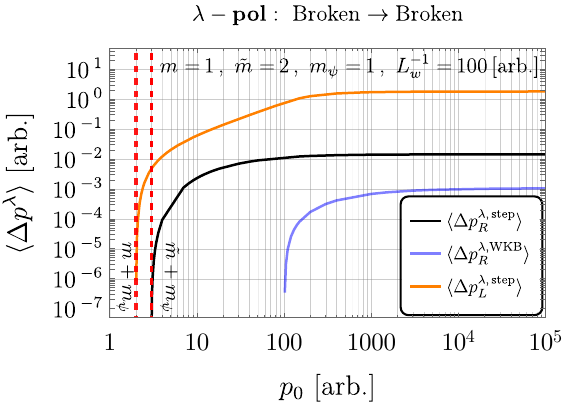}
   \includegraphics[width=.48\textwidth]{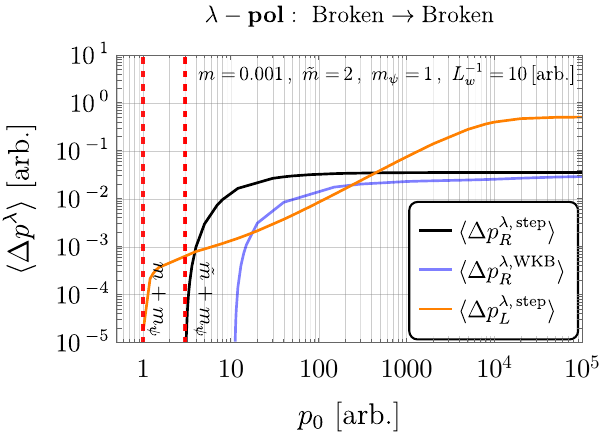}
   \caption{\underline{Broken $\to$ Broken}. \textbf{Left}: Averaged momentum exchange for $\lambda$ polarisation emission for relatively thin wall. The inter-relativistic regime of scaling with $p^0$ lasts till $p^0 \sim L_w^{-1}$. Taking $\Delta m \ll1$ does not qualitatively change anything except for suppressing all values. Notice $m \lesssim \tilde m$. \textbf{Right}: Similar to previous, but in the case $m \ll \tilde m$. Here we see that the inter-relativistic regime is distorted at low $p^0$ and also extended up to $p_0 \sim L_w^{-1} m_\psi /m$. }
   \label{fig: brok to brok}
\end{figure}

\begin{figure}
    \centering
    \includegraphics[width=.48\textwidth]{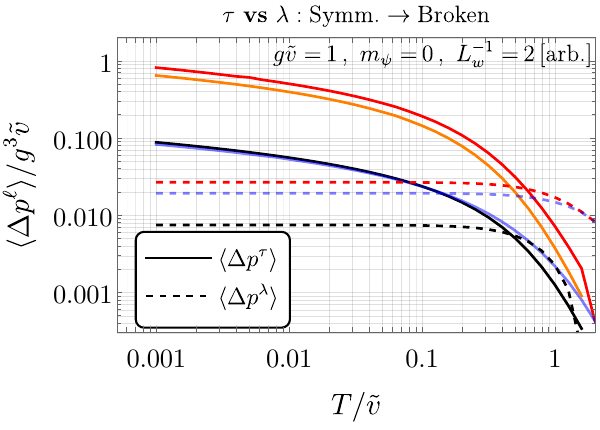}\includegraphics[width=.48\textwidth]{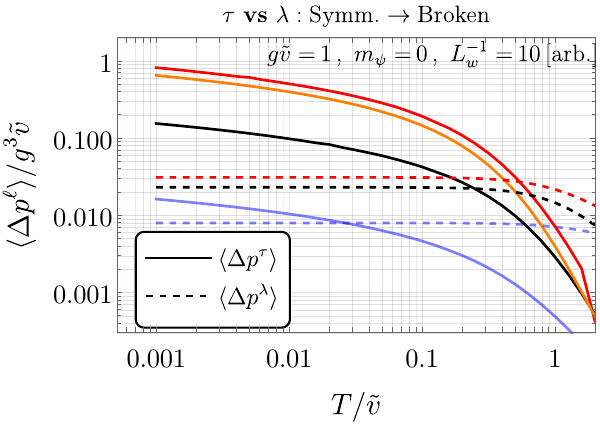}\\
    \includegraphics[width=.48\textwidth]{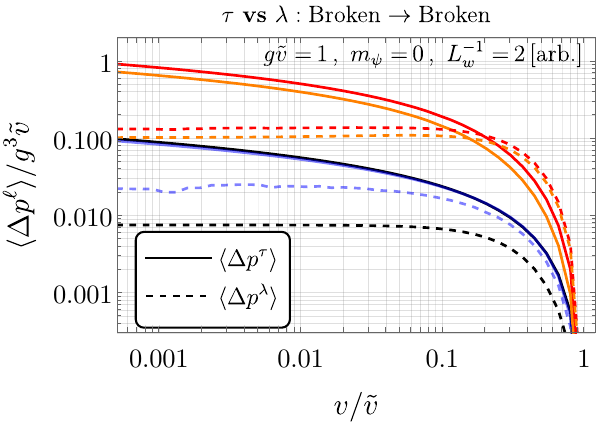}\includegraphics[width=.48\textwidth]{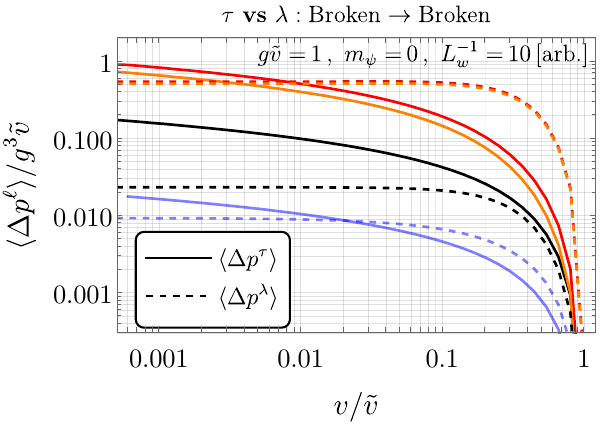}
    \caption{Here is presented the same plot as in \cref{fig:punchline1}, but this time disentangling all the contributions. Solid lines refer to $\tau$ polarisation, while dashed to $\lambda$. The red lines are the total sum, for each polarisation.  \textbf{Top}: Symmetric $\to$ Broken. It is shown two cases with different wall widths and it is worth noticing that while the total contribution for $\tau$ and $\lambda$ does not change appreciably, the single contributions do, like $R$ step wall versus WKB. \textbf{Bottom}: Broken $\to$ Broken. Here it happens the same as in the top panels, but, since in this case, $L$ step contribution from $\lambda$ polarisation has to be taken into account, we see that, depending on the value of $L_w$ and the hierarchy between the different vevs, can be the most relevant contribution. 
    }
    \label{fig:comparison vac}
\end{figure}

We now report and comment on our results for the average NLO momentum transfer $\vev{\Delta p}$ due to transition radiation from a $\psi$ particle travelling across the wall and thereby compute the total pressure on the bubble wall. We show a break down of \cref{fig:punchline1,fig:punchline 2} into all their contributions, as well as provide some analytical formulae. A comprehensive comparison (via numerical integration - see \cref{app:evaluation_int} for some analytical evaluation of phase space integrals) of all the different parts in \cref{eq:contr-list} is shown in \cref{fig: symm to brok,fig: brok to brok}, as a function of the energy of the incoming emitter particle, $\psi$. In the asymptotic $p_0 \rightarrow\infty$ limit all contributions are constant. Their relative importance is displayed in \cref{fig:comparison vac}.
We now discuss the two cases of interest $v=0$ and $v\neq 0$ separately.

\paragraph{Symmetric to broken case ($v=0$).} 

We observe in \cref{fig: symm to brok} that when the emitting particle is $m_\psi \lesssim m$ the contribution from $\tau$ saturates quickly to the constant $p_0\to \infty$ value (top left in \cref{fig: symm to brok}), but if there is a hierarchy $m \ll m_\psi$, we observe an inter-relativistic regime of logarithmic dependence on $p_0$ up until $p_0\sim \mt m_\psi / m$. This can be traced to a collinear log divergence of phase space integration in the limit $m=0$ \& $p_0 \rightarrow \infty$ as explained in \cref{app:evaluation_int}.
This behaviour is not present in contributions from $\lambda$ polarisation emission, which are insensitive to $m_\psi$ for relativistic $p^0$, even when the symmetric side mass is set to zero $m_{h,s}=0$.

In the asymptotic $p^0 \rightarrow \infty$ limit $\tau$ contributions depend significantly only on the ratio \footnote{See for example the exact evaluation of the dominant contribution \cref{eq:DeltaPtauLStep}.} $m/\mt$, which can be translated to $T/\tilde v$. While individual $\lambda$ contributions depend also on  $L_w^{-1}$ their sum is constant.
The total momentum transfer (summing over all contributions) can be fitted by the following expression
\begin{align}
\label{eq:limit-pinf}
    \lim_{p_0\to \infty }\vev{{\Delta p}^{\rm total}}\simeq  g^3 \tilde v
    \l[ 0.135 \log\left(\frac{\tilde v }{ T} + 2.26 \right) 
    - 0.085
    - 0.2 \frac{\log \left(\frac{\tilde v}{T} + 2.26 \right)}{\tilde v/T}+\frac{0.19}{\tilde v/T}\r] \,.
\end{align}
We recall that to obtain this expression we cut the phase space integrals in the IR at energies  $\sim g T$. The expression in \cref{eq:limit-pinf} becomes valid for the energies of initial particle $p_0 \gtrsim \frac{ \tilde v}{ T}\times  m_{\psi }$, which in the case of massless emitter $m_\psi\to 0$ becomes $p_0 \gtrsim \tilde v$.   {At last we would like to remind the reader that the expression above was obtained just for the single vector emission, and one cannot trust it for very large values of log.
}

 The contribution of the longitudinal modes is sub-leading except perhaps for mild super-cooling $\tilde{v}/T \sim $few,
and is equal approximately to
\bea
\label{eq:LambdaPressureMainText}
\lim_{p_0\to\infty }\vev{{\Delta p}^\lambda} \sim g^3 \tilde v \times c_\lambda(L_w,T,\tilde m, m_\psi) ,~~~c_{\lambda}\in [0.02,0.03].
\eea
  {This result qualitatively agrees with the estimate in the Ref.\cite{Bodeker:2017cim}.} An analytical form for the function  $c_\lambda$ is given in \cref{eq:clambdaAnalytic}.

So far we have been calculating the momentum transfer from individual collisions. 
In order to find the pressure acting on the bubble wall we need to perform the 
integration over the flux of incoming particles.   This can be easily done in the thermal case since we know the distributions
\bea
{\cal P}=\int \frac{d^3 p}{(2\pi)^3}f_\psi(T,\gamma_w)\times\frac{p^z}{p_0} \vev{\Delta p}.
\eea
If the average momentum transfer is a constant the integration is simple and we find, since in the ultra-relativistic case $p^z/p_0\to 1$,
\bea
{\cal P}= n_\psi(T,\gamma_w)\vev{\Delta p}= \gamma_w n_\psi(T) \vev{\Delta p},
\eea
where $n_\psi$ is the density of emitters $\psi$ defined in the plasma frame. Then, for the symmetric to broken transition we obtain the following expression for the pressure:
\begin{align}
\label{eq:ThermalPressureFittedFinal}
    \lim_{\gamma_w\to \infty }{\cal P}_{\rm th.}\simeq & \; \frac{\zeta(3) \gamma_w T^3}{\pi^2}\times \\ 
    & g^3 \tilde v  
    \l[ 0.135 \log\left(\frac{\tilde v }{ T} + 2.26 \right) 
    - 0.085
    - 0.2 \frac{\log \left(\frac{\tilde v}{T} + 2.26 \right)}{\tilde v/T}+\frac{0.19}{\tilde v/T}\r]  \nonumber \  ,
\end{align}
where $\rm th.$ stands for thermal.

\paragraph{Broken to broken case ($v\neq0$).}

In \cref{fig: brok to brok} we show the evolution of $\vev{\Delta p}(p_0)$ for a broken to broken transition. We focus only on $\lambda$ contributions since the ones from $\tau$ are essentially the same as in \cref{fig: symm to brok} with suitable re-interpretation of what $m,\mt$ mean (see \cref{sec:Phase_space_vector}). Again the curves eventually saturate to a constant value but we highlight the strongly $L_w-$dependent novel contribution from $L-$mover emission, which easily comes to dominate in the thin wall regime. As in \cref{fig:punchline 2}, we highlight that in the left panel, we can clearly distinguish the inter-relativistic region where this last contribution develops a linear growth in $p_0$.
In \cref{fig:comparison vac} we see the dependence of the saturation value of the averaged exchanged momentum (in the limit $p_0\to \infty)$ on the ratio between vevs (lower panels).

Broken to broken transitions were recently studied at  leading order by \cite{GarciaGarcia:2022yqb} and it was  found that reflection of the longitudinal vectors is efficient  for the energies below $L_w^{-1}$
(inverse width of the wall).
This in its turn leads to the pressure scaling as $\gamma_w^2$, {{as long as $p_0 < L_w^{-1}$}}. 
We find that a reminiscent effect happens at NLO level, the main 
difference is that the momentum of the vector is not 
fixed by the speed of the bubble expansion and is 
always integrated over all possible values. We find that the momentum transfer is dominated by left-mover modes and for the values of the energies of incoming particle  $ p_0 < L_w^{-1}{\rm Max}[1,m_\psi/m]$ it is proportional to  $\propto p_0$
\bea
\lim_{p_0 < L^{-1}_w{\rm Max}[1,m_\psi/m]}  \vev{{\Delta p}_{v \neq 0}}\sim  10^{-2} \times  g^2 \frac{(v^2-\tilde v^2)^2}{(v^2+\tilde v^2)^2}\times  \frac{p_0}{{\rm Max}[1,m_\psi/m]} \ .
\eea
In the case of large hierarchy $0<m\ll m_\psi$, not only is the saturation point delayed, but a further slight distortion occurs at low $p^0$, as shown in \cref{fig: brok to brok}.
Once the energy of the initial particle becomes larger than $\sim L_w^{-1}{\rm Max}[1,m_\psi/m]$, we have 
\bea
\label{eq:vnonzerolambda}
  {\lim_{p_0\to \infty, L_w\to 0}}   \vev{{\Delta p}_{v\neq 0}}\simeq  0.05g^2 \frac{(v^2-\tilde v^2)^2}{(v^2+\tilde v^2)^2} L^{-1}_w.
\eea
Momentum transfer stops growing and reaches the saturation value. Note that the maximal value of this pressure is controlled by $L_w^{-1}$ and not by the mass of the 
vector $g \tilde v $. This is related to the fact that at high energies Goldstone Boson equivalence theorem 
relates the longitudinal vectors to the Goldstone bosons, and the strength of their interaction with 
the bubble wall (Higgs field) is controlled by the mass of the Higgs (wall width).
Consequently, in the case of broken to broken transition, there is 
an additional contribution to the pressure which scales as 
\bea
\lim_{\gamma_w\to \infty}  {\cal P}^\lambda_{v \neq 0}\sim   0.05  \frac{\zeta(3) \gamma_w T^3}{\pi^2}\times  g^2 \frac{(v^2-\tilde v^2)^2}{(v^2+\tilde v^2)^2}\times  L_w^{-1}.
\eea
with an intermediate regime scaling as ${\cal P}^\lambda\propto \gamma_w^2$  for the values of boost factor $\gamma_w < (L_wT)^{-1}$.

\section{Summary}
\label{sec:summary}

We conclude by summarising the 
main results of our work. We 
analyzed in detail the phenomena of transition 
radiation in the presence of 
domain walls. We quantised from first 
principles scalar and vector 
theories on a translation-violating 
background and identified the correct asymptotic states. We split emission into soft and UV regimes and used a step wall and WKB approximation respectively to compute the desired matrix elements for transition radiation. Quantisation of vector field theories was naturally performed via the 
introduction of new degrees of freedom which do not coincide with the traditional transverse and longitudinal polarisation but are a convenient admixture.
In this way, we have resolved 
some puzzles regarding the inclusion of longitudinal 
polarisations in the calculation of transition radiation. 

We applied these results to calculate the pressure experienced by the bubble wall 
during the ultra-relativistic expansion. 
For the phase transitions with spontaneous breaking of the gauge symmetries in the regime of strong supercooling, we find the pressure which scales as 
\bea
{\cal  P}\propto \gamma_w g^3 \tilde v T^3 \log \frac{\tilde v}{T},~~~~ \text{for} \quad \tilde v/T\gg 1, 
\eea
and is dominated by the emission back out of the wall of transverse-like polarisations with momenta $k^z \in (0,\Delta m)$. 
This result qualitatively agrees with the previous literature on the subject. 
For moderate ratio $\tilde v/T \sim $ few we find that the contribution from the longitudinal-like polarisations can lead to significant corrections. We provide an updated fitted formula for the total pressure in \cref{eq:ThermalPressureFittedFinal}.

We also analyzed the pressure in the case of transition between two vacua with broken gauge symmetry. Interestingly we find in this case the contribution from the longitudinal-like polarisation can easily become dominant for thin walls, with the asymptotic value controlled by the inverse wall width ${\cal P}^{\rm max}\propto \gamma_w g^2 T^3 L_w^{-1}$. Moreover, we find a transient intermediate regime of ${\cal P} \propto \gamma_w^2$ scaling for $p_0 \lesssim L_w^{-1}$.

Our results make an advance in 
understanding the balance between bubble acceleration and friction which plays a crucial role in determining most phenomenological consequences of FOPTs as well as their detection prospect at upcoming gravitational wave detectors. 

\paragraph{Future outlook:}
The work can be improved and generalised in several ways. An important remaining question is the inclusion of finite temperature effects in a robust first principles fashion (see for example the discussion in \cref{sec:Phase_space_vector}). It is relatively easy, though cumbersome, to allow also for the emitting particle to feel the wall $m_{\psi}\rightarrow m_{\psi}(z)$. Though in the $\gamma_w \rightarrow \infty$ limit, any dependence on $m_{\psi}(z)$ should drop out, we saw how $m_\psi$ can distort the intermediate shape of $\mathcal{P}^{\rm NLO}(\gamma_w)$ and ultimately finding the equilibrium velocity $v_w$ will require full knowledge of this curve. Similarly one could rigorously quantise fermionic fields that change mass across the wall.

A possibly important direction is the analysis of multiple vector emissions, particularly in regimes with large logs, possible IR enhancements, and back-reaction effects coming from the overdensity of soft vector bosons around the wall (see discussion in \cite{Gouttenoire:2021kjv}). 
Furthermore, it would be interesting to compare our wall-shape-independent results with a full numerical calculation using a specific smooth wall ansatz (for example $tanh$).
Finally, with some tweaks, our expressions can be used to analyse pressure in qualitatively different types of FOPTs, such as spontaneous breaking of \textit{global} symmetries, or even symmetry \textit{restoring} transitions.

While our interest here was in friction, we emphasise that our set up is useful for rigorously computing any process / Feynman diagram\footnote{At tree or even loop level.} in the presence of background walls which are not treated as a perturbation. Just as an example, one can easily re-purpose our expressions to compute the number of particles of a given species produced from collisions between an expanding bubble and surrounding particles \footnote{These could be heavy, Dark matter candidates.}. 
The spontaneous breaking of Lorentz/translational symmetry in the early universe results in a rich phenomenology that is only starting to be explored systematically.

\color{black}


\section*{Acknowledgements}
 AA  is supported by the MUR contract 2017L5W2PT.
 MV is supported by the ``Excellence of Science - EOS" - be.h project n.30820817, and by the Strategic Research Program High-Energy Physics of the Vrije Universiteit Brussel. AA and GB would like to thank M. Serone for discussions.
 RPB is very grateful to Giacomo Koszegi for numerous discussions at the preliminary stages of the project and wishes him all the best in his new career. RPB also thanks Giovanni Villadoro and Mehrdad Mirbabayi for illuminating discussions. 
 We thank Anson Hook and Isabel Garcia Garcia for feedback on the draft.

\newpage

\appendix

\section{Wavepackets and asymptotic states}
\label{app:Wavepackets}

\paragraph{No wall:}
Let us recall a few things in the usual manifestly translation-invariant case (no wall background). For clarity, we will first focus on $1+1$ dimensions. For a scalar field theory, the field operator $\psi(t,z)$ is interpreted as creating a particle at position $z$ at time $t=0$. The position space wavefunction of a state is thus given by its inner product with $\langle 0|\psi $. The wavefunction of a single particle eigenstate of momentum $|k\rangle = \sqrt{2k^0}a_k^\dagger |0\rangle$ is, as expected,
\begin{align}
    \langle 0 | \psi(t,z) | k \rangle &=e^{-i k_\mu x^\mu } \ .
\end{align}
To derive formulae for observables such as scattering cross sections or emission probabilities involving realistic particles we have to crucially go through appropriately defined wavepackets that describe those asymptotic states, taking a limit of sharp momentum only at the end. 
A wavepacket state describing a particle with momentum peaked around $p$ and localised in space is given by
\begin{align}
\label{eq:Psi_wavepacket}
    | \Psi_p \rangle \equiv \int_{-\infty}^{\infty} { d k^z \over (2 \pi)\sqrt{2 k^0} } f_p(k^z) |k\rangle \ , \qquad \langle \Psi_p | \Psi_p \rangle=\int {dk^z \over (2 \pi)} |f_p(k^z)|^2 = 1 \ . 
\end{align}
where $f_p(k^z)$ is like a sharp Gaussian in $k^z$ peaked at $p^z$. Note there is no spacetime dependence in this expression (we are always in the Heisenberg picture). The limit to recover a momentum eigenmode is $f_p(k^z) \longrightarrow (2\pi)^3 \sqrt{2k^0} \delta(p^z-k^z)$.
However, the more appropriate limit used in the derivation of physical rates makes use of the normalisation condition above 
\begin{equation}
\label{eq:PeakedMomentumLimit}
    |f_p(k^z)|^2 \longrightarrow (2\pi) \delta(p^z-k^z) \ , \quad \text{(peaked momentum limit)} \ .
\end{equation}
To see the localisation we can look at the
 wavefunction of the wavepacket state:
\begin{align}
    \langle 0 | \psi(t,z) | \Psi_p \rangle &= \int_{-\infty}^{\infty} { d k^z \over (2 \pi)\sqrt{2 k^0} } f_p(k^z) e^{- i k_\mu x^\mu}  \ .
\end{align}
At $t=0$ this is an oscillating function of $z$ (with wavelength controlled by $p$) with a Gaussian envelope so that it is indeed localised at $z=0$.  As a function of time, the wavepacket moves in the direction of sign$(p)$. Because each mode has a slightly different dispersion relation, the spatial width of the wavepacket tends to widen in time (dispersion) but this can be counteracted by making $f_p(k)$ a sharper Gaussian, taking an appropriate order of limits.

\subsection{Asymptotic states in the wall background}
In the presence of the wall we should still define asymptotic particle states as appropriate wavepackets to compute formulae for physical rates. A slight complication comes from the fact that the one particle states $|k_{L,R} \rangle$ we quantise are not eigenstates of momentum. Moreover, in general, the same state describes different types of particles in different regions of space\footnote{This can be simply because the mass changes or, as in the case of $\lambda(x)$ in a symmetric to broken transition, the field interpolates even between particles of different spin.}.  Defining the asymptotic state carefully will get rid of any ambiguity.

We now consider an operator that feels the wall $\phi(t,\vec{x})$ expanded in left and right mover modes as in \cref{eq:fieldExpansion}. As before, the action of this on the vacuum should be thought of as creating a particle localised at $x^\mu=0$. Contracting a state with $\langle 0|\phi $ still gives the wavefunction understood in the usual sense. In fact
\begin{align}
    \langle 0 | \phi(t,\vec{x}) | k_I \rangle &= e^{-i k_n x^n} \chi_{I,k}(z) \ ,
\end{align}
with $I={R,L}$. To gain physical intuition of the one particle states $|k_I \rangle$, consider constructing wavepackets by superimposing exclusively right (left) movers:
\begin{align}
    | \Phi^{\rm in}_{I,p} \rangle &\equiv \int_{0,\:\Delta m}^{\infty} { d k^z \over (2 \pi)\sqrt{2 k^0} } f_p(k^z) |k_I\rangle \ ,
\end{align}
where again we focus on $1+1$ dimensions  and
where the lower limit is $0$ $(\Delta m)$ for R (L) movers respectively. The `in' labels will become clear shortly.
Their wavefunctions are respectively 
\begin{align}
   \langle 0 | \phi(t,z) | \Phi^{\rm in}_{I,p} \rangle &\equiv \int_{0,\:\Delta m}^{\infty} { d k^z \over (2 \pi)\sqrt{2 k^0} } f_p(k^z) e^{-i k^0 t}\chi_{I,k}(z) \ . \label{eq:WaveFnIMover}
\end{align}
Ignoring the slight dispersion mentioned at the end of the previous subsection, the time evolution of \cref{eq:WaveFnIMover} for $I=R (L)$ describes an isolated localised wavepacket travelling in towards the wall from $z < 0$ $(z>0)$ when $t<0$. This wavemode scatters off the wall and for $t>0$ splits into reflected and transmitted wavemodes travelling in opposite directions. Thus, $| \phi^{\rm in}_{R,p} \rangle$ ($| \phi^{\rm in}_{R,p} \rangle$) is a good asymptotic state for an \textit{incoming} particle with positive (negative) $z-$momentum. It cannot however be used as an asymptotic state for any single \textit{outgoing} asymptotic particle since at late times it describes a superposition.

So what about single localised \textit{outgoing} particles? Clearly, when the wave equation is reduced to Schrodinger-like form with a real potential, these can be obtained by complex conjugation of the spacial part of the wavefunction. In other words we require states $| \Phi^{\rm out}_{I,p} \rangle$ such that
\begin{align}
   \begin{split}
       &\langle 0 | \phi(t,z) | \Phi^{\rm out}_{I,p} \rangle \equiv \int_{\Delta m, \:0}^{\infty} { d k^z \over (2 \pi)\sqrt{2 k^0} } f_p(k^z) e^{-i k^0 t}\zeta_{I,k}(z) \ ,\label{eq:WaveFnIMoverOut}\\
   &\text{with} \quad \zeta_{R, L}(z) = \chi_{L,R}^*(z) \ .
   \end{split}
\end{align}
In time, these correspond to two waves coming from opposite sides of the wall, hitting the wall around $t\approx 0$ and interfering in just the right way so that at late time there is only one wavemode travelling towards $z\rightarrow -\infty$ or $z\rightarrow \infty$ respectively.
Notice the swap in labels. We adopt the convention that the right-mover $R$ (left-mover $L$) label always denotes a particle of positive (negative) $z-$momentum. 
It is an easy exercise to write the $\zeta_{R,L}$ in terms of a linear composition of the complete basis $\chi_{R,L}$ as explicitly done in \cref{eq:LoutWavefn,eq:RoutWavefn}. The appropriate wavepacket for an outgoing state is thus
\begin{align}
    | \Phi^{\rm out}_{I,p} \rangle &\equiv \int_{\Delta m, \:0}^{\infty} { d k^z \over (2 \pi)\sqrt{2 k^0} } f_p(k^z) |k^{\rm out}_I \rangle  \ , 
\end{align}
where  $|k^{\rm out}_I \rangle $ are the one-particle states with wavefunctions $\zeta_{I,k}(z)$ and are related to the $|k_I \rangle $ by
\begin{align}
| k^{\rm out}_L \rangle &= r_{R,k}^*|k_R\rangle + t_{R,k}^* \sqrt{\tilde{k}^z / k^z}\; |k_L\rangle \; \Theta(k^z-\Delta m)  \ , 
\\
| k^{\rm out}_R \rangle  &=  t_{L,k}^* \sqrt{k^z/\tilde{k}^z}\; |k_R\rangle + r_{L,k}^* |k_L\rangle  \ , 
\end{align}
where the reflection and transmission coefficients were defined in general for scalar d.o.f. in \cref{sec:basis}, and given explicit form in the case of a step wall for fundamental scalars also in \cref{sec:basis}  and for different vector polarisations in \cref{sec:stepwall}.

\paragraph{Reflection and transmission probabilities:}
As a consistency check, we can compute reflection and transmission probabilities in this language. Focusing on the identity part of the S matrix $S=1+\cancel{i\mathcal{T}}$, the amplitude for a particle incoming from $z<0$ to reflect and transmit should be, respectively:
\begin{align}
 \langle \phi^{\rm out}_{L,p}  | \phi^{\rm in}_{R,p} \rangle &= \int_{0}^\infty {dk \over 2\pi} | f_p(k) |^2 \; r_{R,k}  \quad \rightarrow \quad   r_{R,p} \ , \\
  \langle \phi^{\rm out}_{R,p}  | \phi^{\rm in}_{R,p} \rangle &= \int_{\Delta m}^\infty {dk \over 2\pi} |f_p(k)|^2 \; t_{L,k} \sqrt{\tilde{k}^z / k^z} \quad \rightarrow 
  \begin{cases} 
  t_{L,p} \sqrt{p/\tilde{p} } \ , & p > \Delta m \\
  0 \ , & p < \Delta m
  \end{cases} \ ,
\end{align}
where the arrow denotes taking the peaked momentum limit \cref{eq:PeakedMomentumLimit} in the end.
So the relative probability for a peaked-around-$p$ incoming mode to reflect/transmit into a peaked-around-$p$ outgoing mode is what one could have already guessed
\begin{align}
{\rm Reflection \ probability} = |\langle \phi^{\rm out}_{L,p}  | \phi^{\rm in}_{R,p} \rangle|^2 &\rightarrow |r_{R,p}|^2 \ ,  \\
{\rm Transmission \  probability} = |\langle \phi^{\rm out}_{R,p}  | \phi^{\rm in}_{R,p} \rangle|^2 &\rightarrow 
\begin{cases} 
  |t_{L,p}|^2 p/\tilde{p} \ , & p > \Delta m \\
  0 \ , & p < \Delta m
  \end{cases} \ .
\end{align}

\subsection{Phase space derivation from wavepackets}
\label{sec:phase_space_derivation}
In this section we will obtain formulae for the averaged momentum exchanged by transition radiation processes in the background of the wall, deriving \cref{eq:PSI-RIGHT} in the main text.
Treating the incoming particle as a wavepacket as in \cref{eq:Psi_wavepacket}, we find the amplitude squared for splitting into $\ell$ particles, which may or may not feel the wall. These outgoing states are late-time eigenstates of momentum, obtained as limits of their own wave-packet forms. We have 
\begin{align}
    |\langle k_1^{\rm out} \dots k_\ell^{\rm out} | \Phi_p\rangle|^2 
    &=\int {d^3p_1d^3p_2 \over \sqrt{4 p^0_1 p_2^0}}
    f_p(p_1)f_p^*(p_2)
   \delta^{(3)}\left( p_1 -  \underset{{i=1}}{\overset{\ell}{\Sigma}} k_i \right) 
   \delta^{(3)}\left( p_2 -  \underset{{j=1}}{\overset{\ell}{\Sigma}} k_i \right) 
   |\mathcal{M}^{(3)}_{1\rightarrow \ell}|^2 \nonumber \\
   &=\int {d^3p_1dp^z_2 \over  \sqrt{4 p^0_1 p_2^0}}
    f_p(p_1)f_p^*(p_2)
   \delta^{(3)}\left( p_1 -  \underset{{i=1}}{\overset{\ell}{\Sigma}} k_i \right) {p^0_1\over|p^z_1|} \delta\left( p_1^z -  p_2^z \right) |\mathcal{M}^{(3)}_{1\rightarrow \ell}|^2 \nonumber \\
   & = \int {d^3p_1\over  2 |p^z_1|}
    |f_p(p_1)|^2
   \delta^{(3)}\left( p_1 -  \underset{{i=1}}{\overset{\ell}{\Sigma}} k_i \right) |\mathcal{M}^{(3)}_{1\rightarrow \ell}|^2 \nonumber \\
   & \longrightarrow \frac{(2\pi)^3}{2|p^z|} \delta^{(3)}\left( p -  \underset{{i=1}}{\overset{\ell}{\Sigma}} k_i \right)  |\mathcal{M}^{(3)}_{1\rightarrow \ell}|^2 \ ,
   \label{eq:WavePacketAmplSq}
\end{align}
where the temporary label $(3)$ on $\mathcal{M}$ emphasises only three of $4-$momentum are conserved.  In the last step, we take the peaked momentum limit of \cref{eq:PeakedMomentumLimit}.
For the particular case of a $1\rightarrow2$ process as discussed in this work, we have
\begin{align}
   \langle \Delta p^z_{I=R,L} \rangle \equiv \int \mathcal{P}_{p\rightarrow q k_I^{\rm out}}\Delta p^z_I 
     &= \int \frac{d^3k}{(2\pi)^3 2 k^0}\int \frac{d^3q}{(2\pi)^3 2 q^0}  |\langle q \: k_I^{\rm out}  | \Phi_p\rangle|^2 \nonumber \Delta p^z_I\\
     &=\int {d^3k d^3q \over (2\pi)^3 8p^z k^0 q^0 } 
     \delta^{(3)}\left( p -  k - q \right)  |\mathcal{M}_I^{(3)}|^2 \Delta p^z_I \nonumber \\
     &=\int {d^3k  \over (2\pi)^3 8 p^z  k^0 |q^z_k|} 
     \left[ |\mathcal{M}_I^{(3)}|^2 \Delta p^z_I \right]_{q^z =\pm q^z_k} 
\end{align}
where in going to the last line we have used 
\begin{align}
\begin{split}
    \delta\left( p^0 - k^0 - q^0 \right) = {q^0 \over |q^z_k|}\left[ \delta(q^z-q^z_k) +  \delta(q^z + q^z_k) \right] \\ 
    \text{with} \quad q^z_k = \sqrt{(p^0-k^0)^2-\kp^2-m_q^2}
\end{split}
\end{align}
Thus, in principle, one should sum contributions from both signs of $q^z$ to obtain the full integrated rate. In practice, the $q^z<0$ branch will be highly sub-leading at large energies. 

\paragraph{Comparison with decay formula:}
One might wonder how to recover the familiar decay formula in the limit of no wall. The latter can be derived in terms of the full $4-$momentum conserving matrix element $\mathcal{M}^{(4)}$ as
\begin{align}
    |\langle k_1 \dots k_\ell | p\rangle|^2 
    &=\int {d^3p_1d^3p_2  (2\pi)^2 \over  \sqrt{4 p^0_1 p_2^0}}
    f_p(p_1)f_p^*(p_2)
   \delta^{(4)}\left( p_1 -  \underset{{i=1}}{\overset{\ell}{\Sigma}} k_i \right) \delta^{(4)}\left( p_2 -  p_2\right) |\mathcal{M}^{(4)}_{1\rightarrow \ell}|^2 \nonumber \\
   & = \int {d^3p_1 (2\pi)^2\over  2 p_1^0}
    |f_p(p_1)|^2
   \delta^{(4)}\left( p_1 -  \underset{{i=1}}{\overset{\ell}{\Sigma}} k_i \right) \underbrace{\delta\left(0\right)}_{T/2\pi} |\mathcal{M}^{(4)}_{1\rightarrow \ell}|^2 \nonumber\\
   & \longrightarrow \frac{(2\pi)^4}{2p_0} \mathbb{T} \; \delta^{(4)}\left( p -  \underset{{i=1}}{\overset{\ell}{\Sigma}} k_i \right)  |\mathcal{M}^{(4)}_{1\rightarrow \ell}|^2 \ ,
   \label{eq:DecayFormula}
\end{align}
 where $\mathbb{T}$ is total time (one can only define a decay probability per unit time).
We can see how \cref{eq:WavePacketAmplSq} reduces to this result. In the absence of the wall, the matrix element gives
\begin{align}
    |\mathcal{M}^{(3)}_{1\rightarrow \ell}|^2 &=  2\pi \: \mathbb{L}^z \delta \left(p^z-\underset{{i=1}}{\overset{\ell}{\Sigma}}\right) |\mathcal{M}^{(4)}_{1\rightarrow \ell}|^2
\end{align}
where $\mathbb{L}^z$ is the distance traversed in the $z$ direction. Finally, using that $\mathbb{L}^z = \mathbb{T} \;p^z/p^0$, \cref{eq:WavePacketAmplSq} reduces   to \cref{eq:DecayFormula}.

\section{Current conservation in the presence of the wall }
\label{app:current}
In the main text, we have discussed the modifications of the Ward identities in the presence of the wall and that in general the conserved current 
 coupled to the total derivative must give zero matrix element,
 \bea
J^\mu \d_\mu f ~~~\Rightarrow  \langle  final   |\: \mathcal{S}_f \:| initial \rangle =0 \ .
\eea
Let us discuss the effects of various choices of the gauge transformation function $f$.
For example  let us consider $f=\chi_{1,2}^\tau(z)$, where $\chi_{1,2}^\tau$ are the wave functions for the $\tau$ polarisations (see section \ref{sec:Global_degrees_of_freedom}). Then the matrix element will be equal to
\bea
J^{\mu}\propto (p+q)^\mu ~~~~\Rightarrow~~~
&&\mathcal{M}=\frac{(p+q)_\mu k^\mu}{\Delta p}+r^\tau_k\frac{(p+q)_\mu k^\mu_{r}}{\Delta p_{r}}-t^\tau_k\frac{(p+q)_\mu \tilde k^\mu}{\Delta \tilde p}\ ,\nn
&&k_{r}^\mu\equiv (k^m, -k^z),~~~\tilde k^\mu\equiv (k^m,\tilde k^z) \ .
\eea
We can simplify the amplitude using the following identities:
\begin{align}
(p+q)_\mu k^\mu&= (p+q)_mk^m-(p+q)_z k^z
\nn
&\overset{k_m = (p-q)_m}{=} (p+q)_m(p-k)^m-(p+q)_z k^z\nn
&=p_z^2 + m_\psi^2 -q_z^2 - \tilde m_\psi^2-(p+q)_zk^z
\nn
&\overset{ m_\psi^2= \tilde m_\psi^2}{=} (p+q)_z (p-q-k)^z =  (p+q)_z \Delta p \ .
\end{align}
Performing similar manipulations for all of the terms we get:
\bea
\mathcal{M}=(p+q)_z (1+r^\tau_k-t^\tau_k)=0 \ ,
\eea
where we used the fact that $1+r^\tau_k-t^\tau_k=0$ for $\tau$ polarisations, as expected. Similarly we can choose $f=\alpha(z)$, of the $\lambda$. Then from the eq. \eqref{eq: lambda R} we can get
\begin{align}
    \alpha|_{z<0}&=\frac{ik^z}{g v E}\l(e^{i k^z z }-r^\lambda_k e^{-i k^z z}\r) \ ,\nn
\alpha|_{z>0}&=it^\lambda_k\times \frac{\tilde k^z}{g E \tilde{v}}e^{i \tilde k^z z} \ ,
\end{align}
where we have used that  $\alpha = \frac{1}{ g v^2 E}\partial_z(v \lambda) \to \frac{1}{g E v}\partial_z \lambda$ outside of the wall. Using the expression for reflection and transmission coefficients from \cref{eq:rt-lambda}
\bea
r^\lambda_k = \frac{\tilde{v}^2 k^z- v^2 \tilde k^z}{\tilde{v}^2 k^z+v^2 \tilde k^z} \ , \qquad t^\lambda_k = \frac{2 k^z v \tilde{v}}{\tilde{v}^2 k^z+v^2 \tilde k^z}\ ,
\eea 
we can compute the amplitude for the processes $J^\mu \to \chi_\lambda^\mu$
corresponding to the interaction
$J_\mu \chi_\lambda^\mu$.
The computation goes as follows 
\bea 
\mathcal{M}&=&\frac{k^z}{g E v}\frac{(p+q)_\mu k^\mu}{\Delta p}- \frac{k^z}{E gv}r^\lambda_k\frac{(p+q)_\mu k^\mu_{r}}{\Delta p_{r}}-\frac{\tilde k^z}{g E \tilde{v}}t^\lambda_k\frac{(p+q)_\mu \tilde k^\mu}{\Delta \tilde p}
\nn 
&=& \frac{(p+q)_z}{E} \underbrace{\bigg(\frac{k^z}{ g v}- \frac{k^z}{g v}r^\lambda_k- \frac{\tilde k^z}{ g\tilde{v}}t^\lambda_k\bigg)}_{= 0}=0 \ ,
\eea 
the last expression in brackets is the 
matching condition for the $\lambda$ field (\cref{eq:matching-lambda})
which must be satisfied.
Note that terms cancelling each 
other in the brackets are growing 
in energy, which makes crucially important the calculation of exact 
values of reflection and 
transmission coefficients.

\color{black}
\section{WKB regime in the case of current non-conservation}
\label{sec:wkb-noncons}
In the main text we focus only on the transition radiation from the conserved current. How we can perform a similar calculation in the case when the current is not conserved? Let us consider the following example with scalar fields 
\bea
\label{eq:current-nc}
{\cal L} &= -\frac{1}{4} F_{\mu \nu} F^{\mu \nu}+| D_\mu H|^2 -V(|H|) + | D_\mu \phi|^2 -m_\psi^2 |\phi|^2+\l(\kappa \phi^2 H+ h.c.\r).\nn
\eea
The charges under the gauged $U(1)$ symmetry are as follows: $Q_{U(1)} (H)=1, Q_{U(1)}(\phi)=-1/2$. In the section \ref{sec:WKB-vector}, in order to perform the WKB calculation and get rid of interactions that can potentially lead to the divergences we have used the current conservation equations to modify the expression for the matrix elements. In the case  of the system in \cref{eq:current-nc}  the divergence of the current  becomes equal to:
\bea
\d_\mu J_\phi^\mu= \sqrt{2} v(z) \l( \kappa^* \phi^{* 2}- \kappa \phi^2\r)\ ,~~~J_\phi^\mu=i(\phi^*\d^\mu \phi-\phi \d^\mu \phi^*) \ .
\eea
The interaction between the $\lambda$ polarisation and $J_\phi^\mu$
 and  can be written as follows:
\begin{align}
    g Q_\phi J_\phi^\mu A_\mu^{(\lambda)}\to Q_\phi\l[-\sqrt 2 \l( \kappa^* \phi^{* 2}- \kappa \phi^2\r) \frac{1}{E v(z)}\d_z \l(  v(z) \lambda(z)\r)
-\frac{g^2 v(z)}{E}\lambda(z) J_z \r] \ .
\end{align}
We can see that on top of the term $\lambda J_z$  present in the conserved current case,  there is an additional interaction. However, this interaction is not growing with energy, and in the limit $v(z)\to 0$, it is finite (see discussion in the appendix
\ref{sec:LambdaProof}), thus the calculation of the vector emission becomes straightforward.

\section{Properties of the potential for $\lambda$ field}
\label{sec:LambdaProof}
In the main text we have shown that $\lambda$ field satisfies the following equation of motion \cref{eq:EOM_lambda}$-$\eqref{eq:Potential_lambda}
\bea
&&\left(-E^2-\d_z^2 +  U_\lambda(z) \right)\lambda=0\ ,\nn
 &&  U_\lambda(z) =  g^2 v^2(z)  +2 \left( \d_z v \over v\right)^2-{\d_z^2 v  \over v}\ .  
\eea
Let us investigate the properties of the function $U_{\lambda_z}$.
\paragraph{Broken$\to$ Broken}In this case
\bea
v_{z\to \pm \infty }\neq 0\ ,
\eea
and the potential has the limits :
\bea
U_\lambda(z)|_{z\to \pm\infty}= g^2 v^2(z\to \pm \infty)\ ,
\eea
where we have used that $v', v''\to 0$ outside of the wall.  Physics wise this result is expected since $\lambda$ mode must have the mass of the vector $g v$ outside of the wall.
\paragraph{Symmetric$\to$Broken} Here we will assume
\bea
v(z\to -\infty)=0\ ,~~~v(z\to \infty)=\tilde v\neq 0 \ .
\eea
On the broken side as expected 
\bea
U_\lambda(z\to \infty)=g^2 v^2(z\to \infty)= g^2 \tilde v^2 \ ,
\eea
the potential becomes equal to the mass square of the vector boson. To find its limit on the symmetric side we need to look at the equation defining $v(z)$:
\bea 
\label{eq: wall eom}
\d_z^2 v(z)= V'(v(z)) \ ,
\eea 
where the prime stands for a derivative with respect to $v(z)$. Integrating this equation we get
\bea 
\label{eq: wall eom integrated}
{1 \over 2}\; (\partial_z v(z))^2 = V(v(z)) - C \ ,
\eea 
where $C =-(\partial_z v(z_0))^2 + V(v(z_0)) $ is a `constant of integration'. We can choose $z_0\to -\infty$, then using $v', v \to 0$ we get $C=V(v(- \infty))=V(0)$, which need not be zero. Now we look at the limits as $z\rightarrow -\infty$.
At this point, using eqs. \eqref{eq: wall eom} and \eqref{eq: wall eom integrated} we can write the various terms of the $U_{\lambda}$ function in terms of the potential $V$ and its derivatives $V'$:
\bea 
{\d_z^2 v (z) \over v(z)} &=&{\d_z^2 v \over v} ={V'(v)\over v} \overset{v \to 0}{\longrightarrow}V''(0)\ , \nn
\left( \d_z v (z) \over v(z) \right)^2&=&\left( \d_z v \over v \right)^2=2{V(v)-V(0)\over v^2} \overset{v \to 0}{\longrightarrow}V''(0) \ .
\eea
However, in the limit $z\to -\infty, v\to 0$,  we know 
 that the first derivative of the potential $V$ must be equal to zero at this point
\bea
\frac{\d V}{\d v}\biggr|_{v\to 0}=0\ ,
\eea
since there is a local minimum at $v=0$. Thus in the region $v\to 0$ we can write down:
\bea
V(v)|_{v\to 0}&=&\frac{V''(0) v^2}{2}+ V(0)\Rightarrow\nn
\frac{V'(v))}{v}\biggr|_{v\to 0}&=&V''(0)\ ,~~~2{V(v)-V(0)\over v^2} =V''(0)\ .
\eea
Combining these results we can see that
\bea 
\lim\limits_{z\to -\infty} U_\lambda(z)=V''(0)=m_{h,s}^2\ ,
\eea 
where $m_{h,s}$ is the mass of the scalar on the symmetric side. We can see that as was discussed in the 
section \ref{sec:Global_degrees_of_freedom} $U_{\lambda}$ has the
correct properties for a potential of  a d.o.f 
interpolating between the scalar field and $\lambda$ 
polarisation on different sides of the wall.

\section{Evaluation of phase space Integrals}
\label{app:evaluation_int}
In this appendix, we provide some details regarding the evaluation of the phase space integrals, and how to derive analytical expressions. We will always be interested in the limit of large incoming energy $p^0 \gg $ all masses. The types of integrals we deal with are all of the form
\begin{align}
    I(p^0)= \int^{l_1(p^0)}_{\rm const}dk \int^{l_2(p^0,k)}_{\rm const'} d\kp^2 \; {\rm Int}(p^0,k,\kp^2) \ ,
\end{align}
where the dependence on couplings and masses $m_i$ is implicit. When $I(p^0)$ admits an expansion around infinity, this can be obtained in principle straightforwardly as
\begin{align}
\label{eq:p0Taylor}
     I(p^0)= \left. I(1/x)\right|_{x \rightarrow 0} + \left.{d \over d x }I(1/x)\right|_{x \rightarrow 0}\; {1 \over p^0} +  \; \dots 
\end{align}
and one can sometimes take this `exact' approach. However, such an expansion does not always exist \footnote{For example, in vector emission in symmetry breaking transitions, for $m=0$ we find logarithmic growth.}, or if it does, its coefficients might in practice be difficult to evaluate. 
Moreover, in the presence of large mass hierarchies, the regime where  \cref{eq:p0Taylor} is actually a good approximation can begin at arbitrarily high energies
 \footnote{We saw this for example when there are large scale other than $p^0$ such as $\mt m_\psi/ m$ in plots in the main text \cref{fig:punchline 2,fig:PSI,fig: symm to brok,fig: brok to brok}.}.
One can often successfully use a different expansion instead, which we call `collinear expansion', of $p^0,k^z \gg k_\perp, m_i$ where $m_i$ stands for all masses. Keeping the leading term, reduces integrands to the form 
\begin{align}
\label{eq:collinear}
   {\rm Int}(p^0,k,\kp^2) =   {N(k,\kp,m_i) \over \left[(k^{IR}_{\perp,m})^2+\kp^2\right]^2} \qquad \text{for} \quad p^0,k^z \gg k_\perp, m_i  , 
\end{align}
where 
\begin{align}
\label{eq:kperpIR}
    k_\perp^2 \lesssim (k^{IR}_{\perp,m})^2 \equiv {k_z^2 m_\psi^2 - k^z m^2 p_0 \over (p_0)^2} + m^2 > 0\, .
\end{align}
The function $N$ depends on the contribution in question but we wish to highlight the factor in denominator. This form makes most manifest the properties of the phase space structure in particular for $\tau$ polarisation (and scalars), most importantly when $m$ (and possibly $m_\psi$) is very small. For a given $k^z$, the integrand is peaked for $\kp^2 \lesssim (k^{IR}_{\perp,m})^2$. Notice the role of $k^{IR}_{\perp,m}$ - its presence regulates an otherwise logarithmically IR divergent integral. Indeed the momentum transfer diverges in the limit of $m,m_\psi \rightarrow 0$.
It is easy to prove that  $k^{IR}_{\perp,m}<k^z$ for small $m$, justifying the expansion. When $m/m_\psi$ is very small, clearly there are two relativistic regimes, the first given by 
\begin{equation}
    p^0 \lesssim k^z  m_\psi / m\, ,
\end{equation}
where $k^z$ should be taken in the dominant region. Typically $k^z\sim \mt$ for step function contributions and $\sim L_w^{-1}$ for WKB. Above this scale instead one reaches the true asymptotic behaviour where \cref{eq:p0Taylor} is valid.

\subsection{Scalars}
\label{app:Scalars}

\paragraph{Scalar $R$-mover emission (WKB)} We begin with the most dominant contribution to scalar emission at the highest of energies $p^0\rightarrow \infty$. The amplitude is the simplest that one can encounter,
\bea 
\mathcal{M}_R^{\rm wkb}=  \left( {y \over i\Delta p} + {\tilde{y} \over -i\Delta \tilde{p}} \right) \ ,
\eea 
and
\begin{align}
    \langle \Delta p^{\rm wkb}_{R}\rangle&= \int_{L^{-1}_w}^{k^z_{\rm max}} { \;dk^z  } \int^{k_{\perp, cut}^{2}}_{0}{d{k_\perp^2} \over 2} 
    \frac{|\mathcal{M}_R^{\rm wkb}|^2}{(2\pi)^28  p^zq^zk_0} \Delta p_R\ .
\end{align}
For $m\neq0$, properly evaluating the expansion \cref{eq:p0Taylor} for this case is tricky
\footnote{The coefficient of the $1/p^0$ term in \cref{eq:p0Taylor} is the first non-zero contribution as expected from \cref{fig:PSI}. Out of the three terms to evaluate only one is non-zero in the $x\rightarrow0 $ limit. Said limit cannot be taken before the integral is evaluated however and such a task proves unnecessarily difficult.}. 
However, the collinear expansion  \cref{eq:collinear} works extremely well, giving
\begin{align}
\label{eq:Rfancy}
   \langle \Delta p^{\rm wkb}_{R}\rangle  & \approx { y^2 \Delta m^4 \over 32 \pi^2}  \int_{L^{-1}_w}^{k^z_{\rm max}} { \;dk^z  } \int^{k_{\perp, cut}^{2}}_{0}dk_\perp^2  
    {(k^z-p_0)^2/(p_0)^4 \over \left[(k^{IR}_{\perp,m})^2+\kp^2\right]^2\left[(k^{IR}_{\perp,\mt})^2+\kp^2\right]}
    \ ,
\end{align}
where $k^{IR}_{\perp}$ is given by \cref{eq:kperpIR}.
We see indeed that for large $p^0$ the integrand is relatively flat in $k^z$ while $\kp$ is strongly IR dominated as anticipated in the preamble above, justifying the expansion. 
The integrals can now be computed exactly and we obtain \cref{eq:WKBScalars}. We can see clearly however that there is also an intermediate regime for small $m/m_\psi$ characterised by a plateau:
\begin{align}
    \langle \Delta p^{\rm wkb}_{R}\rangle &\approx  
    \begin{cases}
        {y^2\mt^2 L_w \over 32 \pi^2 m_\psi^2}\ ,  & L_w^{-1 } \ll p^0 \lesssim  L^{-1}_w {m_\psi \over  m} \\
        \cref{eq:WKBScalars} \ , &  p^0 \gtrsim L^{-1}_w {m_\psi \over  m}
    \end{cases}
     , 
\end{align}
where we remind ourselves that there is no WKB contribution at all unless we can emit modes greater than the inverse wall length. The value of the plateau is easily obtained by setting $m=0$ at the start and computing the leading order in \cref{eq:p0Taylor}. However, in this regime, we find the step function contributions are more important and we move on there now.


\paragraph{Scalar $R$ and $L$-mover emission (step wall)}
The step wall case is an interesting theory exercise in its own right, since everything can be done exactly. For scalars the contributions are
\begin{align}
    \langle \Delta p^{\rm step}_{R}\rangle&= \int_{\Delta m} ^{L^{-1}_w} { \;dk^z  } \int^{k_{\perp,\rm max}^2}_{0}{d{k_\perp^2} \over 2} 
    \frac{|\mathcal{M}_R^{\rm step}|^2}{(2\pi)^28  p^zq^zk_0} \Delta p_R\ . \\
    \langle \Delta p^{\rm step}_{L}\rangle&= \int_{0} ^{L^{-1}_w} { \;dk^z  } \int^{k_{\perp,\rm max}^2}_{0}{d{k_\perp^2} \over 2} 
    \frac{|\mathcal{M}_L^{\rm step}|^2}{(2\pi)^2 8  p^zq^zk_0}  \Delta p_L\ .
\end{align}
where the matrix elements are given by \cref{eq:amplitudeMscalarsR,eq:amplitudeMscalarsL}. 
Evaluating these in the asymptotic limit we obtain
\begin{align}
\label{eq:ScalarL step}
     \langle \Delta p_{L}^{\rm step}\rangle &\approx 
     {y^2 \mt \over 32\pi^2 m_\psi^2} 
 \begin{cases}
     4 \ , & p_0 \lesssim \mt m_\psi / m
     \\
      {4 m_\psi^2 \Delta m \over 3 \mt m^2 p_0^2}\left[ m^2 \left( \ln {m^2 \over \mt^2}-1\right)+\mt^2\right] \ , & p_0 \gtrsim \mt m_\psi / m
 \end{cases}\\
 \langle \Delta p_{R}^{\rm step}\rangle &\approx
{y^2 \mt \over 32 \pi^2 m_\psi^2 }
\begin{cases}
1- \mt L_w \ , & p_0 \lesssim  4 \: \mt m_\psi / m   \\
{m_\psi^2( L_w^{-1}-\Delta m) \over 4 \mt m^2 p_0^2} \left[ \Delta m^2 + 2 m^2 \ln\left(m \over\mt\right)\right] \ , & p_0 \gtrsim  L_w^{-1} m_\psi /m 
\end{cases}
\label{eq:ScalarR step}
\end{align}
For the $R-$contribution there is an intermediate regime well approximated by the WKB asymptotic formula in \eqref{eq:WKBScalars}, as can be seen from \cref{fig:PSI} (bottom right panel).
\subsection{Vectors: $\tau$ emission }
The three relevant amplitudes are repeated here again
\begin{align}
\begin{split}
        \mathcal{M}^{\rm step}_{\tau,L}  &= - i g \epsilon_{\tau_2 }^\mu (p+q)_\mu \left( \frac{1}{\Delta p_r } + \frac{r_{\tau, k}}{\Delta p} +  \frac{t_{\tau, k}}{-\Delta \tilde{p}_r} \right) \ ,
\\
   \mathcal{M}^{\rm step}_{\tau,R}  &= - i g  \sqrt{ k^z \over \tilde k^z} \epsilon_{\tau_2 }^\mu (p+q)_\mu \left[ \frac{\tilde k^z}{k^z} \frac{t_k}{\Delta p} -   \frac{1}{\Delta \tilde{p}} + \frac{r_k}{\Delta \tilde p_r}\right] \ ,
\\
    \mathcal{M}^{\rm wkb}_{\tau}  &= - i g \epsilon_{\tau_2 }^\mu (p+q)_\mu \left( \frac{1}{\Delta p_r } + \frac{1}{-\Delta \tilde{p}} \right) \ .
\end{split}
\end{align}
For $\tau$ polarisation we can treat both types of transitions of interest at the same time by interpreting the $m,\mt$ correctly.

\subsection{$m\neq 0$ asymptotic $p^0$ regime}

We present now the dominant contribution  $\mathcal{M}^{\rm step}_{\tau,L}$ in detail 
\footnote{Similar asymptotic expressions can be derived for all contributions but we do not believe it useful to fill the paper with multiple complicated formulae. We deem the total fitted formulae in the main text of more practical use given the overall uncertainties in the physics of FOPTs in the early universe.}.   The amplitude squared is
\begin{align}
\left|\mathcal{M}^{\rm step}_{\tau,L}\right|^2 &= \frac{16 g^2 k_z^2 k_\perp^2 p_0^2}{(k_z^2 + m^2) (k_z^2 - (p^z - q^z)^2)^2}  
\begin{cases}
    \frac{\Delta m^2}{\Delta m^2 - k_z^2 + (p^z - q^z)^2} \ , & k^z < \Delta m\\
    \frac{(k^z - \kt^z)^2}{(\kt^z + p^z - q^z)^2} \ , & k^z > \Delta m 
\end{cases} \ ,
\end{align}
where, as usual, we make a distinction between two branches distinguished by $\kt$ being imaginary or real.
We can explicitly evaluate the leading term in the large $p^0$ expansion \cref{eq:p0Taylor} for the total momentum exchanged. We report only the first, dominant, branch
\begin{align}
\label{eq:DeltaPtauLStep}
    \langle {\Delta p_{L}^{\tau, \rm step}} \rangle \longrightarrow {g^2 \mt \over 8 \pi^2} F^{\rm step}_{\tau,L}(r \equiv \mt/m) \ , \qquad k^z < \Delta m \quad \text{branch} \ , 
\end{align}
where the dimensionless function defined is
\begin{align}
&F^{\rm step}_{\tau,L}(r) =
    \frac{1}{ (  r^2-1)} \left\{  2 \left(\pi (1 + r)^2 - (1 + r^2) \csc^{-1}[r]\right) \ln[r - \sqrt{r^2-1}] - 2 \tan^{-1}[\sqrt{r^2-1}] \right. \nonumber \\
  & - 4(1 + r)^2\left(G - \Im  {\rm Li_2}\left[ i (r+\sqrt{r^2-1}) \right]\right) - 2 (r^2-1 ) (  r-1 + \sqrt{r^2-1}) + 2 \pi r \cosh^{-1}[r]   \nonumber \\
  &\left. - r^2 \left(8 \sqrt{ r^2-1} \coth^{-1}\left[1 - \frac{2r}{\sqrt{r^2-1}}\right] + \tan^{-1}[\sqrt{r^2-1}] ( 4 \sqrt{r^2-1} + 4 \log[r]-2)\right)
  \right\} \nonumber \ .
\end{align}
$G\approx 0.916$ is Catalan's constant and $\Im$ stands for imaginary part. We caution the reader that, while the analytical evaluation of integrals is a fun endeavour, we believe that, at least at the present time, the numerical fits presented in the main text are more useful.

\subsection{The $m\rightarrow0$ regime}
\label{sec: Analytics Tau m=0}
The inter-relativistic regime of growth $\propto \log \left(p_0 \right)$ discussed in the main text when $m/m_\psi \ll 1$ can be found analytically by simply setting $m=0$, keeping $m_\psi \neq 0$, and using the collinear expansion \cref{eq:collinear} to evaluate the phase space integrals. We obtain
\bea 
\label{eq:vecmomT}
\vev{{\Delta p_{L}^{\tau, \rm step}}}\simeq&&\dfrac{  {g^2}\tilde m}{2 \pi^2} \left( 1+2\ln\left( \dfrac{p_0 }{ m_\psi}\right) \right),\\
\vev{\Delta p_{R}^{\tau, \rm step}}\simeq&& \dfrac{  {g^2}\tilde m}{1800 \pi^2}\left[ 526-900 G-195\pi+480 \ln \left( \dfrac{\sqrt{2}p_0}{m_\psi} \right) \right]\nn
&& +\dfrac{  {g^2}\tilde m^2 L_w}{8\pi^2}\left[3- 2 \ln \left( \dfrac{\tilde m L_w p_0}{m_\psi}\right) \right] + \mathcal{O}\left((\tilde m L)^2\right) ,\\
 \vev{\Delta p_{R}^{\tau, \rm wkb} }\simeq&& {  {g^2}\tilde m^2 L_w \over 8 \pi^2}\left[ 2\ln\left( \tilde m L_w p_0 \over m_\psi \right) - 3\right],
\eea 
where $G\approx 0.916$ is Catalan's constant. The total contribution is dominated by $L$ emission and then scales like
\bea 
\vev{\Delta p^{\tau, \rm \, tot}}\simeq\vev{\Delta p^{\tau, \rm \, step}_L}\simeq \dfrac{  {g^2}\tilde m}{2 \pi^2} \left( 1+2\ln\left( \dfrac{p_0 }{ m_\psi}\right) \right).
\eea
  {We remind the reader that these are the results for the single vector emission, to the expression above cannot be trusted for the large values of the log.
}
\begin{figure}
    \centering
    \includegraphics[width=.6\textwidth]{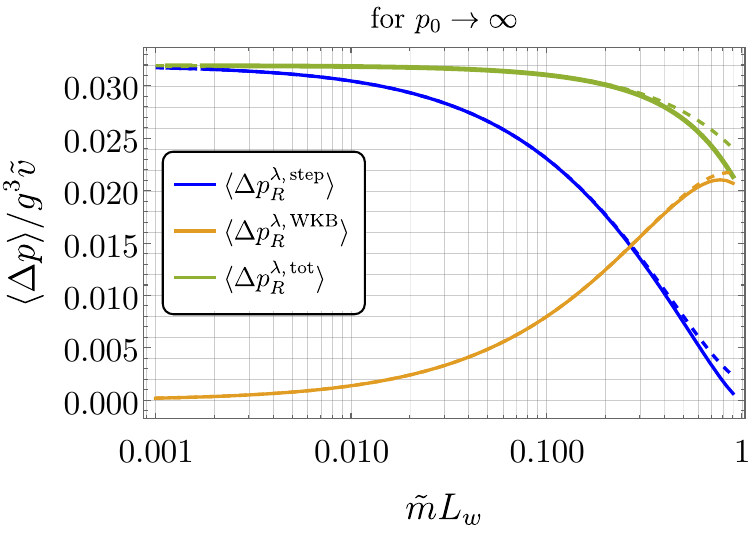}
    \caption{Analytically derived exact formulae for the exchanged momentum in the asymptotic limit $p^0\rightarrow \infty$ from the emission of $\lambda$ polarisations. Dashed lines correspond to the small $\mt L_w$ approximations reported in  \cref{eq:DeltaPlambdaStep}. The sum of the two different contributions is roughly constant.}
    \label{fig:lambdaContributionsAsFunctionOFmtL}
\end{figure}

\subsection{Vectors: $\lambda$ emission} 
We report again the relevant amplitudes here with general $v\neq 0$ :
\begin{align}
     \mathcal{M}^{\rm step}_{\lambda,L} &= - i  
     \frac{g}{E} (p^z+q^z) \left[ g v \left( \frac{1}{\Delta p_r} + \frac{r^\lambda_k}{\Delta p}\right) -
     g \tilde{v} \frac{t^\lambda_k}{\Delta \tilde{p}_r} \right] \ ,
\\
   \mathcal{M}^{\rm step}_{\lambda,R}  &= - i  \frac{g}{E}(p^z+q^z) \sqrt{ k^z \over \tilde k^z} \left[ g v \frac{\tilde k^z}{k^z} \frac{t^\lambda_k}{\Delta p} - g \tilde{v} \left( \frac{1}{\Delta \tilde{p}} - \frac{r^\lambda_k}{\Delta \tilde p_r}\right)\right] \ ,
\\
    \mathcal{M}^{\rm wkb\ red.}_{\lambda} &= - i  \frac{g }{E} (p^z+q^z) \left[{ g v\over   {\Delta p} }-{ g\tilde v\over \Delta\tilde{p} } \right] \ .
\end{align}
Unlike the case of $\tau$ we must consider the two transitions of interest separately.
\paragraph{Symmetric to Broken ($v=0$):}
The symmetric side mass of the $\lambda$ degree of freedom is $m_{h,s}$ $-$  the mass of the Higgs in that phase. We find however that taking $m_{h,s}=0$ incurs an error which is at most $\sim 10 \%$ when $m_{h,s}\sim {\rm few} \: \times \mt$.
One can rigorously prove that in the limit $p^0\rightarrow \infty$ we have
\begin{align}
\label{eq:DeltaPlambdaStep}
\langle \Delta p_{R}^{\lambda, \rm step} \rangle \longrightarrow& \quad {g^4 \tilde v^2 \over 8\pi^2 \mt}\; F^{\rm step}_{\lambda,R}(\mt L_w)   \\
=& \quad {g^4 \tilde v^2\over 8\pi^2 \mt }\left[{2+4 G - \pi} + \mt L_w \left( 2\ln(\tilde m L_w)-2\ln(2) -1 \right)\right] + \mathcal{O}\left((\tilde m L_w)^3\right)\ ,
 \nonumber\\
 \label{eq:DeltaPlambdaWKB}
    \langle \Delta p_{R}^{\lambda, \rm wkb} \rangle \longrightarrow& \quad {g^4 \tilde v^2 \over 8\pi^2 \mt}\left[ \tan^{-1}\left(\tilde m L_w \over \sqrt{1-\tilde m^2L_w^2}\right) - \tilde m L_w \ln\left( 1- \sqrt{1-\tilde m^2 L_w^2} \right)\right] \ ,
\end{align}
where $G\approx 0.916$ is Catalan's constant. $F^{\rm step}_{\lambda,R}$ has a closed form in terms of (hyperbolic) trig functions and the dilogarithm, but we have deemed it more useful to explicitly report only its small $\mt L_w$ limit - an excellent approximation, as can be seen in \cref{fig:lambdaContributionsAsFunctionOFmtL}. There we also see that the sum of the two contributions is roughly constant, giving
\begin{align}
   \langle \Delta p^{\lambda, \rm total} \rangle \approx \langle\Delta p_{R}^{\lambda, \rm step}\rangle + \langle \Delta p_{R}^{\lambda, \rm wkb}\rangle \simeq {  {g^4} v^2 \over 8 \pi^2 \mt} \left[2 + 4 G - \pi - \mt L_w \ln(2)\right] \ ,
   \label{eq:clambdaAnalytic}
\end{align}
to leading orders in $\mt L_w$. A more accurate expression when $\mt L_w \sim 1$ is \cref{eq:DeltaPlambdaWKB}.

\section{Pressure in the EW phase transition}

In this work, we computed the pressure in the context of an Abelian toy model, where the $U(1)$ gauge boson was emitted by a complex scalar. 
Emission from fermions will not significantly change our result per degree of freedom. Moreover, the emission can be straightforwardly expanded to the non-Abelian case. 

For the SM case, the pressure at $1\to 2$ level originates from the vertices inducing $\psi \to Z \psi$ (where the first particle in the final state is soft and $\psi$ is some fermion of the SM), $\psi \to W^{\pm} \psi$ for the gauge bosons emitted from fermions $H \to  W^{\pm} H$, $H \to  Z H$ for the gauge bosons emitted from the Higgs and $A \to WW, W \to WA$, $Z \to WW, W \to WZ, W \to ZW$ for gauge bosons emitting gauge bosons. The careful counting of all the processes involved in the pressure was presented in Appendix C of \cite{Azatov:2022tii}.  
This leads to a final pressure 
of the form 
\begin{align}
\mathcal{P}_{\text{SM}, 1 \to 2}
\approx &157 \times  \alpha_{\rm em} M_Z \frac{ 4 \gamma_w \zeta (3) T^3}{\pi} \bigg[ 0.135 \log\left(\frac{\tilde v }{ T} + 2.26 \right) \nonumber
\\
   & - 0.085
    - 0.2 \frac{\log \left(\frac{\tilde v}{T} + 2.26 \right)}{\tilde v/T}+\frac{0.19}{\tilde v/T}  \bigg] \ ,
\end{align}
where every contribution is normalised to the fine structure constant $\alpha_{\rm em}$ and the $Z$ boson mass $M_Z$ in the broken phase. 
We caution the reader that this expression is still an estimate and may incur future revision, for example, from a better understanding of finite temperature corrections -  see \cref{sec:Phase_space_vector}.

\section{Sensitivity to wall width}
\label{sec:L dependence}
\label{app:cut_off}
In this work we separated the phase space of particles according to when the step wall and WKB approximations are justified 
\bit
\item
$k^z<L_w^{-1}$,~~~ {Step wall } 
\item $k^z> L_w^{-1}$,~~~ WKB  
\eit
where $L_w$ is the width of the wall. However, this is a somewhat arbitrary quantity, significant up to some order 1 factor. Although we have discussed the sensitivity of our results to $L_w$ in several places, we summarise it in this dedicated appendix. 
 
\begin{figure}
  \centering
    \includegraphics[width=.48\textwidth]{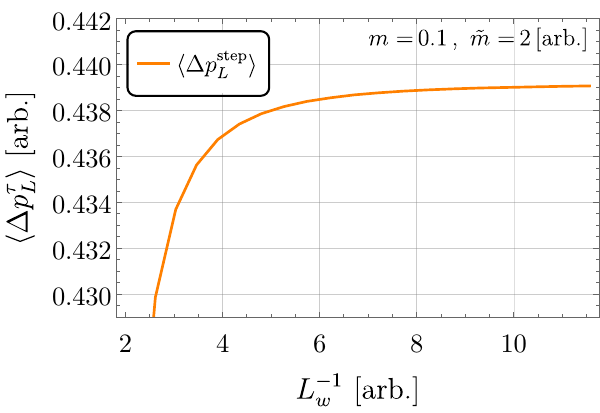}
   \includegraphics[width=.48\textwidth]{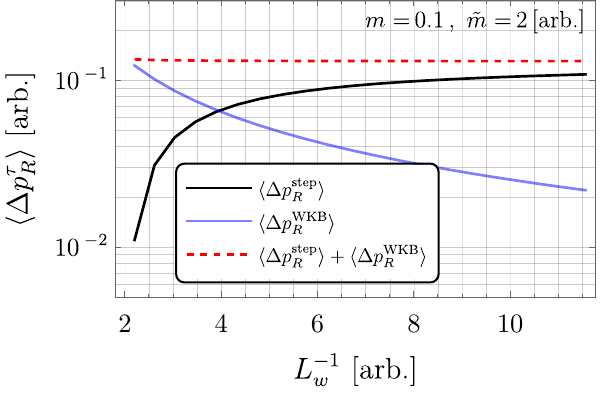}\\
     \includegraphics[width=.465\textwidth]{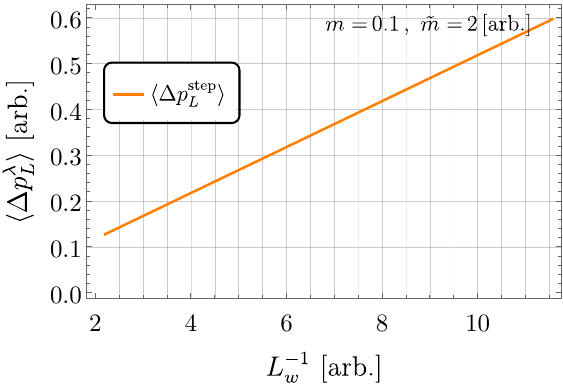}
   \includegraphics[width=.48\textwidth]{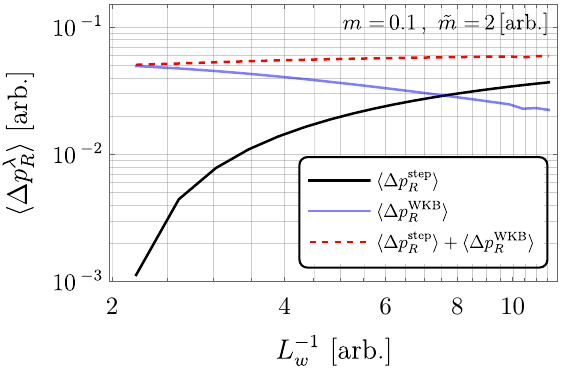}
    \caption{
    Dependence of various contributions to average momentum transfer, due to the emission of a vector boson, on the inverse wall width $L_w^{-1}$ in the limit $p_0 \to \infty$.  Top (Bottom) panels refer to $\tau$ ($\lambda$) emission, while left (right) column refers to $L$ ($R$) emission.
    In this limit, any dependence on the mass of the emitter ($m_\psi$) vanishes.  }
    \label{fig:cutoff}
    \end{figure}

We study numerically how pressure changes with wall width. For scalars, the dominant contribution comes from the WKB regime and dependence on $L_w$ vanishes, as can be seen in explicitly by the analytical formulae   \cref{eq:WKBScalars,eq:ScalarL step}.
The results for vector emission are highlighted in \cref{fig:cutoff}. The left and right columns show left and right mover emission respectively, while top and lower panels correspond to $\tau$ and $\lambda$ polarisations. 
The first thing to notice is that $\vev{\Delta p_{L}^{\tau, \,\rm step}}$, which is the dominant contribution for super-cooled symmetric to broken transitions is largely insensitive to $L_w$. Secondly, $\vev{\Delta p_{R}^{\lambda, \,\rm step}}$ and $\vev{\Delta p_{R}^{\lambda,\,\rm wkb}}$ show almost linear dependence around $L_w \sim {\rm few} \; \mt$ but their \textit{sum} is largely constant. This point was made also in \cref{fig:lambdaContributionsAsFunctionOFmtL} using analytical asymptotic formulae.
Instead, $\vev{\Delta p_{L}^{\lambda,\,\rm step}}$, which exists only in broken to broken transitions is linearly dependent on the cut-off (see eq.~(\ref{eq:vnonzerolambda})).



\section{The suppressed region $\Delta p_z L_w \gg 1$: the Fourier constraint}
\label{app:suppression}

In section \ref{sec:emission_energ} and afterward, we stated that in WKB approximation the region $\Delta p_z L_w \gg 1$  should have a very suppressed contribution to the pressure. In this appendix, we bring some arguments to this claim (see also Appendix B.1 of \cite{Gouttenoire:2021kjv} and Section V of \cite{Bodeker:2017cim} for previous discussion). 
The function that we have to study is typically the following integral 
\bea 
\mathcal{M} \approx \int^{\infty}_{- \infty} dz V(z) e^{ i \int^z_{-\infty} \Delta p_z (z') dz' }\, ,
\eea 
where  far from the wall both $\Delta p(z) $ and $V(z)$ are constant. In this case, we can always absorb $V(z)=V_0\exp[\int_0^z (V'/V) dz]$ inside the exponent (redefining $\Delta p$) and thus we can focus only on the integrals where $V(z)= V_0$
\bea 
\mathcal{M} \approx V_0 \int^{\infty}_{- \infty} dz e^{ i \int^z_{0} \Delta p_z (z') dz' }\,.
\eea 
In general, these integrals must be evaluated numerically for various wall shapes. However, for particular choices like
\bea 
\label{eq:choice_tanh}
\Delta p_z(z) = \Delta p_z + \frac{\epsilon}{2} \tanh{z/L_w}, 
\eea 
we can evaluate the integrals analytically. In \cref{eq:choice_tanh}, $\epsilon$ parameterises the change of the phase across the wall. This leads to 
\begin{align}
\mathcal{M}&=V_0\int^{\infty}_{- \infty} dz \exp \l[i\Delta p_z z+ i \epsilon f(z)\r], \nn
\\
f(z) &= \frac{ L_w}{2} \log \cosh{\frac{z}{L_w}} \, ,
\end{align}
and give finally 
\begin{align}
|\mathcal{M}|^2 &= \frac{\pi \epsilon L_w |V_0|^2}{2\Delta p_z^2 - \epsilon^2/2}\times  \frac{\sinh{(\pi \epsilon L_w/2)}}{\sinh{(\pi (\Delta p_z - \epsilon/2 )L_w/2)}\sinh{(\pi (\Delta p_z + \epsilon/2 )L_w/2)}}\nn
\\
& \approx   \frac{\pi \epsilon L_w |V_0|^2}{2\Delta p_z^2}\times  \frac{\sinh{(\pi \epsilon L_w/2)}}{\sinh^2{(\pi \Delta p_z L_w/2)}}\, .
\end{align}
For $\epsilon L_w \lesssim 1, \Delta p_z L_w \gg 1$, we obtain 
\bea 
|\mathcal{M}|^2 \approx \frac{\pi \epsilon L_w |V_0|^2}{4\Delta p_z^2}\times  \pi \epsilon L_w e^{-\pi \Delta p_z L_w} = \frac{(\pi \epsilon L_w)^2 |V_0|^2}{4\Delta p_z^2}\times  e^{-\pi \Delta p_z L_w}
\eea 
which shows that $|\mathcal{M}|^2 \propto e^{-\pi \Delta p_z L_w}$ and then decay exponentially with $\Delta p_z L_w \gg 1$. We have also checked numerically the behaviour of the amplitude for other wall shapes, with a similar behaviour (for example Erf function) and always find exponential suppression.

\bibliographystyle{JHEP}
{\footnotesize
\bibliography{biblio}}

\end{document}